\documentclass[11pt]{article}

\usepackage[english]{babel}
\usepackage{graphicx}
\usepackage{float}
\usepackage{lmodern}
\usepackage{booktabs}
\usepackage{amsmath,amsfonts,bbm,bm,commath,mathtools,relsize,exscale}
\usepackage{dsfont}
\usepackage{nicefrac}       
\usepackage{makecell}
\usepackage{tabularx}
\usepackage{microtype}      
\usepackage[OT1]{fontenc}
\usepackage[utf8]{inputenc}
\usepackage[inline,shortlabels]{enumitem}
\usepackage{caption}
\usepackage{subcaption}
\usepackage[flushleft]{threeparttable}
\usepackage{algorithm}
\usepackage{algpseudocode}
\usepackage{soul}
\usepackage{adjustbox}
\usepackage{fancyhdr}
\usepackage{array}
\usepackage{arydshln}

\graphicspath{ {figs/} }

\usepackage{refcount,nameref,zref-xr,zref-user} 
\zxrsetup{toltxlabel} 
\usepackage{url}

\let\href\undefined
\usepackage[colorlinks,citecolor=blue,urlcolor=blue]{hyperref}

\usepackage{pgf,tikz}
\usetikzlibrary{shapes,decorations,arrows,calc,arrows.meta,fit,positioning}
\tikzset{
    -Latex,auto,node distance =1 cm and 1 cm,semithick,
    state/.style ={ellipse, draw, minimum width = 0.7 cm},
    point/.style = {circle, draw, inner sep=0.04cm,fill,node contents={}},
    bidirected/.style={Latex-Latex,dashed},
    el/.style = {inner sep=2pt, align=left, sloped}
}
\usetikzlibrary{arrows,shapes.arrows,shapes.geometric,shapes.multipart,decorations.pathmorphing,positioning,swigs}

\makeatletter%
\@ifclassloaded{biometrika}{%

  \AtEndDocument{\printhistory}
  
  \def\rm{}

  \jname{Biometrika}

  \markboth{JIANG ET AL.}{Causal mediation analysis with informatively
  left-censored mediators}

  \newcommand{\authorlist}{
      \author{C.~JIANG}
      \affil{Department of Biostatistics
        Harvard Chan School of Public Health\\
        665 Huntington Avenue, Boston, MA 02115
      \email{cjiang@hsph.harvard.edu}}

      \author{M.D.~HUGHES}
      \affil{Department of Biostatistics
        Harvard Chan School of Public Health\\
        665 Huntington Avenue, Boston, MA 02115
      \email{mhughes@sdac.harvard.edu}}

      \author{N.S.~HEJAZI}
      \affil{Department of Biostatistics
        Harvard Chan School of Public Health\\
        665 Huntington Avenue, Boston, MA 02115
      \email{nhejazi@hsph.harvard.edu}}

  }
}{%
  \usepackage{arxiv}
  \usepackage{standalone}
  \usepackage{multirow}
  \usepackage{setspace}
  \usepackage{amsthm}
  \usepackage[round]{natbib}

  \newtheorem{theorem}{Theorem}

  {\theoremstyle{definition}\newtheorem{assumption}{}}
  {\theoremstyle{definition}}
  {\theoremstyle{definition}}

\newcommand{\authorlist}{
    Cong Jiang \\
    Department of Biostatistics\\
    Harvard T.H.~Chan School of Public Health, and\\
    Department of Epidemiology and Biostatistics\\
    SUNY Downstate Health Sciences University\\
    \texttt{Cong.Jiang@downstate.edu}
    \And
    Michael D.~Hughes \\
    Department of Biostatistics, and\\
    Center for Biostatistics in AIDS Research\\
    Harvard T.H. Chan School of Public Health\\
    \texttt{mhughes@sdac.harvard.edu}
    \And
    Nima S.~Hejazi \\
    Department of Biostatistics\\
    Harvard T.H. Chan School of Public Health\\
    \texttt{nhejazi@hsph.harvard.edu}
    }
}
\makeatother

\newcommand{\titlepaper}{
Evaluating causal indirect effects when mediators are left-censored by assay
limit of quantification
}

\AtEndDocument{\refstepcounter{theorem}\label{finalthm}}
\AtEndDocument{\refstepcounter{equation}\label{finaleq}}
\AtEndDocument{\refstepcounter{lemma}\label{finallemma}}

\newcommand{\E}{\mathbb{E}}

\newcommand{\M}{\mathcal{P}}
\newcommand{\R}{\mathbb{R}}

\renewcommand{\P}{\mathbb{P}}

\newcommand{\Pf}{\mathsf{P}}

\newcommand{\indep}{\mbox{$\perp\!\!\!\perp$}}

\usepackage{xcolor}

\newcommand{\biostats}[1]{\textcolor{black}{#1}}

\zexternaldocument*[supp:]{sm}
\makeatletter
\@ifclassloaded{biometrika}{%
  \authorlist
}{%
  \author{\authorlist}
  \date{\today}
}
\makeatother
\title{\titlepaper}

\begin{document}
\maketitle

\begin{abstract}
  Causal mediation analysis is essential for disentangling the mechanisms by
  which investigational therapeutic and preventive agents impact clinical
  outcomes. However, the measurement of biological mediators is often subject
  to left-censoring by technical measurement limitations, most commonly an
  assay's limit of quantification. This form of censoring can pose severe
  challenges for both identification and estimation of causal mediation
  estimands, particularly when the censoring mechanism is deterministic and the
  resulting missingness is missing not at random (MNAR) or nonignorable.
  Motivated by the question of assessing the role of viral RNA in the action
  mechanism of monoclonal antibody therapies for COVID-19 in the Accelerating
  COVID-19 Therapeutics and Vaccine (ACTIV)-2 platform trial, we develop a
  semi-parametric framework for estimation of the natural direct and indirect
  effects when the mediator of interest is partially subject to this form of
  left-censoring. Our proposed strategy combines fractional imputation with a
  semi-parametric EM algorithm to flexibly estimate key components of the
  factorized data likelihood. Applying the proposed strategy to circumvent the
  left-censoring, we discuss both traditional plug-in and asymptotically
  efficient estimators of the direct and indirect effect estimands, introducing
  a data-adaptive $m$-out-of-$n$ bootstrap for robust inference under the
  imputation procedure. We demonstrate in numerical experiments that our
  approach significantly reduces bias and allows for reliable inference. An
  application to data from the ACTIV-2 platform trial confirms that monoclonal
  antibody therapies reduce the risk of hospitalization and death due to
  COVID-19, while suggesting that changes in viral RNA mediate only a modest
  proportion of the overall treatment effect.
\end{abstract}

\section{Introduction}\label{intro}
Disentangling and better understanding \textit{how} an investigational
therapeutic or preventive agent exerts its clinical effect is a central goal in many
biomedical and public health studies. Questions of mechanism are ubiquitous in
the biological sciences, as improved characterizations of specific biological
pathways through which an observed effect may operate can help to confirm the
action mechanism of an investigational agent or reveal new targets for improved
versions of contemporary therapeutic and preventive agents. Investigational agents
can be said to \textit{act indirectly} by their action on some biological
pathway or process, for example, by activating an immune response or promoting
pathogen clearance. Causal mediation analysis provides a formal analytic
framework for quantifying direct and indirect causal effects, identifying how
much of an investigational agents' total effect is mediated through intermediate
variables (mediators). Causal mediation analysis has been the subject of a great
deal of attention recently, with innovations that have opened the door to
handling flexible hypothetical intervention schemes~\citep{diaz2020causal,
hejazi2022nonparametric}, complex forms of post-treatment
confounding~\citep{diaz2020nonparametric, rudolph2023efficient,
rudolph2024practical}, and time-varying
treatment-mediator-confounder feedback in longitudinal
settings~\citep{diaz2023efficient}.

\biostats{
In applied science settings in which putative mediators are measures of
biological processes (e.g., viral RNA), measurement of the mediator may be
subject to technical limitations. For example, when measuring a biological
mediator, an assay's lower limit of quantification (LLoQ) may induce
left-censoring of the mediator measurements, typically resulting in those
values of the candidate mediator that fall below the assay limit being
imprecisely or unreliably measured and thus reported as being below the LLoQ
\citep{hughes2000analysis, keizer2015incorporation}. While such issues do not
impact evaluation of the total effect of an investigational agent, they pose a
significant obstacle to developing a deeper understanding of the underlying
mechanism. Common strategies for handling such issues include imputing
left-censored values with a fixed number (most commonly, LLoQ/2), though these
approaches may introduce bias in downstream estimates \citep{jin2011analysis,
jain2016consequence}. In this work, we propose a general framework for causal
mediation analysis when the mediator is subject to left-censoring due to an
assay's LLoQ. The method places minimal assumptions on the mediator's
distribution and encompasses commonly used substitution rules as special cases;
in particular, the LLoQ/2 approach arises when the proposal distribution is
degenerate, concentrating all mass at a single imputed value.
}

Our work is motivated by the Accelerating COVID-19 Therapeutics and Vaccine
(ACTIV)-2 or AIDS Clinical Trials Group (ACTG) A5401 platform clinical trial,
which evaluated the efficacy of novel investigational agents, including
monoclonal antibody (mAb) therapies for the treatment of non-hospitalized
individuals with symptomatic SARS-CoV-2 infection~\citep{evering2023safety,
giganti2025implementation}. The therapeutic action mechanism of mAb agents is
thought to involve neutralization of SARS-CoV-2 and  clearance of circulating
viral RNA \citep{dougan2021bamlanivimab, taylor2021neutralizing}. A central
mechanistic question therefore concerns the extent to which mAb agents reduce
the risk of severe clinical outcomes by accelerating clearance of viral RNA.
Within ACTIV-2, we focus on evaluating the indirect effect of mAb agents
mediated through viral RNA clearance, measured in anterior nasal swabs
(self-administered daily by study participants), on the primary clinical
endpoint, defined as hospitalization or death through day 28
post-randomization.

Quantification of viral RNA is subject to left-censoring at the assay lower
limit of quantification, in part because nasal swabs typically collect
substantially less biological material than deeper nasopharyngeal
swabs~\citep{moser2023statistical}. Importantly, values below the LLoQ are not
merely a statistical nuisance arising from measurement error; rather, in many
infectious disease settings, including the present study, they often reflect
biologically meaningful states such as recovery or effective viral clearance.
As a result, the distribution of viral RNA may exhibit a complex structure,
combining a continuous component for detectable viral loads with a discrete
mass at or near zero for undetectable values. This phenomenon is well
recognized in other scientific domains, notably environmental and chemical
exposure studies, where concentrations below detection limits similarly arise
and require specialized methodological treatment~\citep{hornung1990estimation,
susmann2025non}. Consequently, correctly specifying the mediator density is
challenging in practice, motivating the development of assumption-lean,
data-driven approaches for the causal mediation estimation and inference that
explicitly accommodate left-censoring at the LLoQ.

A few recent advances in causal mediation analysis have considered challenges
that arise from the mediator being subject to left-censoring or missingness. In
particular, \cite{chernofsky2024causal} proposed a semi-parametric estimation
strategy for the organic indirect effect~\citep{lok2021causal} for candidate
biological mediators constrained by an assay's lower limit of quantification.
Their two methods, which respectively leverage extrapolation and numerical
integration of the observed data likelihood, improved upon conventional
imputation approaches and demonstrated practical utility when applied to data
from an HIV study. In a contemporaneous contribution, \cite{zuo2024mediation}
explored identifiability of natural direct and indirect effects when both the
mediator and the outcome may be missing not at random (MNAR), introducing a
completeness condition that is necessary for non-parametric identification.
These authors investigated identifiability under interpretable MNAR mechanisms
and developed inference procedures tailored to such forms of censoring,
motivated by an application to the National Job Corps Study. In this work, we
propose a new framework for performing causal mediation analysis when mediator
values are LLoQ left-censored. To introduce our approach, we review
identification of the natural direct and indirect
effects~\citep{robins1992identifiability, pearl2001direct}, with a particular
focus on the failure of non-parametric identification of these estimands under
LLoQ left-censoring. To address this challenge, we impose a parametric
structure that renders the natural direct and indirect effects identifiable
from the observed data. Concerning estimation, given the continuous nature of
our mediator in the ACTIV-2 study, we develop a strategy that adopts fractional
imputation~\citep{kalton1984some,kim2011parametric}, which, when combined with
an expectation-maximization (EM) algorithm, allows for the construction of
estimators of the direct and indirect effect estimands of interest. Our
imputation strategy introduces a degree of non-regularity, due to the
dependency structure of the imputed data, for which we propose a data-adaptive
$m$-out-of-$n$ bootstrap to construct confidence intervals that remain valid
for both traditional plug-in and efficient one-step estimators.


\textbf{Contributions:} We develop a unified framework for causal mediation
analysis in settings in which a continuous mediator is subject to censoring, as
in biomarker studies with lower limits of quantification. First, under the
missing-not-at-random mechanism and violation of \citet{zuo2024mediation}'s
completeness condition, we establish identification in a parametric framework
that ensures the natural direct and indirect effects are uniquely determined
from the (incomplete) observed data. Second, we propose an estimation strategy
that integrates fractional imputation~\citep{kalton1984some, kim2011parametric}
with an EM algorithm, enabling the use of maximum likelihood to repair the
incomplete data in such a way that the causal mediation estimands of interest
can be learned from the data. Unlike fully parametric approaches, our method
allows for flexible, semi-parametric modeling of the conditional mediator
density, thus decoupling estimation strategies from identification assumptions.
Finally, we make a methodological contribution to causal mediation inference.
We develop valid large-sample inference procedures for both conventional
plug-in estimators and asymptotically efficient one-step estimators of
mediation effects. To address non-regularity introduced by the imputation step,
we propose a robust, data-adaptive $m$-out-of-$n$ bootstrap. For the one-step
estimator, we further design a computationally efficient multiplier bootstrap
that remains asymptotically valid despite substantially reduced computational
cost.

\biostats{Beyond targeting a distinct causal mediation estimand, necessitating
a different identification strategy, our approach provides an alternative
estimation strategy to the likelihood-based approach of
\cite{chernofsky2024causal}. Their numerical integration approach directly
evaluates and optimizes a parametrically specified observed-data likelihood.
Under common parametric specifications of the mediator and outcome models, our
fractional-imputation EM procedure instead approximates the conditional
expectations associated with the corresponding observed-data likelihood using
importance-weighted candidate mediator values; as the number of imputations
increases, the resulting estimator converges to the corresponding maximum
likelihood solution. This formulation is particularly useful because it
accommodates flexible semi-parametric estimation of the conditional mediator
density, thereby reducing reliance on restrictive parametric specifications.
Our approach is also related to earlier Monte Carlo EM methods for censored
data, such as that of \cite{hughes1999mixed} but differs computationally in
that candidate mediator values are generated once and retained across EM
iterations, with only their fractional weights updated. This avoids repeated
Monte Carlo sampling while providing a natural framework for estimation of the
causal mediation parameters of interest.}


The remainder of this manuscript is organized as follows. In
Section~\ref{sec:back}, we present our
identification framework, stressing how the setting with LLoQ left-censored
mediators differs from the standard setting and limits opportunities for
non-parametric identification. Then, section~\ref{sec:meth}
introduces our proposed methodology for
estimating the direct and indirect effects under LLoQ censoring of the
mediator, including a fractional imputation procedure that incorporates
flexible semi-parametric estimation of the conditional mediator density and a
data-driven $m$-out-of-$n$ bootstrap for valid inference after imputation.
Section~\ref{sec:sim} presents numerical
experiments in which we evaluate the performance of the proposed strategy in
comparison to conventional approaches, indicating favorable performance of our
proposal. In
Section~\ref{sec:realdata},
we apply our proposed approach to estimate the degree to which the effects of
monoclonal antibody agents are mediated by viral RNA clearance in the ACTIV-2
platform trial. We conclude by addressing limitations and outlining directions
for future investigation in Section~\ref{sec:disc}.

\section{Identification framework}\label{sec:back}
We begin by reviewing identification frameworks for causal mediation analysis
without censored mediators, including key assumptions necessary for
identification of the natural direct and indirect effects. We build on this
discussion in Section~\ref{sec:realdata}
by applying the interventionist framework~\citep{robins2022interventionist},
which we adopt within the framing of the ACTIV-2 platform trial, our applied
science case-study. We then extend our discussion of identification to address
how the assumptions must be adapted in the presence of left-censored
mediators.

\subsection{Identification without censored mediators}

Let $\boldsymbol{L}$ denote a vector of observed baseline covariates (i.e.,
potential confounders), $A \in \{0, 1\}$ a binary treatment or exposure, $M \in
\R$ the candidate mediator, and $Y$ the outcome of interest. We assume
that the data unit $O = (\boldsymbol{L}, A, M, Y) \sim \Pf$ is drawn from a
distribution $\Pf \in \M$, with the non-parametric statistical model $\M$
placing no restrictions on the form of $O$. The available data $O_1, \ldots,
O_n$ consists of $n$ iid copies of the data unit $O$. Using the potential
outcomes framework~\citep{imbens2015causal}, let $M(a)$ and $Y(a)$ denote the
potential (or counterfactual) mediator and outcome under hypothetical
assignment to treatment $A = a$. Further, define $Y(a, m)$ as the potential
outcome under a joint intervention assigning treatment $A = a$ and mediator $M
= m$. Noting that $Y(a) = Y(a, M(a))$ by the composition
assumption~\citep{vanderweele2015explanation}, the average treatment effect
(ATE), defined $\theta_{\text{ATE}} = \E[Y(A = 1) - Y(A = 0)]$, can be
re-expressed $\theta_{\text{ATE}} = \E[Y(A = 1, M = M(1)) - Y(A = 0, M =
M(0))]$, which suggests a decomposition into the natural direct and indirect
effects~\citep{pearl2001direct, pearl2022direct}:
\begin{align*}
  \theta_{\text{ATE}} = \underbrace{\E\left[Y(A = 1) - Y(A = 0)
    \right]}_{\text{average treatment effect (ATE)}}
    =&~~\E\left[Y(A = 1, M = M(1)) - Y(A = 0, M = M(0))\right] \\
    =&~~\underbrace{\E\left[Y(A = 1, M = M(1)) - Y(A = 1, M = M(0))
      \right]}_{\text{natural indirect effect (NIE)}} \\
    &+ \underbrace{\E\left[Y(A = 1, M = M(0)) - Y(A = 0, M = M(0))
      \right]}_{\text{natural direct effect (NDE) }} \\
    \coloneqq&~~\theta_{\text{NIE}} + \theta_{\text{NDE}} \ .
\end{align*}
The definitions of the natural direct effect (NDE) and natural indirect effect
(NIE) rely on a cross-world (or nested) counterfactual of the form $Y(A = 1,
M = M(0))$, which involves two mutually incompatible interventions: assigning
treatment $A = 1$ along the non-mediated (direct) pathway and $A = 0$ along the
mediated (indirect) pathway, the latter of which leads to $M = M(0)$. The
counterfactual $Y(A = 1, M = M(0))$ may be interpreted as the outcome under
treatment $A = 1$, had the mediator taken the value it would naturally have
taken under control $A = 0$; thus, the treatment and mediator are realized in
different, mutually incompatible ``counterfactual worlds,'' leading to the
labeling of this counterfactual quantity as
``cross-world''~\citep{andrews2020insights}.

Absent censoring of putative mediators, the natural direct and
indirect effects may be identified via either the structural causal
modeling~\citep{pearl2001direct, pearl2022direct} or potential
outcomes~\citep{robins1992identifiability} frameworks. We will review standard
identification shortly but note in advance that in
Section~\ref{sec:realdata},
where we review our case study in the ACTIV-2 platform trial, we further adopt the
recently developed \textit{interventionist} framework
of~\cite{robins2022interventionist}, which replaces the much-critiqued and
untestable cross-world counterfactual independence assumption with one of
treatment component separability, enabling identification through
experimentally testable, scientifically interpretable contrasts. The standard
approach to identification of the NDE and NIE requires several untestable
assumptions~\citep{pearl2001direct, imai2010general, imai2010identification},
which we summarize below.
\begin{enumerate}
  \item Consistency: The observed mediator and outcome values correspond to
    their counterfactual counterparts under assignment of the appropriate
    treatment regimen; i.e., for (a) \textit{mediators:} $A = a \implies M =
    M(a)$; and (b) \textit{outcome:} $(A, M) = (a, m) \implies Y = Y(a, m);$
    for $a \in \{0, 1\}$ and $m \in \R$.
  \item Positivity: All treatment and mediator levels have non-zero probability
    of occurring within strata of the confounders; i.e., of (a)
    \textit{treatment:} $\mathbb{P}(A = a \mid \boldsymbol{L}) > 0$ with
    probability one; and (b) \textit{mediators:} $\mathbb{P}(M = m\mid A = 1,
    \boldsymbol{L}) > 0$ with probability one only whenever $\mathbb{P}(M = m
    \mid A = 0, \boldsymbol{L}) > 0$.
  \item Exchangeability: No unmeasured confounding of the
    treatment--outcome, mediator--outcome, and treatment--mediator
    relationships, conditional on putative confounders, which are (a)
    \textit{treatment-outcome exchangeability:} $Y(a, m) \indep A \mid
    \boldsymbol{L}$; (b) \textit{mediator-outcome exchangeability:} $Y(a, m)
    \indep M \mid A, \boldsymbol{L}$; (c) \textit{treatment-mediator
    exchangeability:} $M(a) \indep A \mid \boldsymbol{L}$; and (d)
    \textit{cross-world exchangeability:} $Y(A = 1, M = M(1) = m) \indep M(0)
    \mid A, \boldsymbol{L}$ for all $A \in \{0, 1\}$ and $m \in \R$.
  \item No mediator--outcome confounders affected by treatment; formally, for
    all $a \in \{0, 1\}$ and $m \in \R$,
    $$Y(A = a, M = m) \indep M(1-a) \mid \boldsymbol{L} \ .$$ \label{assumption4}
\end{enumerate}
Under these assumptions, the expected counterfactual outcome under treatment
strategy $A = 1$ and with mediator $M = M(0)$---that is, $\E[Y(A = 1, M =
M(0))]$---can be identified via the g-formula as
\begin{align}\label{medfor}
  \E\left[Y(A = 1, M = M(0))\right]
    &= \int_{\mathcal{L}} \int_{\mathcal{M}} \E(Y \mid A=1, M=m,
      \boldsymbol{L}=\boldsymbol{l}) f_{M \mid A, \boldsymbol{L}}
      (m \mid A=0, \boldsymbol{L}=\boldsymbol{l})
      f_{\boldsymbol{L}}(\boldsymbol{l}) \, dm \, d\boldsymbol{l} \ ,
\end{align}
and thus simplifies for identification of $\E[Y(A = a, M = M(a))]$ to
\begin{align*}
\E[Y(A = a, M = M(a))]
  &= \int_{\mathcal{L}} \int_{\mathcal{M}} \E(Y \mid A=a, M=m,
  \boldsymbol{L}=\boldsymbol{l}) f_{M \mid A, \boldsymbol{L}}
  (m \mid A=a, \boldsymbol{L}=\boldsymbol{l})
  f_{\boldsymbol{L}}(\boldsymbol{l}) \, dm \, d\boldsymbol{l} \ .
\end{align*}
Then, the mediation causal estimand $\theta_{\text{NDE}}$ and $\theta_{\text{NIE}}$ are identified respectively by the following statistical functionals:
\begin{align}\label{medest}
  \Psi_{\text{NDE}}(\Pf) &:= \E_{\boldsymbol{L}} \left[ \E_{M \mid A=0,
    \boldsymbol{L}} \left\{ \E(Y \mid A=1, M, \boldsymbol{L}) - \E(Y \mid A=0,
     M, L) \mid A=0, \boldsymbol{L} \right\} \right] \\
  \Psi_{\text{NIE}}(\Pf) &:= \E_{\boldsymbol{L}} \left[ \E_{M \mid A=1,
    \boldsymbol{L}} \left\{ \E(Y \mid A=1, M, \boldsymbol{L}) \mid A=1,
    \boldsymbol{L} \right\} - \E_{M \mid A=0, \boldsymbol{L}}
    \left\{ \E(Y \mid A=1, M, \boldsymbol{L}) \mid A=0, \boldsymbol{L}
    \right\} \right] \nonumber \ .
\end{align}
Equation~\eqref{medfor}, often referred to as the
\textit{mediation formula}~\citep{pearl2022direct}, provides the statistical
functionals that identify the NDE and NIE under the aforementioned assumptions.
These same functionals are preserved under the alternative identification
strategy of~\cite{robins2022interventionist},  a point we note in Section~\ref{sec:realdata}, with technical details provided in the corresponding Supplementary Material. Among the four sets of identification assumptions outlined above, the most
challenging to justify, and consequently the most heavily critiqued, is that of
cross-world counterfactual exchangeability, i.e., $Y(A = 1, M = M(1) = m)
\indep M(0) \mid A, \boldsymbol{L}$ for all $A \in \{0, 1\}$ and $m \in
\mathbb{R}$. This identification assumption stands out from others because it
supposes the independence of counterfactual terms that arise under distinct,
incompatible interventions and therefore have no empirical
analog~\citep{andrews2020insights}; moreover, it is impossible to enforce
cross-world exchangeability by design, even in randomized
experiments~\citep{diaz2020causal, robins2022interventionist}. Despite these
limitations, the NDE and NIE remain in widespread use as \textit{descriptive},
rather than prescriptive, causal mediation estimands, and we focus on them in
the present work precisely because of their ubiquity in causal mediation
analysis. Our proposed strategy, as highlighted in
Section~\ref{sec:meth}, can be
generalized beyond the NDE and NIE.

\subsection{Identification with left-censored mediators}

We now turn to considering how the observed data differ when the mediator of
interest is subject to a censoring process. Consider a biological
mediator (e.g., viral RNA, neutralizing antibody) measured by an assay with a
known lower limit of quantification (LLoQ), denoted by $\lambda_{\text{LLoQ}}$. The observed data unit is denoted as $O = (\boldsymbol{L}, A, M, Y, C),$ where the mediator $ M = \max(M^{\dagger}, \lambda_{\text{LLoQ}}) $ is subject to left-censoring at the assay’s lower limit of quantification, and $M^{\dagger}$ represents the latent true mediator value prior to censoring. The censoring indicator is defined as $C = \mathbb{I}(M^{\dagger} > \lambda_{\text{LLoQ}}) ,$ indicating whether the latent mediator value exceeds the LLoQ; equivalently, $ M^{\dagger} $ is observed if and only if $ C = 1,$ corresponding to cases where the mediator’s value lies above the assay’s quantification threshold. We choose this definition because it yields $CM = M^{\dagger}$ when $C=1,$ and simplifies the notation throughout.

In the context of our motivating case-study, the left
censoring process induced by the assay's lower quantification limit is both
\textit{missing not at random (MNAR)} and \textit{deterministic}---that is, censoring by LLoQ
induces MNAR (or non-ignorable) left-censoring, where censoring depends on
the underlying value of the mediator $M^{\dagger}$, even after conditioning on observed variables. Formally,
$$
  \P(C = 1 \mid M^{\dagger}, A, \boldsymbol{L}) \neq
  \P(C = 1 \mid A, \boldsymbol{L}),\  \text{or, equivalently,}\ f(M^{\dagger} \mid  A,
  \boldsymbol{L}) \neq f(M^{\dagger} \mid A, \boldsymbol{L}, C = 1) \ ,
$$
indicating that the problem falls outside the standard missing-at-random (MAR)
paradigm. Since, by definition, $C$ is fully determined by $M^{\dagger}$ and the fixed
quantification limit $\lambda_{\text{LLoQ}}$, given either the observed mediator ($M$) or the latent true mediator ($M^{\dagger}$), the censoring
mechanism simplifies as $\P(C = c \mid M) = 1$ or $\P(C = c \mid M^{\dagger}) = 1$ for $c \in \{0, 1\}$. In other
words, censoring occurs solely based on whether the potentially unobserved
value of the mediator exceeds the assay LLoQ. The former property---that
censoring is missing not at random---complicates both identification and estimation of
NDE and NIE; however, the latter property---that the censoring mechanism is
deterministic---can simplify aspects of modeling and estimation, as we will
demonstrate.

In the presence of a LLoQ left-censored mediator, identification of the
NDE and NIE depends, in turn, on the identification of $\P(Y \mid M^{\dagger}, A,
\boldsymbol{L})$ and $f(M^{\dagger} \mid A,\boldsymbol{L})$, or, equivalently,
on the joint distribution $f(Y, M^{\dagger} \mid A, \boldsymbol{L})$ based on
the observed data. By assuming that the censoring indicator $C$ is
conditionally independent of the outcome $Y$, given the true mediator
$M^{\dagger}$, treatment $A$, and covariates $\boldsymbol{L}$---that is, $C
\indep Y \mid (M^{\dagger}, A, \boldsymbol{L})$---we obtain the simplification
$\P(Y \mid M^{\dagger}, A, \boldsymbol{L}) = \P(Y \mid M^{\dagger}, A,
\boldsymbol{L}, C = 1) = \P(Y \mid M, A, \boldsymbol{L}, C = 1)$, specifically,
that the conditional probability of the outcome, given mediators, treatment,
and covariates, $\P(Y \mid M^{\dagger}, A, \boldsymbol{L})$, can be
consistently estimated by complete-case analysis. Notably, in our motivating
case-study, the censoring mechanism is a deterministic function of
$M^{\dagger}$, that is, $C \indep Y \mid M^{\dagger}$, satisfying the
aforementioned assumption. Building on~\cite{zuo2024mediation} (cf.~Theorem 1),
we note that the distribution $f(Y, M^{\dagger} \mid A, \boldsymbol{L})$, or of
$f(M^{\dagger} \mid A, \boldsymbol{L})$, is not identifiable non-parametrically
from the observed data in the case that $M$ is continuous and $Y$ is binary due
to a violation of the \textit{completeness} condition that is necessary for
non-parametric identification. This is exactly the case that shows up in our ACTIV-2 study. Briefly, the completeness assumption of $f(Y, M
\mid A, \boldsymbol{L})$ in $Y$ requires that, conditional on $(A,
\boldsymbol{L})$, any variability in the partly observed $M$ is captured by
variability in $Y$; or, formally,
\begin{assumption}[\textit{Completeness}]
For any square-integrable function $g$ and for any $(a, \boldsymbol{l})$,
$\int g(M) f(Y, M \mid A = a, \boldsymbol{L} = \boldsymbol{l}) d\mu(M) = 0$
almost surely if and only if $g(M) = 0$ almost surely, for $\mu(\cdot)$
an appropriate measure (e.g., Lebesgue).
\end{assumption}

Specifically, to establish identification when $M^{\dagger}$ is MNAR,
\citet{zuo2024mediation} consider that $M^{\dagger}$ and $Y$ are discrete and show that the distribution $f(Y, M^{\dagger} \mid A,
\boldsymbol{L})$ is identifiable if the ratio of the missingness probabilities,
i.e., $\P(R^M = 0 \mid M^{\dagger} = m, A = a, \boldsymbol{L} =
\boldsymbol{l})/ \P(R^M = 1 \mid M^{\dagger} = m, A = a, \boldsymbol{L} =
\boldsymbol{l})$ is identifiable, where $R^M = 1$ if $M^{\dagger}$ is observed
and $R^M = 0$ otherwise. Intuitively, identification of $f(Y, M^{\dagger} \mid A, \boldsymbol{L})$
corresponds to recovering a joint density in which the mediator component is
subject to missingness or censoring, i.e., $f(M^{\dagger}, Y, R^M = 0 \mid A,
\boldsymbol{L})$. In this context, the uniqueness of $f(M^{\dagger}, Y, R^M = 0
\mid A, \boldsymbol{L})$ is equivalent to the uniqueness of solutions to the
missingness ratio integral equation, which in turn requires that $f(Y,
M^{\dagger}, R^M = 1 \mid A = a, \boldsymbol{L} = \boldsymbol{l})$ be complete
in $Y$ for all $(a, \boldsymbol{l})$. This can alternatively be viewed as an
identification problem in which $M^{\dagger}$ is subject to non-ignorable
missingness, with the outcome $Y$, not itself subject to missingness, serving
as a \textit{shadow variable}~\citep{miao2016varieties}. The completeness
assumption in their framework, interpretable as a non-parametric rank
condition, is most easily illustrated when $M^{\dagger}$ and $Y$ are
categorical with $d_{M^{\dagger}}$ and $d_Y$ levels. In this case, completeness
requires $d_Y \geq d_{M^{\dagger}},$ implying that $Y$ must have at least as
many categories as $ M^{\dagger}$. Otherwise, if $Y$ has fewer levels or is a
coarsening of $M^{\dagger},$ completeness fails, and $f(Y, M^{\dagger} \mid A,
\boldsymbol{L}),$ and thus the NIE or NDE, cannot be uniquely
non-parametrically determined from the incomplete observed data.



In our ACTIV-2 case study with a continuous mediator $M^{\dagger}$ and a binary
outcome $Y$, the joint distribution $f(Y, M^{\dagger} \mid A, \boldsymbol{L})$
is not non-parametrically identifiable under LLoQ left censoring of
$M^{\dagger}$. To address this limitation, we introduce distributional assumptions that enable the identification of
the target causal parameters while accommodating the nonignorable left censoring
mechanism \citep{hansen1982large, chamberlain1987asymptotic,
lotspeich2024making}. It is important to note that the completeness condition
used by \citet{zuo2024mediation} to establish identification holds under
several commonly used parametric families---including exponential families
\citep{newey2003instrumental} and certain location-scale distribution
families~\citep{hu2018nonparametric}---but it does not hold in our setting
because the support of $Y$ not being an open set violates one of the three
conditions~\citep[see Theorem 2.2 of][]{newey2003instrumental} required to
establish completeness in an exponential family. In addition, the positivity
assumption required in~\citet{zuo2024mediation}'s Theorem 1, i.e., $\P(R^M = 1
\mid M^{\dagger} = m, A = a, \boldsymbol{L} = \boldsymbol{l}) > 0$, does not
hold under our LLoQ-based left-censoring setting. This violation arises from the
deterministic LLoQ censoring mechanism discussed above, necessitating a
different identification strategy beyond \citet{zuo2024mediation}'s framework.

Assuming parametric models on several components of the observed data likelihood, we express the complete-data likelihood in terms of
$\P(Y \mid M^{\dagger}, A, \boldsymbol{L}; \boldsymbol{\alpha})$,
$f(M^{\dagger} \mid A, \boldsymbol{L}; \boldsymbol{\beta})$, and $\P(C \mid
M^{\dagger}, A, \boldsymbol{L})$, where the parameters $\boldsymbol{\theta} =
(\boldsymbol{\alpha}, \boldsymbol{\beta})$ determine the outcome mechanism and
the conditional mediator density, respectively. Notably, in this parametric
framework, the problem of distributional identification reduces to the
identification of the parameters $\boldsymbol{\theta}$, owing to the
one-to-one correspondence between the parameter vector and the implied
probability distribution. Since maximum likelihood estimation (MLE) of the
model parameters depends on the existence of unique solutions to score
equations for $\boldsymbol{\theta}$ in the observed-data likelihood,
identifiability of $\boldsymbol{\theta}$ too rests on these same score
equations. First, the complete-data log-likelihood is
\begin{align}\label{llcom}
\begin{split}
\ell_{\text{com}}(\boldsymbol{\theta})=\sum_{i=1}^n  \left[
\log \mathbb{P}\left(Y_i=y_i \mid M^{\dagger}_i=m_i, A_i=a_i, \boldsymbol{L}_i =
  \boldsymbol{l}_i; \boldsymbol{\alpha}\right) + \log f\left(M^{\dagger}_i=m_i \mid
  A_i=a_i, \boldsymbol{L}_i=\boldsymbol{l}_i; \boldsymbol{\beta}\right) 
  \right. \\
  \left. +\log \mathbb{P}\left(C_i =c_i \mid M^{\dagger}_i=m_i, A_i=a_i,
  \boldsymbol{L}_i=\boldsymbol{l}_i\right) \right] \ .
\end{split}
\end{align}
Then, when the mediator $M$ is subject to censoring, the observed data log-likelihood
may be expressed as
\begin{align}\label{llobs}
  \ell_{\text{obs}}(\boldsymbol{\theta}) =& \sum_{i:C_{i}=1} \log
    \mathbb{P}(Y_i \mid M^{\dagger}_i, A_i, \boldsymbol{L}_i; \boldsymbol{\alpha}) + \log
    f(M^{\dagger}_i \mid A_i, \boldsymbol{L}_i; \boldsymbol{\beta}) + \log
    \mathbb{P}\left(C_i =1 \mid M^{\dagger}_i=m_i, A_i, \boldsymbol{L}_i\right)
    \nonumber \\
  &+ \sum_{i:C_{i}=0} \log \int_{-\infty}^{+\infty} \mathbb{P}(Y_i \mid M^{\dagger},
    A_i, \boldsymbol{L}_i; \boldsymbol{\alpha}) f(M^{\dagger} \mid A_i, \boldsymbol{L}_i;
    \boldsymbol{\beta}) \mathbb{P}\left(C_i =0 \mid M^{\dagger}, A_i,
    \boldsymbol{L}_i\right) dM^{\dagger}
    \nonumber \\
  =& \sum_{i:C_{i}=1} \log \mathbb{P}(Y_i \mid M_i, A_i, \boldsymbol{L}_i;
    \boldsymbol{\alpha}) + \log f(M_i \mid A_i, \boldsymbol{L}_i;
    \boldsymbol{\beta})
    \\ \nonumber
  &+ \sum_{i:C_{i}=0} \log \int_{0}^{\lambda_{\text{LLoQ}} } \mathbb{P}(Y_i
    \mid M^{\dagger}, A_i, \boldsymbol{L}_i; \boldsymbol{\alpha}) f(M^{\dagger} \mid A_i,
    \boldsymbol{L}_i; \boldsymbol{\beta}) dM^{\dagger} \ ,
\end{align}
where the latter simplification uses the fact that the censoring mechanism is
deterministic in $M^{\dagger}$, as censoring occurs whenever the value of the
mediator falls below the assay quantification limit $\lambda_{\text{LLoQ}}$.
Following the standard tradition in parametric maximum
likelihood~\citep{van2000asymptotic}, we assume that the score equation for
$\boldsymbol{\theta}$ admits a unique solution---that is,
$U(\boldsymbol{\theta}) = \boldsymbol{0}$ has a unique solution
$\hat{\boldsymbol{\theta}} = (\hat{\boldsymbol{\alpha}},
\hat{\boldsymbol{\beta}})$---which in turn implies that the observed data
log-likelihood admits a unique maximizer.
This parametric identification strategy, imposing constraints on the functional
forms of $\P(Y \mid M^{\dagger}, A, \boldsymbol{L})$ and $f(M^{\dagger} \mid A,
\boldsymbol{L})$ through $\boldsymbol{\theta} = (\boldsymbol{\alpha},
\boldsymbol{\beta})$, contrasts directly with non-identifiability (due to
violation of the completeness condition) of the natural direct and indirect
effects in the non-parametric regime. Further details on identification in the
context of our ACTIV-2 case study appear in Section~\ref{supp:s1_iden} of the
\href{sm}{Supplementary Materials}.

\section{Estimation and inference}\label{sec:meth}
We now develop estimation and inference procedures for NIE and NDE in the presence of left-censored mediators
arising from assay quantification limits. Since $M^{\dagger}$ is left-censored at
the LLoQ for some study participants in the ACTIV-2 platform trial, we employ
the Expectation–Maximization (EM) algorithm~\citep{dempster1977maximum} to
obtain maximum likelihood estimators $\hat{\boldsymbol{\theta}}_{\text{MLE}}$
that maximize Equation~\eqref{llobs}, treating the censored
$M^{\dagger}$ as a latent variable. In the E-step, we evaluate the conditional expectation of the complete-data log-likelihood over the latent mediator distribution given the observed data and current parameter estimates. When $M^{\dagger}$ is continuous, this expectation can be analytically
intractable. To address this, we adopt fractional imputation
\citep{kalton1984some, kim2011parametric}, which leverages importance sampling
and re-weighting to approximate the conditional expectation more efficiently.

\subsection{Fractional imputation and EM algorithm}
Fractional imputation (FI), originally proposed by~\cite{kalton1984some}, is an
alternative to ``single'' imputation that replaces each missing value with
multiple plausible candidate values, with each candidate replacement weighted
according to its approximate probability of occurrence. Parametric FI employs a
parametric model for imputation and uses the EM algorithm for estimation of
the model parameters~\citep{kim2011parametric}. The approach proceeds by first
generating $S$ candidate replacement values for each missing value from a
proposal distribution and then applying importance sampling to compute
fractional weights (i.e., normalized importance weights) assigned to each of
the candidate replacement values. Specifically, for each  censored
mediator  $M_{\text{cen},i}$, we generate $S$ candidate imputed values, denoted
$m_i^{\star(1)}, \dots, m_i^{\star(S)}$ from a \textit{proposal distribution}
$f(m_{i} \mid a_{i}, \boldsymbol{l}_{i})$, using rejection sampling to ensure
that the imputed values remain below the lower limit of quantification
$\lambda_{\text{LLoQ}}$. From these values, we compute normalized importance
weights as
$$
w_{i j} \propto \frac{f(m_{i}^{\star(j)} \mid Y_i, C_i = 0, a_{i},
  \boldsymbol{l}_{i})}{f(m^{\star(j)}_{i} \mid a_{i}, \boldsymbol{l}_{i})} \ ,
  \ \text{s.t.}\ w_{ij}>0\ \ \text{and}\  \sum_{j=1}^{S} w_{ij} = 1 \ ,
$$
where the numerator $f(m_{i}^{\star(j)} \mid Y_i, C_i = 0, a_{i},
\boldsymbol{l}_{i})$ represents the \textit{target distribution} of the censored mediator, that is, the conditional distribution of $M^{\dagger}$,
given $(Y, A, \boldsymbol{L})$ for values below the assay quantification limit.
Rather than explicitly compute this conditional distribution, it suffices to
leverage knowledge of the joint distribution $\P(Y_i, M^{\dagger}_i = m^{*(j)}_{i},
C_i = 0 \mid a_{i}, \boldsymbol{l}_{i})$, based on the following identity:
$$
 \frac{\P(Y_i, M^{\dagger}_i = m^{\star(j)}_{i}, C_i = 0\mid a_{i}, \boldsymbol{l}_{i}) /
 f(m^{\star(j)}_{i} \mid a_{i}, \boldsymbol{l}_{i})}{\sum_{k=1}^S
 \P(Y_i, M^{\dagger}_i = m^{*(k)}_{i}, C_i = 0 \mid a_{i}, \boldsymbol{l}_{i}) /
 f(m^{\star(k)}_{i} \mid a_{i}, \boldsymbol{l}_{i})} =
 \frac{f(m_{i}^{\star(j)} \mid Y_i, C_i = 0, a_{i}, \boldsymbol{l}_{i}) /
   f(m^{\star(j)}_{i} \mid a_{i}, \boldsymbol{l}_{i})}{\sum_{k=1}^S
   f(m_{i}^{\star(k)} \mid Y_i, C_i = 0, a_{i}, \boldsymbol{l}_{i}) /
   f(m^{\star(k)}_{i} \mid a_{i}, \boldsymbol{l}_{i})} \ ,
$$
from which we conclude that the fractional weights may equivalently be expressed
\begin{equation}\label{fwpro}
  w^{}_{ij} \propto \frac{\P(Y_i, M^{\dagger}_i = m^{\star(j)}_{i}, C_i = 0 \mid a_{i},
  \boldsymbol{l}_{i})}{f(m^{\star(j)}_{i} \mid a_{i}, \boldsymbol{l}_{i})} \ ,
  \ \text{s.t.}\ w_{ij}>0\ \ \text{and}\ \sum_{j=1}^{S} w_{ij} = 1 \ .
\end{equation}
We detail the parametric FI procedure in Algorithm~\ref{alg:fiem}.

\begin{algorithm}[!]
\caption{EM algorithm with fractional imputation for LLoQ censored
mediators}\label{alg:fiem}

\textbf{Input:} Conditional density of mediator $f(m_{i} \mid a_{i},
\boldsymbol{l}_{i}; \boldsymbol{\beta}^{(0)})$, the proposal distribution, with
parameter initialized at $\boldsymbol{\beta}^{(0)}$, outcome regression $\P(Y_i
\mid M^{\dagger}_i, A_{i}, \boldsymbol{L}_{i}; \boldsymbol{\alpha})$, and
specified convergence criterion parameters, including (1) maximum number of iterations $n_{\text{iter}},$ (2) the convergence tolerance $\epsilon$.\\

\textbf{1. E-Step}
    \begin{enumerate}[label=(\alph*)]
      \itemsep1pt
      \item \textbf{Imputation step:} Generate $S$ candidate imputed values
        (imputation replicates), restricted to fall below
        $\lambda_{\text{LLoQ}}$, using rejection sampling from the proposal
        distribution $f(m_{i} \mid a_{i}, \boldsymbol{l}_{i};
        \hat{\boldsymbol{\beta}}^{(0)})$ for each  censored
        mediator value $m_{\text{cen},i}$, denoted as $m_i^{\star(1)}, \dots,
        m_i^{\star(S)}$.
      \item \textbf{Weighting step:} Compute fractional imputation weights
        for each of the imputation replicates, using the parameter estimate
        $\hat{\boldsymbol{\theta}}^{(t)} = (\hat{\boldsymbol{\alpha}}^{(t)},
          \hat{\boldsymbol{\beta}}^{(t)})$ for $j=1, 2, \ldots, S$, subject to
          $\sum_{j=1}^{S} w^{(t)}_{ij} = 1$:
          \begin{equation}\label{wetdes}
              w^{(t)}_{ij} \propto \frac{\P(Y_i, M^{\dagger}_i =
              m^{\star(j)}_{i}, C_i = 0
              \mid a_{i}, \boldsymbol{l}_{i}; \hat{\boldsymbol{\theta}}^{(t)})}
              {f(m^{\star(j)}_{i} \mid a_{i}, \boldsymbol{l}_{i};
              \hat{\boldsymbol{\beta}}^{(0)})} = \frac{\P(Y_i \mid C_i = 0, M_i =
              m^{\star(j)}_{i}, a_{i}, \boldsymbol{l}_{i};
              \hat{\boldsymbol{\alpha}}^{(t)})f(m^{\star(j)}_{i} \mid a_{i},
              \boldsymbol{l}_{i}; \hat{\boldsymbol{\beta}}^{(t)})}
              {f(m^{\star(j)}_{i} \mid a_{i}, \boldsymbol{l}_{i};
              \hat{\boldsymbol{\beta}}^{(0)})} \ .
          \end{equation}
    \end{enumerate}

\textbf{2. M-step:} Update $\boldsymbol{\theta}$ by solving the imputed
  weighted score equation:
  \begin{equation}\label{wetscore}
    \sum_{i}^{n} \sum_{j=1}^{S} w^{(t)}_{ij} S(\boldsymbol{\theta}; Y_i,
    m^{\star(j)}_{i}, a_{i}, \boldsymbol{l}_{i}) = 0 \ ,
  \end{equation}
where $S(\boldsymbol{\theta}; Y_i, m^{\star(j)}_{i}, a_{i}, \boldsymbol{l}_{i})
= \partial \ell_{\text{com}}(M^{\dagger}_i = m^{\star(j)}_{i}; \boldsymbol{\theta}) /
\partial \boldsymbol{\theta}$ involves Equation~\eqref{llcom}.\\

\textbf{3. Check (optional):} Monitor the weight distribution using histograms
or summary statistics. If extreme fractional weights are detected, modify the
proposal distribution to a more suitable form, for example, by setting
$f(m_{i} \mid a_{i}, \boldsymbol{l}_{i}; \hat{\boldsymbol{\beta}}^{(0)}) =
f(m_{i} \mid a_{i}, \boldsymbol{l}_{i}; \hat{\boldsymbol{\beta}}^{(t)})$,
and revert to the imputation step 1(a) above.\\

\textbf{4. Iteration:} Set $t = t + 1$ and return to the weighting step (1(b))
if the convergence criterion, $\lVert \hat{\boldsymbol{\theta}}^{(t+1)} -
\hat{\boldsymbol{\theta}}^{(t)} \rVert \geq \epsilon$, has not been met or the
maximum number of iterations $n_{\text{iter}}$ has not been reached.\\

\textbf{Output:} $S$ imputed values are generated for each 
censored mediator, i.e., $M_{\text{cen},i} = (m_i^{\star(1)}, \dots,
m_i^{\star(S)})$, and corresponding fractional weights (i.e., $w^{}_{i1},
\dots,w^{}_{iS}$) are assigned to the imputed values.
\end{algorithm}

\textbf{Remark 1:} The output of Algorithm \ref{alg:fiem}
consists of $S$ imputed values for each mediator that is left-censored due to
the LLoQ, along with their corresponding fractional weights. Specifically, for
each censored mediator, we generate $M_{\text{cen},i} =
(m_i^{\star(1)}, \dots, m_i^{\star(S)}),$ with associated fractional weights
$w_i = (w_{i1}, \dots, w_{iS})$, satisfying $\sum_{j=1}^{S} w_{ij} = 1$ for
each individual $i$. We treat observed mediators $M_i$ as having a
single observed value with weight $w_i = 1$. Consequently, after fractional
imputation, the complete dataset can be viewed as representing the latent
mediator through the observed and imputed values with their corresponding
weights. Formally, $\widetilde{M}^{\dagger}_i = M_i\mathbb{I}(C_i=1) +
M_{\text{cen},i}\mathbb{I}(C_i=0),$ and $W_i = \mathbb{I}(C_i=1) +
w_i\mathbb{I}(C_i=0).$

\textbf{Remark 2:} In specifying the proposal distribution for the mediator,
that is, the form of $f(m_{i} \mid a_{i}, \boldsymbol{l}_{i};
\boldsymbol{\beta}^{(0)})$, we suggest remaining agnostic about the form of
the underlying true density, acknowledging that a restrictive, parametric
choice of the proposal distribution is likely to result in model
misspecification. It is this robustness consideration that motivates our
semi-parametric conditional density estimation method, detailed next in
Section~\ref{sec:cde-semipar}.

\textbf{Remark 3:}\label{remark3} Our proposed Algorithm \ref{alg:fiem}
provides a generalization of established imputation techniques. The conventional LLoQ/2, or LLoQ/$\sqrt{2}$, substitution method can be viewed as a
special case of our proposed framework, obtained when the proposal distribution
is degenerate, concentrating all probability mass at the single point LLoQ/2 or
LLoQ/$\sqrt{2}$. In this way, our proposal extends and unifies existing and
frequently used ad-hoc approaches.


The imputation and weighting steps (i.e., 1(a) and 1(b) in Algorithm
\ref{alg:fiem})
constitute the E-step of the EM algorithm, following the general FI-EM
framework \citep{kim2011parametric, yang2016fractional}. For computational
efficiency, we factorize the joint distribution in the weighting step using the
fractional weights from Equation~\eqref{fwpro}, as shown in
Equation~\eqref{wetdes}. In the M-step, the fractional weights
improve the Monte Carlo approximation of the conditional expectation, which
becomes more accurate as $S$ increases: $\E\left\{\ell_{\text {com},i}
\left(M^{\dagger}_i=m; \boldsymbol{\theta}\right) \mid Y_i, C_i=0, a_i,
\boldsymbol{l}_{i}; \hat{\boldsymbol{\theta}}^{(t)}\right\} \approx
\sum_{j=1}^S \ell_{\text{com}, i}\left(M^{\dagger}_i=m_i^{*(j)} ;
\boldsymbol{\theta}\right) w_{ij},$ with the complete-data log likelihood
$\ell_{\text {com}, i}\left(M^{\dagger}_i=m; \boldsymbol{\theta}\right)$ in
Equation~\eqref{llcom}. The corresponding Q-function of the EM
Algorithm~\ref{alg:fiem}
is
$$
Q\left(\boldsymbol{\theta} \mid \hat{\boldsymbol{\theta}}^{(t)}\right) =
\mathbb{E}\left(\ell_{\text{com}}\left( \boldsymbol{\theta}\right) \mid Y, M^{\dagger},
a, \boldsymbol{l}, \hat{\boldsymbol{\theta}}^{(t)}\right) \ .
$$
The weighted log-likelihood function based on fractional imputation, related to
Equation (\ref{wetscore}) in the M-step, is denoted as
\begin{equation}\label{QSTAR}
    Q^{\star}\left(\boldsymbol{\theta} \mid
      \hat{\boldsymbol{\theta}}^{(t)}\right) =
      \sum_{i=1}^n \sum_{j=1}^S w_{i j}^{(t)} \ell_{\text {com }}
      \left(M^{\dagger}=m_i^{\star(j)}; \boldsymbol{\theta}\right) \ .
\end{equation}
Notably, for a sufficiently large number of imputations $S$,
$Q\left(\boldsymbol{\theta} \mid \hat{\boldsymbol{\theta}}^{(t)}\right)$ is
well approximated by $Q^{\star}\left(\boldsymbol{\theta} \mid
\hat{\boldsymbol{\theta}}^{(t)}\right)$; therefore, in the M-step, the
Q-function is maximized so as to update the parameter estimates, yielding
$\hat{\boldsymbol{\theta}}^{(t+1)}$. The proposed EM framework is extended in
Section~\ref{supp:generic-inf-cens} of the \href{sm}{Supplementary Materials}
to handle generic censoring without the deterministic property. Whereas LLoQ censoring is
deterministic, these more general mechanisms are probabilistic and thus
necessitate the modeling of censoring probabilities, which we are able to
forego presently.

Unlike Monte Carlo EM methods, where imputed values are resampled at each
iteration, the FI approach retains the same imputed values throughout
iterations while only updating the fractional weights. This offers notable
computational advantages---importance sampling is performed only once, reducing
computational overhead, and provides guaranteed convergence as demonstrated in
Theorem 1 of \cite{kim2011parametric}.  We further extend the theoretical
results of fractional imputation from \cite{kim2011parametric} to our framework
of causal mediation analysis with a censored mediator, and present the
following theorem.

\begin{theorem}\label{thm:effem}
Under suitable regularity conditions, for a sufficiently large number of
iterations $t$ in Algorithm
\ref{alg:fiem},
the estimated parameter $\hat{\boldsymbol{\theta}}^{(t)}$ converges to its
asymptotic limit $\hat{\boldsymbol{\theta}}^{\star}_{S}$, which is a stationary
point of $Q^{\star}$ in Equation (\ref{QSTAR}) for a fixed $S$;
that is,
$\hat{\boldsymbol{\theta}}^{(t)} \to \hat{\boldsymbol{\theta}}^{\star}_{S}\
\text{as}\ t \to \infty$. Then, for a sufficiently large number of imputations
$S$, we have $\hat{\boldsymbol{\theta}}^{\star}_{S} \to
\hat{\boldsymbol{\theta}}_{\text{MLE}}$.
\end{theorem}
The proof of Theorem~\ref{thm:effem} is provided in
Section~\ref{supp:s1:proofofTH1} of the \href{sm}{Supplementary Materials}.
Theorem~\ref{thm:effem} indicates that, for any fixed number of
imputations $S$, the sequence of estimators ${\hat{\boldsymbol{\theta}}^{(t)}}$
produced by Algorithm~\ref{alg:fiem} converges to a stationary point
$\hat{\boldsymbol{\theta}}^{\star}_{S}$ of the $Q^*$-function associated with
the weighted log-likelihood. This convergence holds under a set of standard
regularity conditions, including the continuity and differentiability of the
expected complete-data log-likelihood function $Q^{\star}(\boldsymbol{\theta}
\mid \hat{\boldsymbol{\theta}}^{(t)})$ in both arguments and the compactness of
the parameter space, conditions formally established by
\cite{wu1983convergence}. The specific regularity assumptions used in our
setting are listed in Section~\ref{supp:s1:proofofTH1} of the
\href{sm}{Supplementary Materials} for completeness. In addition, as the number
of imputations $S$ increases, the stationary point
$\hat{\boldsymbol{\theta}}^{\star}_{S}$ converges in probability to the true
maximum likelihood estimator of $\boldsymbol{\theta}$. Consequently, the
limiting estimator $\hat{\boldsymbol{\theta}}^{\star}_{S}$ inherits the
desirable large-sample properties of the MLE, including consistency and
asymptotic efficiency. \biostats{In addition, a formal connection between the proposed FI-EM procedure and direct numerical optimization of the observed-data likelihood, including the approach of \cite{chernofsky2024causal}, is provided in Section~\ref{supp:chernofsky-connection} of the \href{sm}{Supplementary Materials}.}

\subsection{Semi-parametric estimation of the conditional mediator
density}\label{sec:cde-semipar}

A key complication of the problem we face is the need for estimation of the
\textit{conditional density of the mediator}, a nuisance quantity required for
our proposed EM procedure in Algorithm~\ref{alg:fiem},
as it plays a crucial role in both the imputation step (i.e., as specified in
Equation~\eqref{wetdes}) and in estimation of the natural direct and indirect effect estimands. 

Since assuming the conditional mediator density
follows a simple parametric form may be more restrictive than is justified by
available domain knowledge, use of such a technique may prove insufficient to
capture the underlying complexity and thereby lead to model misspecification.
Instead, and in the absence of domain knowledge suggesting a specific
parametric form, we opt for more flexible, semi-parametric modeling
strategies---specifically, we adopt a \textit{semi-parametric location-scale
conditional density estimator}, which we pair with a non-parametric regression
estimator, the highly adaptive lasso (HAL), for which optimality
properties~\citep{van2017generally, van2017uniform} have been established under
relatively mild global smoothness assumptions (a finite global variation norm).
Specifically, HAL provides a non-parametric approach for approximating
infinite-dimensional target functionals, such as conditional means, hazards, or
densities, using a linear combination of basis functions, allowing for flexible
adaptation to complex and possibly high-dimensional data structures. We provide
more details on HAL in Section~\ref{supp:sec:hal} of the
\href{sm}{Supplementary Materials}. Within our proposed approach, HAL is
employed to approximate nuisance functions that arise within our
semi-parametric conditional density estimation strategy.

\begin{algorithm}[!]
\caption{Semi-parametric estimation of the conditional mediator density via a
location-scale restriction}\label{alg:cde}

\textbf{Inputs:} Imputed dataset with fractional imputation weights $\{(Y_i, \widetilde{M}^{\dagger}_i,
  A_i, \boldsymbol{L}_i, W_i)\}_{i=1}^n$; regression estimator $f_\mu$ of
  $\mu(A, \boldsymbol{L}) = \E[M^{\dagger} \mid A, \boldsymbol{L}]$; regression estimator
  $f_\sigma$ of $\sigma^2(A, \boldsymbol{L}) = \E[(M^{\dagger} -
  \mu(A, \boldsymbol{L}))^2 \mid A, \boldsymbol{L}]$ (optionally, for the
  heteroscedastic variant); and kernel function $K$ with bandwidth $h$ for
  marginal density estimation.\\

\textbf{1. Estimate the conditional mean:}
Fit $\hat{\mu}(A, \boldsymbol{L}) = f_\mu(A, \boldsymbol{L})$ to predict
the mean of $M^{\dagger}$ from $A, \boldsymbol{L}$.\\

\textbf{2. Estimate the conditional variance:}
\begin{itemize}
  \itemsep1pt
  \item For the homoscedastic variant: let $\hat{\sigma}^2 =
    \frac{1}{n}\sum_{i=1}^n \sum_{j=1}^{S}w_{ij}(\widetilde{M}^{\dagger}_i - \hat{\mu}(A_i, \boldsymbol{L}_i))^2$
  \item For the heteroscedastic variant: fit $\hat{\sigma}^2(A, \boldsymbol{L})
    = f_{\sigma}(A, \boldsymbol{L})$ to predict the mean of
    $(M^{\dagger} - \hat{\mu}(A, \boldsymbol{L}))^2$ from $A, \boldsymbol{L}$\\
\end{itemize}

\textbf{3. Estimate the standardized residual density:}
\begin{enumerate}[label=(\alph*)]
  \itemsep1pt
  \item Compute the standardized residual $Z_i = W_i(\widetilde{M}^{\dagger}_i - \hat{\mu}(A_i,
    \boldsymbol{L}_i))/\hat{\sigma}(A_i, \boldsymbol{L}_i)$
  \item Apply kernel density estimation, with kernel function $K$ and bandwidth
    $h$, to the standardized residual:
    $$
      \{Z_i\}_{i=1}^n: \hat{f}_0(z) = \frac{1}{n h}
        \sum_{i=1}^n K\left(\frac{z-Z_i}{h}\right) \ .
    $$
\end{enumerate}

\textbf{4. Construct the estimated conditional density of the mediator:}
  $$
    \hat{f}_n(M^{\dagger} \mid A, \boldsymbol{L}) = \frac{1}{\hat{\sigma}(A,
    \boldsymbol{L})}\hat{f}_0\left(\frac{M^{\dagger} - \hat{\mu}(A,
    \boldsymbol{L})}{\hat{\sigma}(A, \boldsymbol{L})}\right) \ .
  $$

\textbf{Output:} Estimated density $\hat{f}_n(M^{\dagger} \mid A,
\boldsymbol{L})$ of the mediator $M^{\dagger}$, conditional on exposure $A$ and
baseline covariates $\boldsymbol{L}$.
\end{algorithm}
The semi-parametric location-scale conditional density estimation approach
provides a flexible strategy for estimating conditional density functions by
re-formulating the estimation problem in terms of the estimation of a limited
set of regression functions, the conditional mean $\mu(A, \boldsymbol{L})$ and,
optionally, the conditional variance $\sigma^2(A, \boldsymbol{L})$, which are
accompanied by non-parametric estimation of the marginal density,
$f_0$, of the standardized residuals corresponding to these regression steps.
The modular structure of this strategy enables straightforward use of a large
suite of flexible, non-parametric regression and machine learning techniques
for estimation of the nuisance functions while employing kernel density
estimation to map nuisance regression estimates into estimates of the target
conditional density function. Assuming that the target density function belongs
to a location-scale family or, equivalently, that the standardized residuals
$(\widetilde{M}^{\dagger} - \mu(A, \boldsymbol{L})) / \sigma(A,
\boldsymbol{L})$ follow a common marginal distribution, we reduce the
fundamentally challenging problem of density estimation into the more readily
achievable task of estimating the regression objects $\mu(A, \boldsymbol{L})$
and $\sigma^2(A, \boldsymbol{L})$, a task to which arbitrarily flexible
regression algorithms (e.g., lasso, random forests) can be applied. We argue
that this degree of flexibility is especially valuable in applied biomedical
science settings, in which domain knowledge is often unavailable to rule out
the possibility of nonlinear, high-dimensional relationships between relevant
covariates. We summarize our proposal as
Algorithm~\ref{alg:cde},
presented next. We note that while we are not the first to discuss such a
strategy, its use as part of a broader imputation strategy appears novel.%

While less restrictive than assuming a parametric form, the approach of
Algorithm~\ref{alg:cde}
is of a restrictive semi-parametric variety---its validity hinges on the target
density function belonging to a location-scale family, with deviation from such
a family being likely to introduce misspecification bias. On the other hand,
the proposal's merit lies in allowing use of flexible regression procedures for
nuisance estimation and in its computational efficiency, which it obtains by
reducing density estimation to a regression problem. In cases in which the
location-scale restriction may be invalid, alternative non-parametric methods,
including those based on the highly adaptive lasso~\citep{hejazi2022haldensify,
hejazi2025efficient}, may be employed instead. In finite-sample settings,
kernel density estimation of $f_0$ can be sensitive to the choice of bandwidth
$h$, particularly with heavy-tailed residuals; however, cross-validation may be
used to select an optimal choice of bandwidth from a large set of candidates in
practice. Relatedly, by allowing for the consideration of heteroscedasticity
via estimation of $\sigma(A, \boldsymbol{L})$, the proposed approach benefits
from a degree of robustness. For implementation, we recommend cross-validated
selection of the tuning parameters of regression algorithms used for nuisance
estimation (e.g., the global variation norm in HAL) of the nuisance parameters
$\mu(A, \boldsymbol{L})$ and $\sigma(A, \boldsymbol{L})$.

Stepping back, our proposed strategy combines semi-parametric estimation of the
conditional mediator density with fractional imputation of the left-censored
mediator (including a tailored EM algorithm) and flexible estimation of those
nuisance regression functions needed for estimation of the conditional density.
By estimating $\mu(A, \boldsymbol{L})$ and $\sigma^2(A, \boldsymbol{L})$
without specifying a parametric form for the conditional mediator density $f(M^{\dagger}
\mid A, \boldsymbol{L})$, we limit reliance on restrictive assumptions while
simultaneously avoiding the computational challenges inherent to non-parametric
estimation of the conditional density function \citep{qin1998inferences, cheng2004semiparametric, sugiyama2012density, reisach2025transforming}. Our approach
accommodates complex, nonlinear relationships between baseline covariates,
exposure, and the mediator, and supports the use of fractional importance
weights in the EM algorithm used to recover an approximation of the underlying
true density of the left-censored mediator.

\subsection{Estimation of target estimands after fractional imputation of the
mediator}

Having obtain imputed values of the left-censored mediator, we now review the
use of standard, previously studied and documented estimation strategies for
the natural direct and indirect effect estimands, including both plug-in (or
g-computation) and inverse probability weighted (IPW)
estimation~\citep{imai2010general, imai2010identification}. Notably,
semi-parametric estimation strategies leveraging the efficient influence
function (EIF) have also been developed for asymptotically efficient
estimation~\citep{tchetgen2012semiparametric, zheng2012targeted} of the target
statistical functionals. We provide only an abbreviated review of a subset of
these strategies, with a specific focus on the natural direct effect, referring
the interested reader to the voluminous literature on causal mediation analysis
for further details. The derivation for the natural indirect effect is
analogous, and we direct the reader to the  literature on causal mediation
analysis for further details.

In our simulation experiments, presented in Section~\ref{sec:sim},
we examine both the plug-in estimator and the asymptotically efficient one-step
estimator (derived from the EIF), pairing these with the fractionally imputed
values of the left-censored mediator obtained from our procedure documented to
this point. To illustrate, we briefly review construction of the plug-in
estimator of the natural direct effect, which typically proceeds as follows.
\begin{enumerate}
  \item Obtain an estimate of the conditional expectation of the outcome, given
    mediator, exposure, and baseline covariates, i.e., $\hat{\E}[Y \mid A,
    M^{\dagger}, \boldsymbol{L}]$, using an appropriate parametric regression
    procedure.
  \item Compute pseudo-outcomes under two counterfactual scenarios described by
    $A \in \{a_1, a_2\}$: $\hat{\E}[Y \mid M^{\dagger}, A = a_1,
    \boldsymbol{L}]$ and $\hat{\E}[Y \mid M^{\dagger}, A = a_0,
    \boldsymbol{L}]$. Compute their difference and regress this pseudo-outcome
    on exposure and baseline covariates to obtain $\hat{\E}_{M^{\dagger}
      \mid A, \boldsymbol{L}} \left\{\hat{\E}(Y
    \mid M^{\dagger}, A = a_1, \boldsymbol{L}) - \hat{\E}(Y \mid M^{\dagger},
    A = a_0, \boldsymbol{L}) \mid A, \boldsymbol{L} \right\}$.
  \item Using its empirical distribution, compute an estimate of the marginal
    distribution of $\boldsymbol{L}$, i.e.,
    $\P(\boldsymbol{L} = \boldsymbol{l})$.
  \item Compute a plug-in estimator of the natural direct effect estimand:\\
    $\Psi^{\text{g-{comp}}}_{\text{NDE}}(\hat{\Pf}) = \hat{\E}_{L} \left[\hat{\E}_{M^{\dagger}
      \mid A = a_0, \boldsymbol{L}}
    \left\{\hat{\E}(Y \mid M^{\dagger}, A=a_1, \boldsymbol{L}) - \hat{\E}(Y
    \mid M^{\dagger}, A=a_0, \boldsymbol{L}) \mid A=a_0, \boldsymbol{L}
    \right\} \right]$.
\end{enumerate}

Building on the above, with fractionally imputed values of the mediator $\widetilde{M}^{\dagger}_i =
(M_{i}, M_{i,\text{cen}})$, where $M_{\text{cen},i} =
m_i^{\star(1)}, \dots, m_i^{\star(S)}$ (i.e., those with $C_i = 0$) and
fractional weights $w_{ij}$, the plug-in estimator of the NDE is obtained as
\begin{align*}
  \sum_{i}^{n}\sum_{j=1}^{S} w_{ij}
  \left[\Psi^{\text{g-{comp}}}_{\text{NDE}}(\hat{\Pf}) - \hat{\E}_{M^{\dagger}
      \mid A = a_0, \boldsymbol{L}}
  \left\{\hat{\E}(Y \mid M^{\dagger}, A=a_1, \boldsymbol{L}) - \hat{\E}(Y \mid
  M^{\dagger}, A=a_0, \boldsymbol{L}) \mid A=a_0, \boldsymbol{L} \right\}
  \right] = 0 \ .
\end{align*}
The plug-in estimator restricts nuisance estimation, including those regression
steps involving the pseudo-outcome above, to the use of parametric strategies.
When such a procedure is correctly specified, the plug-in estimator can be
asymptotically efficient (since the nuisance estimators will then be MLEs of
the corresponding nuisance functions); however, correct specification of the
nuisance functions requires, in practice, a great degree of domain knowledge,
which may not be available in practice. As an alternative, one may employ
asymptotically efficient estimators of the natural direct and indirect effect
estimands, which are derived from the corresponding EIF.

One such asymptotically efficient estimator, the one-step de-biased estimator,
is derived from the EIF and constructed by updating the plug-in estimator with
the empirical mean of the estimated EIF. The EIF consists of three components:
a so-called plug-in term (corresponding to the plug-in estimator) and two
centered correction terms derived based on IPW. Let $\phi^{\text{EIF}}(\cdot)$
denote the EIF for the NDE estimand, the asymptotically efficient one-step
estimator is obtained as the solution to an estimating equation based on the
EIF, incorporating fractional imputation weights $w_{ij}$:
\begin{align}\label{eqn:fi-eif-ose}
  \sum_{i}^{n} \sum_{j=1}^{S} w_{ij} \left[
  \hat{\Psi}_{\text{NDE}}^{\text{EIF}}(\Pf) - \phi^{\text{EIF}}(\hat{\Pf})(O_i)
  \right] = 0 \ ,
\end{align}
where the EIF $\phi^{\text{EIF}}(\hat{\Pf})(O_i)$, evaluated at an augmented
empirical distribution $\hat{\Pf}$ and for a given data unit $O_i$ is
\begin{align*}
  \phi^{\text{EIF}}(\hat{\Pf})(O_i) =& \underbrace{\left\{\frac{f_{M^{\dagger} \mid A =
    a_0, \boldsymbol{L}}(\widetilde{M}^{\dagger}_i)}{f_{M^{\dagger} \mid A = a_1, \boldsymbol{L}}(\widetilde{M}^{\dagger}_i)}
    \left(Y_i - \E[Y \mid \widetilde{M}^{\dagger}_i, A_i=a_1, \boldsymbol{L}_i] \right)
    \right\}}_{\text{(1): centered re-weighted (IPW) term}} \\
    &+ \underbrace{\left\{\E[Y \mid \widetilde{M}^{\dagger}_i, A_i=a_1, \boldsymbol{L}_i] - \E_{M^{\dagger}
      \mid A_i = a_0, \boldsymbol{L}_i}\left[\E[Y \mid M^{\dagger}_i, A_i=a_1, \boldsymbol{L}_i] \right]
      \right\}}_{\text{(2): centered plug-in term}} \\
    &+ \underbrace{\left\{\E_{M^{\dagger} \mid A_i = a_0, \boldsymbol{L}_i}
      \left[\E[Y \mid M^{\dagger}_i, A_i = a_1, \boldsymbol{L}_i] \right] -
      \E_{\boldsymbol{L}}\left[\E_{M^{\dagger} \mid A = a_0, L}\left[\E[Y \mid M^{\dagger}_i, A_i=a_1,
      \boldsymbol{L}_i] \right] \right]
      \right\}}_{\text{(3): plug-in (g-computation) estimator}} \ ,
\end{align*}
where the term (3) corresponds to the plug-in (g-computation) estimator
previously described, the term (2) is a centered version of a plug-in (nested
expectation) component, and the term (1) is a centered version of an IPW
correction term.

A practical challenge in the construction of asymptotically efficient
estimators based on the EIF of the NDE estimand is accurate estimation of the
ratio of the conditional densities of the mediator given exposure and
covariates under differing counterfactual contrasts of the exposure.
To address this, previous works \citep[e.g.,][]{diaz2020nonparametric,
hejazi2022nonparametric} have considered re-parametrization of the density ratio
in terms of conditional probabilities of the exposure, given mediator and
baseline covariates, applying Bayes' rule:
$f_{M^{\dagger} \mid A = a_0, \boldsymbol{L}}(\widetilde{M}^{\dagger}_i) /
f_{M^{\dagger} \mid A = a_1, \boldsymbol{L}} (\widetilde{M}^{\dagger}_i) = \P(A
= a_0 \mid M^{\dagger}, \boldsymbol{L}) \P(A = a_1 \mid \boldsymbol{L}) / \P(A
= a_1 \mid M^{\dagger}, \boldsymbol{L}) \P(A = a_0 \mid \boldsymbol{L})$.
In our simulation experiments and illustrative data analysis, we rely
on the \texttt{medoutcon}~\citep{hejazi2022medoutcon-joss} package for the
\texttt{R} language and environment for statistical computing~\citep{R}, which
implements this re-parametrization strategy.

We have reviewed the construction of plug-in and asymptotically efficient
one-step estimators of the NDE estimand, and corresponding estimators of the
NIE estimand can be constructed in an analogous manner from input data that has
been ``repaired'' via our proposed fractional imputation strategy. In
particular, the plug-in and one-step estimators of the NDE and NIE estimands
are obtained by evaluating the corresponding statistical functionals at the
maximum likelihood estimates derived from the completed data recovered by
Algorithm~\ref{alg:fiem}.
Since the NDE and NIE estimands are smooth functionals of the underlying data
distribution, these estimators inherit the consistency and asymptotic normality
of the MLEs used in their construction.

\subsection{Inference based on a data-driven $m$-out-of-$n$ bootstrap}\label{sec:boot}

Having defined our estimators, we turn to uncertainty quantification. Inference
differs for the plug-in and one-step estimators. For the plug-in estimator,
confidence intervals may be obtained via the nonparametric bootstrap, while for
the one-step estimator they rely on the asymptotic variance of the EIF,
estimated by its empirical variance under standard regularity conditions. Both
approaches assume iid data. In our setting, however, fractional imputation of a
left-censored mediator induces dependence among imputed values, rendering the
``repaired'' dataset non-iid and violating these assumptions. Even simple
schemes, such as imputing a common constant (e.g., LLoQ/2), can induce perfect
correlation, while more general procedures introduce dependence that inflates
variance and invalidates standard variance estimators~\citep{rao1992jackknife,
rubin1978multiple, kim2021statistical}. Although related issues have been
studied in the context of hot-deck imputation~\citep{rao1992jackknife} and
multiple imputation~\citep{rubin1978multiple}, existing approaches are not
directly applicable to causal mediation estimands under fractional imputation.

\biostats{To address this inferential challenge, we propose a data-adaptive
\(m\)-out-of-\(n\) bootstrap~\citep{shao1994bootstrap,bickel2011resampling} to
obtain inference on the NDE and NIE statistical estimands after fractional
imputation of the mediator. The motivation is not that reducing the bootstrap
sample size eliminates the dependence induced by fractional imputation. Rather,
the combination of within-subject dependence among fractional records and
estimation of their fractional weights introduces additional variability and
can lead to irregular finite-sample behavior for which the ordinary
\(n\)-out-of-\(n\) bootstrap may inadequately approximate the sampling
distribution. The \(m\)-out-of-\(n\) bootstrap provides an alternative
resampling strategy in such settings and avoids direct reliance on analytic
variance calculations that would need to account explicitly for the imputation
procedure.}

\biostats{The \(m\)-out-of-\(n\) bootstrap draws resamples of size \(m < n\),
where \(m\) increases with \(n\) but at a slower rate (that is, \(m\to\infty\)
and \(m/n \to 0\) as \(n \to \infty\)). Such procedures have been developed as
alternatives to the ordinary bootstrap in settings where the latter may
inadequately approximate the sampling distribution, including problems
involving non-smooth or nonregular estimators~\citep{shao1994bootstrap,
bickel2011resampling, chakraborty2013inference, xu2015regularized,
simoneau2018non}. While these asymptotic conditions provide limited practical
guidance for choosing \(m\) in finite samples, data-driven methods offering
pragmatic strategies for selecting \(m\) have also been
proposed~\citep{cheung2005iterating, bickel2008choice,
chakraborty2013inference}.}

Intuitively, the choice of $m$ should reflect the degree of non-smoothness
anticipated in the underlying generative model. Correspondingly, greater levels
of censoring are likely to lead to an increased degree of non-regularity in the
fractionally imputed dataset used to estimate the NDE and NIE estimands. As
such, when the proportion of censored units, $p_{\text{cen}}$, is high,
indicating possibly substantial irregularity, the corresponding $m$-out-of-$n$
bootstrap estimator should oscillate at a rate much slower than $n^{-1/2}$. A
reasonable class of resampling sizes is given by $m = \lfloor
n^{f(p_{\text{cen}})} \rfloor$, where $f(p_{\text{cen}})$ is a function of
$p_{\text{cen}}$ satisfying the following two conditions: (i)
$f(p_{\text{cen}})$ is monotone-decreasing in $p_{\text{cen}}$, taking values
in $[0, 1)$ and satisfying $f(0) = 1$; and  (ii) $f(p_{\text{cen}})$ is
continuous with bounded first derivative.

\begin{algorithm}[!ht]
\caption{Adaptive double bootstrap resampling for optimal subsample selection}
\label{alg:dbbootstrap}
\textbf{Input:} Fractionally imputed dataset $D$ with $n$ observations,
censoring rate $p_{\text{cen}}$, number of outer bootstrap replicates $B_1$,
and number of inner bootstrap replicates $B_2$.\\

\textbf{Step 1: Grid selection and initial estimation.}
Define a grid for the sequence of $\gamma$ values, and compute a corresponding
sequence of resampling proportions: $c = (1 + \gamma \cdot
\exp(-p_{\text{cen}})) / (1 + \gamma)$, using these to determine the associated
resample sizes $m = \lfloor n^{c} \rfloor$. Compute initial estimates of the
mediation estimand of interest, e.g., $\hat{\Psi}_{\text{NDE}}$ and
$\hat{\Psi}_{\text{NIE}}$ using the dataset $D$.\\

\textbf{Step 2: Double bootstrap.} 
For each candidate value of $\gamma$: For each $b_1 = 1, \ldots, B_1$, the
outer $n$-out-of-$n$ first-stage bootstrap replicates:
  \begin{enumerate}
    \item Resample $n$ observations from $D$ with replacement, and compute
      the bootstrap NDE and NIE estimates $\hat{\Psi}_{\text{NDE}}^{(b_1)}$
      and $\hat{\Psi}_{\text{NIE}}^{(b_1)}$ using the resampled data.
    \item Conditional on each first-stage bootstrap sample, for $b_2 = 1,
      \ldots, B_2$, the inner $m$-out-of-$n$ bootstrap replicates:
      \begin{enumerate}
          \item Draw $m^{(b_1)}$-out-of-$n$ second-stage (nested) bootstrap
            samples.
          \item Construct the inner $m^{(b_1)}$-out-of-$n$ bootstrap
            confidence interval (e.g., percentile bootstrap or
            double-centered percentile
            bootstrap~\citep{efron1994introduction} confidence intervals).
      \end{enumerate}
    \item Check confidence interval coverage: Verify whether the initial
      estimates $\hat{\Psi}_{\text{NDE}}$, $\hat{\Psi}_{\text{NIE}}$ fall
      within the percentile bootstrap confidence intervals.\\
  \end{enumerate}
\textbf{Step 3: Selection of $m$.}
Estimate the double bootstrap confidence interval coverage rate from all the
first-stage bootstrap data sets. If the coverage rate achieves the desired
level, stop and select the current $c$ and $m$; otherwise, increment $\gamma$
to the next value.\\

\textbf{Output:} Resampling proportion $c^{\text{opt}}$ and resample size
$m^{\text{opt}}$.
\end{algorithm}

Building on the adaptive approach developed by~\cite{chakraborty2013inference}
for estimation in the context of optimal dynamic treatment regimes, we adopt a
data-driven procedure for selecting $m$ that incorporates the censoring rate
$p_{\text{cen}}$ and a sequence of hyperparameters $\gamma$. In view of
satisfying conditions (i) and (ii), we propose the following
\begin{align*}
  c_i = \frac{1 + \gamma_i \cdot \exp(-p_{\text{cen}})}{1 + \gamma_i} \ .
\end{align*}
When $p_{\text{cen}} = 0$, indicating no censoring and no anticipated
irregularity, $m = n$, and the procedure reduces to the standard bootstrap. For
a fixed $n$, the resample size $m$ lies in the interval $(n^{1 / (1 + \gamma)},
n]$, where $\gamma$ controls the smallest allowable resample size. By scaling
$m$ relative to $n$, the approach can mitigate inconsistency of the standard
bootstrap in settings with a meaningful degree of irregularity, helping to
provide more robust inference. To determine an optimal choice of $c_i$, we use,
in practice, a double bootstrap procedure, as outlined in Algorithm
\ref{alg:dbbootstrap}.

The double bootstrap procedure of Algorithm~\ref{alg:dbbootstrap}
begins
by initializing $\gamma$ to a small value, corresponding to a large $c$ and
$m$, and drawing $B_1$ samples of size $n$ from the original fractionally
imputed dataset (i.e., $\{(Y_i, \widetilde{M}^{\dagger}_i, A_i,
\boldsymbol{L}_i, W_i)\}_{i=1}^n$). For each sample $b_1$, estimates of the NDE
and NIE estimands, $\Psi_{\text{NDE}}^{(b_1)}(\hat{\Pf})$ and
$\Psi_{\text{NIE}}^{(b_1)}(\hat{\Pf})$ are computed with a candidate estimator.
An $m$-out-of-$n$ bootstrap is then performed with $B_2$ iterations using the
value of $\gamma$ fixed at the given first-stage step to obtain $m^{(b_1)}$.
The resulting $B_2$ resamples are used to construct confidence intervals via
appropriate methods (e.g., percentile bootstrap, double-centered percentile
bootstrap~\citep{efron1994introduction, chakraborty2013inference}). This
process is repeated for each of the $B_1$ samples and the coverage probability
is assessed by computing the proportion of confidence intervals found to
contain initial estimates of the NDE and NIE estimands from the original
fractionally imputed dataset. If the desired coverage level is achieved, the
corresponding $\gamma$ is selected as $\hat{\gamma}$; otherwise, $\gamma$ is
incremented along the pre-specified sequence, and the procedure is repeated.
The search space for $\gamma$ can be tailored for any given application, by
restricting its maximum value based on the smallest allowable resample size,
for instance. Once $\gamma$ is selected, the final $m$-out-of-$n$ bootstrap is
performed using the estimated optimal $m$.


While the double bootstrap procedure is directly applicable for the plug-in
estimator, it requires minor modification for compatibility with the one-step
estimator. Specifically, in the case of asymptotically efficient estimators
based on the EIF, we propose replacing the inner $m^{(b_1)}$-out-of-$n$ set of
bootstrap replicates with the multiplier bootstrap, a computationally
efficient alternative to the non-parametric bootstrap~\citep{gine1984some,
van1996weak, belloni2018uniformly, kennedy2019nonparametric, diaz2020causal}. To operationalize
the multiplier bootstrap within our adaptive double bootstrap procedure, the
inner $m^{(b_1)}$-out-of-$n$ bootstrap replicates are constructed by applying
resampled Gaussian or Rademacher multipliers to the $m^{(b_1)}$ estimated EIFs
computed from the outer bootstrap resamples, allowing for the construction of
confidence intervals as needed for the second stage. We provide details on this
proposed variant of our algorithm in Section~\ref{supp:MultiplierEIF} of the
\href{sm}{Supplementary Materials}. Briefly, the approach avoids re-estimation
of the nuisance parameters appearing in the EIF, reducing computational cost and
allowing for a larger number of inner bootstrap replicates (i.e., larger $B_2$)
to be used in practice. Meanwhile, fewer outer bootstrap replicates (i.e., lower
$B_1$) may be used to further reduce computational cost without compromising the
overall estimation strategy. We evaluate the performance of this proposal, among
others, in the experiments detailed next.

\section{Numerical studies}\label{sec:sim}
We investigated the performance of our proposed fractional imputation-based
strategy for estimation of the natural direct and indirect effects in a series
of numerical simulation experiments. In this section, we describe the structure
of our experiments and report their results. To generate synthetic data for our
experiments with a mediating variable left-censored by an assay lower limit of
quantification, we employed the following data-generating process (DGP) for the
joint distribution of the components of the observed data unit $O$:
\begin{align*}
L_1 &\sim \operatorname{Bern}(p=0.6) ;
L_2 \sim \operatorname{Bern}(p=0.5) ;
L_3 \sim \operatorname{Bern}(p=0.25) \ ; \\
A \mid \boldsymbol{L} &\sim
  \operatorname{Bern}\left(
    p = \operatorname{expit}(-1 + 0.5 L_1 + 1.25 L_2 + 0.75 L_3 - 1.25 L_1 L_3)
  \right) \ ; \\
\log M^{\dagger} \mid (A, \boldsymbol{L}) &\sim
  \operatorname{Norm}(
    -3 + 1.5 A + 1.75 L_1 + 0.25 A L_1+ 1.5 L_2 - 0.25 L_3,\; 0.25^2
  ) \ ; \\
Y \mid (M, A, \boldsymbol{L}) &\sim \operatorname{Bern}\left(
  p = \operatorname{expit}(-1 + 2.5 A + 1.75 M + 0.5 A M - 2.25 L_1 -
    1.75 L_2 - 1.5 L_3)
  \right) \ ,
\end{align*}
where, for convenience, we define $\operatorname{expit}(x) = \exp(x) / (1 +
\exp(x))$ to refer to the inverse logit transformation. In this DGP, we have
that $\boldsymbol{L} = (L_1, L_2, L_3)$ are binary baseline covariates, $A$ a
binary treatment assignment indicator, $M^{\dagger}$ a continuous mediator, and
$Y$ a binary outcome. Notably, the log-transformed mediator $M^{\dagger}$
represents an assay-derived measurement of a biological variable of interest
(e.g., viral RNA), thought to lie in a pathway between treatment and outcome. \biostats{The log-normal distribution was chosen as a scientifically plausible benchmark for positive, right-skewed biomarker measurements, consistent with previous analyses of assay-limited viral RNA data using Gaussian models on the log-transformed scale \citep{hughes2000, chernofsky2024causal} and qualitatively with the distribution of quantifiable Day 3 AN RNA in our motivating ACTIV-2 analysis (Figure \ref{subfig:distribution}).} 
The mediator is left-censored by an assay-specific LLoQ threshold, denoted
$\lambda_{\text{LLoQ}}$, such that the observed mediator values are reported as
falling at $\lambda_{\text{LLoQ}}$ when $M^{\dagger} < \lambda_{\text{LLoQ}}$ and
retained otherwise. In a given simulation experiment, we set
$\lambda_{\text{LLoQ}}$ as a fixed quantile of $M^{\dagger}$, with the quantile
chosen based on the desired censoring rate (e.g., $30\%$). We generate multiple
experimental scenarios in which the mediator is subject to censoring at one of
three levels $\{25\%, 50\%, 75\%\}$, allowing for the impact of censoring
severity on estimator performance to be gauged. Under the data-generating process described above, the probability of $Y=1$ ranges from 0.47 to 0.52. The true NDE and NIE, expressed on the risk difference (RD) scale, are 0.413 and 0.393, respectively. 

To evaluate the relative performance of differing imputation strategies with
respect to bias mitigation and efficient estimation of the NDE and NIE
estimands, we conducted two distinct sets of numerical studies. In the first,
study 1, we compared five strategies using fractional imputation while assuming
correct specification of relevant nuisance estimators, using these to
implement a plug-in estimator of the NDE and NIE estimands. This study had
purpose to compare the performance of several instantiations of our proposed
FI-EM approach, each with different density estimation strategies, against a
naive imputation strategy that replaces all left-censored mediator values with
LLoQ/2. In the second, study 2, we evaluate a subset of the fractional
imputation strategies from study 1, now outside the idealized context of
correct specification of nuisance estimators and using the asymptotically
efficient one-step de-biased estimators of the NDE and NIE estimands. The
purpose of this study was to evaluate a subset of our proposed strategies in
the more realistic setting of having several nuisance parameters estimated with
flexible machine learning strategies and used in the construction of
asymptotically efficient estimators now commonly employed in practice. Across
both studies, we assess each method using standard performance metrics,
including bias, variance, mean squared error (MSE), and confidence interval
(CI) coverage, with inference based on both standard and our proposed
data-driven $m$-out-of-$n$ bootstrap procedures. We conducted our simulation
experiments across a range of sample sizes: $n \in \{500, 1000, 1500, 2000\}$
and report metrics aggregated across $300$ Monte Carlo replicates. 
\begin{table}[h!]
\centering
\caption{Summary of imputation strategies evaluated in Simulation Study 1.
Methods differ in the specification or estimation of the conditional mediator
density, conditional mediator mean, and conditional outcome model, which
together determine the fractional weights in Equation~\eqref{wetdes}.
EM (1) uses correctly specified nuisance functions; EM (2) retains the
correct mediator density but misspecifies the mediator mean and outcome
models; EM (3) additionally replaces the true log-normal mediator density
with a misspecified Poisson working model; and EM (4) and EM (5) use
semi-parametric conditional density estimation with heteroscedastic and
homoscedastic location-scale specifications.}
\renewcommand{\arraystretch}{1.2}
\begin{tabular}{|>{\raggedright\arraybackslash}p{3.4cm}|
                >{\raggedright\arraybackslash}p{3.7cm}|
                >{\raggedright\arraybackslash}p{3.95cm}|
                >{\raggedright\arraybackslash}p{3.25cm}|}
\hline
\textbf{Methods} & \textbf{Density form of $M^{\dagger}$} & \textbf{Mediator mean model $\E(M^{\dagger} \mid A, \boldsymbol{L})$} & \textbf{Outcome model $\P(Y=1 \mid M^{\dagger}, A, \boldsymbol{L})$} \\
\hline
EM (1): true nuisances & Correctly specified (Log-normal) & Correctly specified & Correctly specified \\
\hline
EM (2) : oracle density, misspecified GLMs & Correctly specified (Log-normal) & Misspecified (omits $L$) & Misspecified (omits $L$) \\
\hline
EM (3): misspecified density and GLMs & Misspecified (Poisson working model) & Misspecified (omits $L$) & Misspecified (omits $L$) \\
\hline
EM (4): heteroscedastic CDE, HAL & Location-scale family
& Conditional mean and variance estimated via HAL & Flexible (HAL-based) \\
\hline
EM (5): homoscedastic CDE, HAL & Location-scale family
& Only conditional mean estimated via HAL & Flexible (HAL-based) \\
\hline
LLoQ/2 imputation & Degenerate point mass at LLoQ/2 & None & None \\
\hline
\end{tabular}
\vspace{0.4em}
\begin{minipage}{0.96\textwidth}
\footnotesize
\raggedright
\footnotesize
\textit{Note:} EM = expectation-maximization; GLM = generalized linear model;
CDE = conditional density estimation; HAL = highly adaptive lasso;
LLoQ = lower limit of quantification.
\end{minipage}
\label{tab:methods}
\end{table}

We evaluated six imputation strategies reflecting a spectrum of model
specifications and estimation approaches. The first, \textbf{EM (1): true
nuisances}, employs a correctly specified log-normal density for the mediator
$M^{\dagger}$, alongside correctly specified models for $\E(M^{\dagger} \mid A,
L)$ and $\E(Y \mid M^{\dagger}, A, L)$. Because this method uses the true
nuisance functions, it represents the best-case scenario, a result of correct
model specification for nuisance functions. \biostats{This correctly specified
parametric setting is most closely related to the modeling regime under which
direct likelihood approaches such as that of~\cite{chernofsky2024causal} are
expected to perform well, although EM (1) uses the true nuisance functions and
therefore represents an \textit{oracle benchmark} rather than an implementation
of their estimator. The remaining EM specifications examine departures from
this idealized setting through nuisance-model misspecification or flexible
semi-parametric estimation.} The second, \textbf{EM (2): oracle density,
misspecified GLMs}, retains the correctly specified log-normal density for
$M^{\dagger}$ but omits some relevant covariates in both the mediator and
outcome nuisance models.
\biostats{The third, \textbf{EM (3): misspecified density and GLMs}, combines
the same misspecified nuisance models with a misspecified Poisson working
distribution for $M^\dagger$. The Poisson distribution was selected as an
unfavorable working model to examine sensitivity to substantial
misspecification of the mediator density. Relative to the true continuous
log-normal distribution, the Poisson working model is discrete and imposes the
restrictive structural assumption of equality of the conditional mean and
variance, thereby substantially misspecifying both the shape and dispersion of
the mediator distribution.} The fourth, \textbf{EM (4): heteroscedastic CDE, HAL}, employs a
semi-parametric location-scale family for conditional density estimation with
heteroscedastic errors, estimating both the conditional mean and variance of
$M^{\dagger}$ using HAL. Finally, the fifth, \textbf{EM (5): homoscedastic CDE,
HAL}, applies the same framework as the fourth but assumes homoscedastic
errors, estimating only the conditional mean of $M^{\dagger}$ via HAL. As a
benchmark, we also consider \textbf{LLoQ/2 imputation}, which replaces
left-censored entries of $M^{\dagger}$ with a fixed value, half the assay limit
of quantification. As emphasized in a previous remark, the conventional ad-hoc
imputation methods (usually choosing LLoQ/2) are special cases of our
framework, with a proposal distribution corresponding to a degenerate
distribution at a single point and the absence of mediator and outcome models
in the imputation. These six strategies constitute an extensive set of the
trade-offs between model complexity, flexibility, and robustness under varying
degrees of nuisance model misspecification.
Table~\ref{tab:methods}
summarizes the candidate approaches.

\begin{figure}[!]
    \centering
    \begin{subfigure}[t]{0.83\textwidth}
        \centering
        \includegraphics[width=\textwidth]{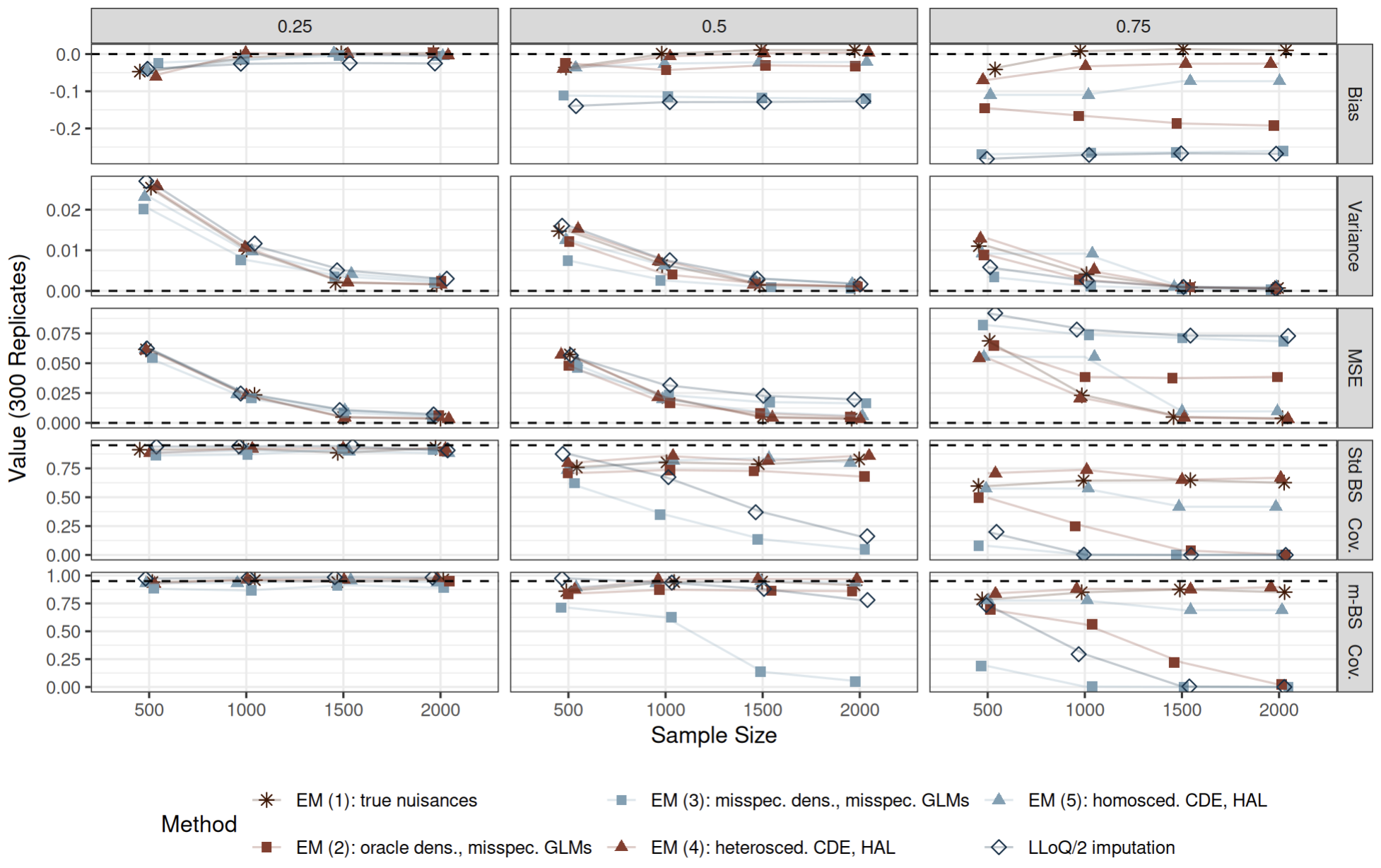}
        \caption{Study 1: Parametric G-computation NIE estimates under
        different imputation methods} \label{study1nde}
    \end{subfigure}
    \vspace{0.05em}
    \begin{subfigure}[t]{0.83\textwidth}
        \centering
        \includegraphics[width=\textwidth]{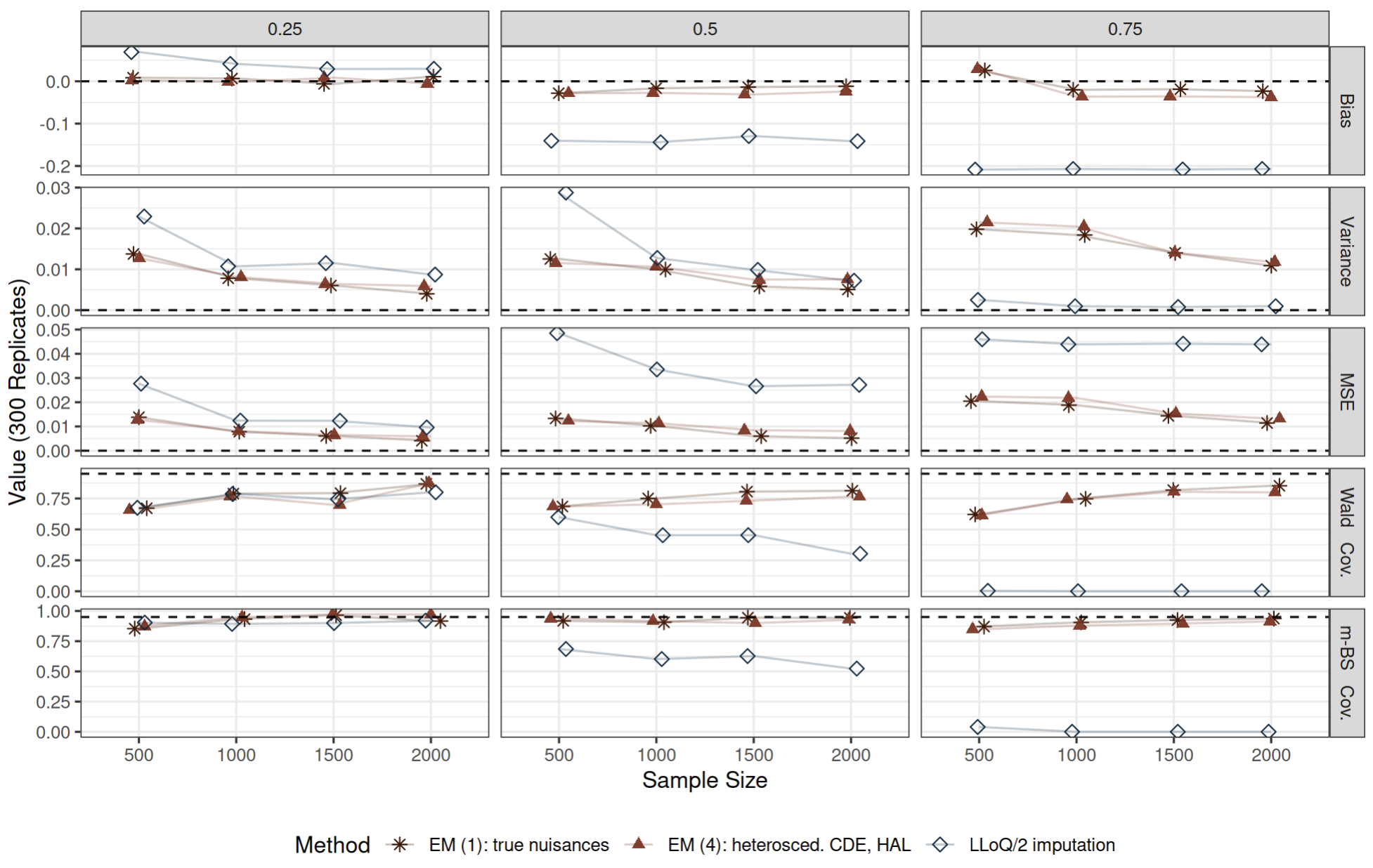}
        \caption{Study 2: Flexible one-step NIE estimates under different
        imputation methods} \label{study2nde}
    \end{subfigure}
\caption{\biostats{Performance of the parametric G-computation (top panel,
    Study 1) and one-step estimators (bottom panel, Study 2) of the NIE with
    several candidate imputation strategies across multiple experimental
    scenarios in which the mediator is subject to censoring at rates of $25\%$,
    $50\%$, or $75\%$, and across sample sizes $n \in \{500, 1000, 1500,
    2000\}$. Each panel reports bias, variance, mean squared error (MSE), and
    empirical $95\%$ confidence interval coverage (Cov.). Two coverage rates
    are reported: one based on the standard method and one based on the
    proposed $m$-out-of-$n$ bootstrap (denoted m-BS Cov.). For plug-in
    estimates, the standard method uses percentile bootstrap intervals (Std
    BS), whereas for one-step estimates, Wald-type confidence intervals are
    used (Wald Cov.). The imputation methods correspond to those described in
    Table~\ref{tab:methods},
    and results are jittered to aid visualization. Results for the NDE are
    provided in the \href{sm}{Supplementary Materials},
    Section~\ref{supp:sm_sims}.}}
    \label{fig:gcom}
\end{figure}

Figures \ref{fig:gcom}
and~\ref{supp:fig:1step} (Section~\ref{supp:sm_sims} of the
\href{sm}{Supplementary Materials}) present simulation results for the NIE and
NDE, respectively. Specifically, Figures~\ref{study1nde} 
and~\ref{supp:study1nie} report results from simulation experiments in Study 1
using parametric G-computation estimators based on correctly specified nuisance
models used for estimation of the NIE and NDE. Figures~\ref{study2nde}
and~\ref{supp:study2nie}
display results from simulation experiments in Study 2 evaluating one-step
estimators of the NIE and NDE under varying degrees of left-censoring. Across
all sample sizes and censoring levels, the LLoQ/2 imputation approach exhibits
substantial bias, particularly under moderate ($50\%$) and severe ($75\%$)
left-censoring of $M^{\dagger}$. Although its bias is less pronounced under
mild censoring ($25\%$), it still underperforms relative to variants of the
proposed strategy, underscoring the importance of addressing left-censoring due
to LLoQ in a data-informed manner.

For the results of Study 1, both the G-computation estimates of the NIE
(Figure~\ref{study1nde}) and
the NDE (Figure~\ref{supp:study1nie}) exhibit similar performance.
Approach EM (3), which combines a misspecified density for $M^{\dagger}$ and
misspecified GLMs for both the mediator and outcome, performs about as poorly
as the naive approach. In contrast, approach EM (2), which uses the correctly
specified density form of $M^{\dagger}$ but misspecified mediator and outcome
models, achieves improved performance, especially under moderate and
severe censoring. 

As expected, approach EM (1), which uses correctly specified nuisance
functions, consistently performs best across all scenarios, achieving minimal
bias and low MSE. \biostats{The flexible semi-parametric strategies implemented
in approaches EM (4) and EM (5) also exhibit strong performance. In particular,
EM (4), which employs HAL for estimation of both the conditional mean and
variance of \(M^{\dagger}\) under a heteroscedastic conditional density model,
performs well throughout all the settings considered. Notably, approach EM (4)
performs comparably to approach EM (5). Taken together with the poor
performance of EM (3), these results illustrate the sensitivity of fully
parametric estimation to mediator-density misspecification and the greater
robustness afforded by flexible conditional density estimation within the
settings considered.}
\biostats{With respect to inference, although the proposed \(m\)-out-of-\(n\)
bootstrap generally improves coverage relative to the standard bootstrap, its
coverage remains below the nominal \(95\%\) level in several settings in
Study 1, with the most pronounced undercoverage occurring under \(75\%\)
mediator censoring. As such, the proposed procedure does not recover reliable
inference at the nominal rate under severe mediator censoring.}

In the case of Study 2, Figures~\ref{study2nde}
and~\ref{supp:study2nie} depict simulation results examining how one-step
estimators of the NIE and NDE perform as the extent of left-censoring
increases. Mirroring study 1, the adhoc LLoQ/2 imputation method exhibits
substantial bias across all sample sizes, with bias worsening noticeably under
$50\%$ and $75\%$ censoring. In contrast, approach EM (1), which assumes fully
correctly specified nuisance functions, performs consistently well across all
settings. Similarly, approach EM (4), which uses HAL to estimate both the
conditional mean and conditional variance of the mediator under a
heteroscedastic conditional density model, also demonstrates strong
performance, yielding negligible bias, low MSE, and approximately nominal
$95\%$ confidence interval coverage with the $m$-out-of-$n$ bootstrap. These
results suggest that when the true data-generating mechanism is unknown, the
one-step estimator combined with EM (4) provides the most reliable and broadly
applicable option among the methods considered.



\section{Illustrative application in the ACTIV-2 platform trial}\label{sec:realdata}
We illustrate our proposed approach using data from the ACTIV-2 platform trial
(see Section~\ref{intro}), focusing on participants
randomized to amubarvimab plus romlusevimab (A/R) or its corresponding placebo.
To align with prior analyses and ensure well-defined mediator measurement, we
restrict attention to participants who remain event-free (i.e., no
hospitalization or death) through Day 3 and have successfully measured Day 3
anterior nasal (AN) SARS-CoV-2 RNA. Motivated by the hypothesis that monoclonal
antibody (mAb) treatment reduces disease progression through early viral
suppression and/or elimination, we posit that Day 3 AN RNA as a mediator of the
effect of mAb treatment on the outcome, defined as hospitalization or death
occurring after Day 3. This framing enables decomposition of the post-Day 3
treatment effect into (i) a natural direct effect, capturing all mechanisms not
operating through Day 3 AN RNA, and (ii) a natural indirect effect, quantifying
the portion of the total effect mediated by early viral suppression and/or
elimination as measured via AN RNA.

\subsection{Organization of ACTIV-2 variables into the proposed framework}\label{activ2setup}

We map the ACTIV-2 trial data into our proposed LLoQ-censored data structure
for causal mediation analysis: $(\boldsymbol{L}, A, M, Y, C)$, where the
continuous mediator is $M = \max(M^{\dagger}, \lambda_{\text{LLoQ}})$ and $C$
is the indicator of mediator censoring by the assay limit of quantification.
Baseline covariates $\boldsymbol{L}$ include age, sex, BMI, race/ethnicity,
country, days since symptom onset, vaccination status, baseline AN RNA, and selected
comorbidities (hypertension, obesity, diabetes). These variables are measured
prior to treatment assignment and are included to account for potential
confounding of the mediator--outcome relationship; note that confounding of the
treatment--mediator and treatment--outcome relationships are addressed by
treatment randomization. Treatment is encoded as a binary indicator $A \in
\{0,1\}$, with $A=1$ denoting active A/R mAb therapy and $A=0$ placebo. The
mediator $M$ is defined as the $\log_{10}$-transformed SARS-CoV-2 RNA level
measured from AN swabs at Day 3, representing the early post-treatment
virologic response; the allowable visit window for this assessment was given a
$\pm 1$ day grace period. The outcome $Y$ is a binary indicator of
hospitalization or death occurring after Day 3 through Day
28. 

\begin{figure}[htbp]
    \centering
    \begin{subfigure}[b]{0.48\textwidth}
        \centering
        \includegraphics[width=\textwidth]{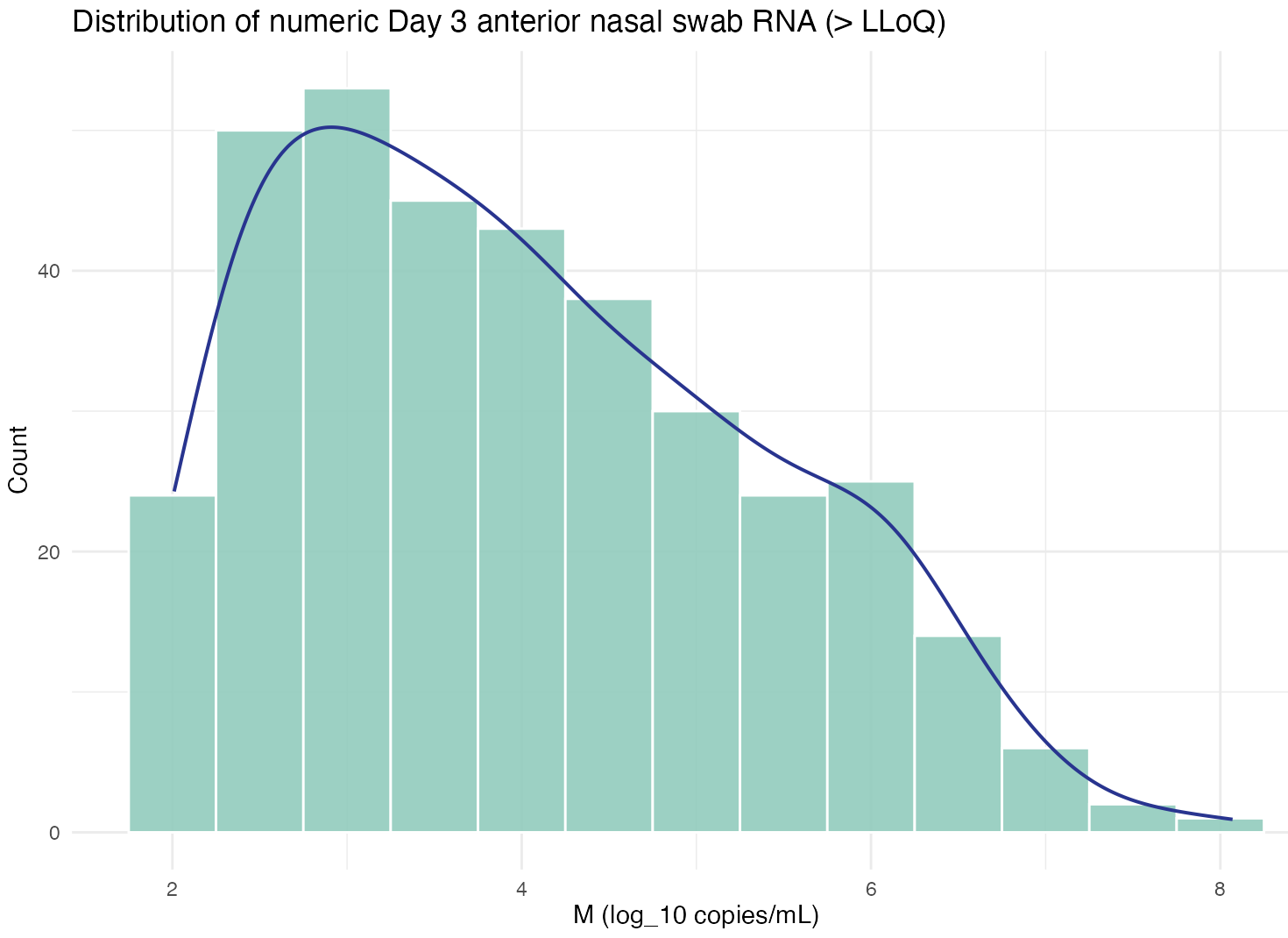}
        \caption{Distribution of quantifiable Day 3 AN RNA levels ($\log_{10}$
        copies/mL) among the 355 participants with values $\ge
        \lambda_{\text{LLoQ}}$. The right-skewed distribution is typical of
        early post-treatment viral load measurements.}
        \label{subfig:distribution}
    \end{subfigure}
    \hfill
    \begin{subfigure}[b]{0.48\textwidth}
        \centering
           \includegraphics[width=\textwidth]{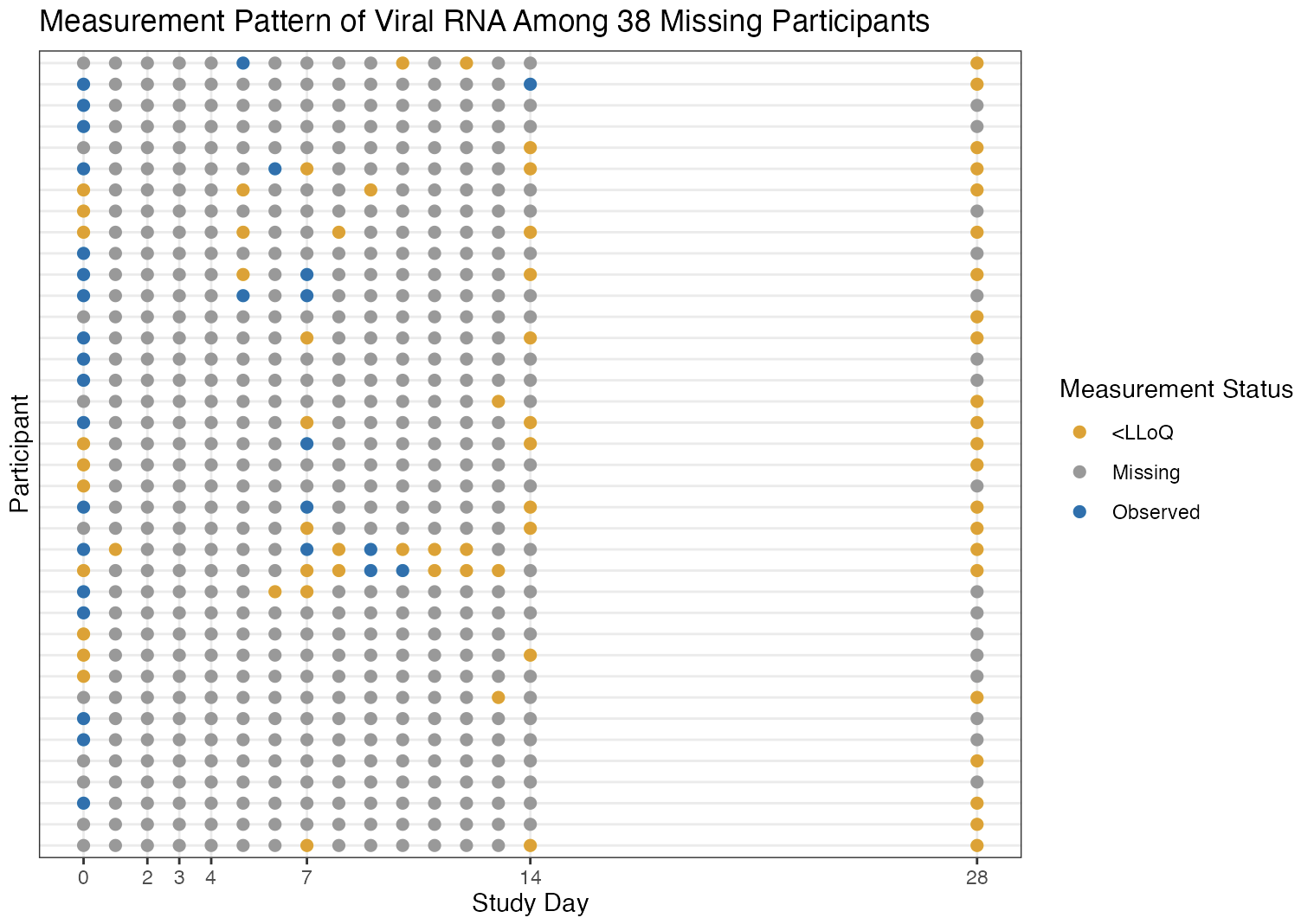}
        \caption{Temporal patterns of AN RNA measurements for the 38
        participants with missing Day 2-4 AN RNA. Each row indicates an
        individual participant's trajectory of AN RNA measurements, with Days
        0, 2, 3, 4, 7, 14, and 28 indicated where available.}
        \label{subfig:missing_patterns}
    \end{subfigure}
    \caption{Mediator data summary for our ACTIV-2 analysis participants. (a)
    Histogram of Day 3 AN SARS-CoV-2 RNA levels among the 355 participants with
    quantifiable values above $\lambda_{\text{LLoQ}}$. (b) Individual viral AN RNA trajectories
    for the participants who had missing Day 3 RNA measurements.}
    \label{fig:rna_distribution_and_missing_patterns}
\end{figure}

Baseline characteristics by treatment group are summarized in
Table~\ref{supp:tab:baseline}, with further details provided in
Section~\ref{supp:real-world} of the \href{sm}{Supplementary Materials}. The
final analysis sample comprises $n=749$ individuals ($n_1=371$ treated, $n_0 =
378$ placebo). Among these, $n_{<\text{LLoQ}} = 356$ participants had Day 3 AN
RNA measurements below the LLoQ, where $\lambda_{\text{LLoQ}} = 2.0\log_{10}$
copies/mL, $n_{\geq\text{LLoQ}} = 355$ had quantifiable levels above the LLoQ,
and $n_{\text{NA}} = 38$ had missing measurements (i.e., no sample collected or
assay failure). The distribution of the $n_{\geq\text{LLoQ}} = 355$
quantifiable $\log_{10}$ SARS-CoV-2 RNA levels is displayed in
Figure~\ref{subfig:distribution}.
The temporal patterns of RNA measurements across Days 0-28 for the 38
participants with missing Day 2-4 AN RNA measurements are presented in
Figure~\ref{subfig:missing_patterns}. \biostats{
To assess the adequacy of a conditional Gaussian approximation
for Day 3 log$_{10}$ AN RNA, we fitted a left-censored Gaussian
regression model adjusting for treatment assignment and
baseline AN RNA. The Q-Q plot of truncation-adjusted residuals
showed no substantial departures from normality over the
quantifiable range (Supplementary
Figure~\ref{supp:fig:qq_conditional}), supporting this
approximation. Our application includes both a parametric
estimator that assumes conditional Gaussianity and a
semiparametric estimator of the conditional mediator density
that retains a location--scale representation but does not
require conditional Gaussianity.
}

Consistent with the analysis previously reported
by~\cite{giganti2023association}, our main analysis is restricted to
participants with observed Day 3 RNA measurements (including values below the
LLoQ), excluding 38 participants with missing Day 3 AN RNA. This results in a
left-censoring rate due to the LLoQ of approximately $50.1\%$ (356 out of 711
observed measurements). As a sensitivity analysis, we implement a model-based
imputation approach in which we model the conditional distribution of RNA given
that it exceeds the LLoQ as a function of baseline covariates, and then impute
values for the 38 missing observations. Under this sensitivity analysis, the
effective left-censoring rate is reduced to approximately $47.53\%$; results
are presented in Table~\ref{supp:tab:sensitivity} and detailed in
Section~\ref{supp:real-world} of the \href{sm}{Supplementary Materials}.

\biostats{Regarding the identification of the natural direct and indirect effects, to
avoid reliance on the much-debated cross-world exchangeability assumption
(Assumption~\ref{assumption4}), we adopt an interventionist (separable
effects) formulation as an alternative identification strategy for the ACTIV-2
analysis. This approach yields the same statistical estimands as the
traditional natural direct and indirect effects, while grounding them in
well-defined, hypothetically manipulable interventions; accordingly, our
estimation procedure remains unchanged. The construction of the separable
interventions in ACTIV-2, the required identification assumptions, and their
representation via edge-expanded graphs and single-world intervention graphs
are provided in Section~\ref{supp:Separable} of the \href{sm}{Supplementary
Materials}. Corresponding results under this alternative identification
strategy are shown in Figure~\ref{supp:fig:swigseparability} therein.}

\subsection{Results of ACTIV-2 data analysis}\label{activ2analysis}

Table~\ref{tab:resACTIV2} presents estimates of the natural direct and indirect
effects (NDE/NIE) of A/R mAb therapy on the risk of hospitalization or death
through Day 28, measured on the risk-difference scale, across multiple varying
candidate estimators and mediator imputation strategies. The estimated total
effect is consistently protective across all specifications. The NDE accounts
for the majority of the treatment effect, with estimates ranging from
approximately $-0.02$ to $-0.04$ and remaining stable across parametric
G-computation, one-step, and targeted maximum likelihood estimation (TMLE). In
contrast, the NIE mediated through Day 3 AN RNA is small, typically between
$-0.0010$ and $-0.0035$.

\biostats{Flexible imputation strategies---including our EM-based parametric
log-normal modeling and semi-parametric heteroscedastic conditional density
estimation (CDE) approaches---yield slightly more negative NIE estimates than
LLoQ/2 imputation, although the absolute differences in the NIE estimates are
small (at most $0.24\%$). The corresponding proportion mediated (PM) estimates
range from approximately \(7\%\) to \(12\%\) when using flexible one-step and
TML estimators with ensemble machine learning (i.e., super learner) for
nuisance estimation, compared with approximately \(4\%\) under parametric
G-computation with GLMs for nuisance estimation.} By contrast, the conventional
approach of imputing values below the LLoQ with \(\mathrm{LLoQ}/2\) produces PM
estimates typically between \(3\%\) and \(5\%\), with less variation across
candidate estimators. Across imputation strategies, EM-based approaches tend to
yield slightly attenuated NDEs and numerically more negative NIEs relative to
imputation by LLoQ/2, with the heteroscedastic CDE approach producing higher PM
estimates when combined with one-step or TML estimators.

Overall, these results indicate that early reduction in AN RNA of SARS-CoV-2
explains only a small fraction of the protective effect of A/R mAb therapy,
with the majority of the well-characterized benefit operating through pathways
not mediated by AN RNA. This finding is consistent with prior results reported
by~\cite{giganti2023association} and further supports the conclusion that AN
viral RNA is unlikely to be a reliable surrogate endpoint for the efficacy of
mAb therapies in COVID-19 therapeutics trials. We note several limitations
of our approach. Firstly, to ensure the temporal ordering required for causal
mediation analysis, we restricted attention to participants who remained
event-free through Day 3; consequently, our estimates pertain to treatment
effects on subsequent hospitalization or death within this risk set defined by
a post-baseline event.
Secondly, in ACTIV-2, AN swabs were self-collected by trial participants;
although prior work suggests strong concordance with provider-collected
samples, some measurement error may attenuate estimates of the indirect effect.
Thirdly, the study population was largely unvaccinated (about 3\%), which may
limit generalizability to settings with widespread vaccination or hybrid
immunity, as now the case globally.


\begin{table}[h]
\centering
\begin{threeparttable}
\caption{Natural direct and anterior nasal viral load-mediated indirect effects
of mAbs on hospitalization/death risk}
\label{tab:resACTIV2}

\setlength{\tabcolsep}{6pt}
\renewcommand{\arraystretch}{1.15}

\begin{tabularx}{\textwidth}{>{\raggedright\arraybackslash}p{2.6cm} *{3}{>{\centering\arraybackslash}X}}
\toprule
& \multicolumn{3}{c}{Imputation strategy} \\
\cmidrule(lr){2-4}
Estimator 
& LLoQ / 2 
& EM: log-normal density fractional imputation
& EM: heteroscedastic
CDE fractional imputation\\
\midrule

G-computation &
\makecell[c]{\textbf{NDE}:  -0.0394 (0.0132) \\ {\small 95\% CI: [-0.0771,\, -0.0010]} \\
\textbf{NIE}: -0.0010 (0.0077) \\ {\small 95\% CI: [-0.0456,\, 0.0095]}\\
\textbf{PM}: 2.57 $\%$ }  &
\makecell[c]{\textbf{NDE}: -0.0305 (0.0123) \\ {\small 95\% CI: [-0.0674,\, 0.0041]} \\
\textbf{NIE}: -0.0013 (0.0100) \\ {\small 95\% CI: [-0.0678,\, 0.0066]}\\
\textbf{PM}:  4.31 $\%$ }  &
\makecell[c]{\textbf{NDE}: -0.0307 (0.0120) \\ {\small 95\% CI: [-0.0645,\, 0.0055]}\\
\textbf{NIE}: -0.0012 (0.0076) \\ {\small 95\% CI: [-0.0611,\, 0.0037]}\\
\textbf{PM}: 3.97 $\%$ }  \\
\hline
\addlinespace[0.35em]

One-step &
\makecell[c]{\textbf{NDE}:  -0.0360 (0.0123) \\ {\small 95\% CI: [-0.0767,\,0.0062]} \\
\textbf{NIE}: -0.0018 (0.0008) \\ {\small 95\% CI: [-0.0045,\,0.0009]}\\
\textbf{PM}: 4.72 $\%$ }  &
\makecell[c]{\textbf{NDE}: -0.0199 (0.0067) \\ {\small 95\% CI: [-0.0451,\,0.0017]} \\
\textbf{NIE}: -0.0024 (0.0009) \\ {\small 95\% CI: [-0.0061,\,0.0013]}\\
\textbf{PM}: 10.81 $\%$ }  &
\makecell[c]{\textbf{NDE}: -0.0198 (0.0062) \\ {\small 95\% CI: [-0.0432,\,0.0036]} \\
\textbf{NIE}: -0.0028 (0.0010) \\ {\small 95\% CI: [-0.0066,\,0.0010]}\\
\textbf{PM}: 12.31 $\%$ }  \\
\hline
\addlinespace[0.35em]

TMLE &
\makecell[c]{\textbf{NDE}: -0.0389 (0.0123) \\ {\small 95\% CI: [-0.0796,\, 0.0031]} \\
\textbf{NIE}: -0.0021 (0.0008) \\ {\small 95\% CI: [-0.0048,\,0.0007]} \\
\textbf{PM}: 5.04 $\%$ } 
&
\makecell[c]{\textbf{NDE}: -0.0274 (0.0067) \\ {\small 95\% CI: [-0.0523,\, -0.0025]} \\
\textbf{NIE}: -0.0021 (0.0007) \\ {\small 95\% CI: [-0.0046,\, 0.0003]}\\
\textbf{PM}: 7.25 $\%$ }  &
\makecell[c]{\textbf{NDE}: -0.0280\ (0.0057) \\ {\small 95\% CI: [-0.0623,\ 0.0063]} \\
\textbf{NIE}: -0.0034\ (0.0004) \\ {\small 95\% CI: [-0.0062,\ -0.0007]}\\
\textbf{PM}: 10.90 $\%$ }  \\

\bottomrule
\end{tabularx}

\begin{tablenotes}[flushleft]
\footnotesize
\item \textit{Notes:} Point estimates are on the risk-difference scale. The
  proposed m-out-of-n bootstrap standard errors (SE) are shown in parentheses,
  and the corresponding confidence intervals are 95\%. Columns correspond to
  increasingly flexible (less parametric) imputation/density modeling for the
  mediator; rows correspond to different estimators, including parametric
  G-computation, one-step, and TMLE. The Proportion Mediated (PM), defined as
  the ratio of the natural indirect effect to the total effect.
\end{tablenotes}
\end{threeparttable}
\end{table}

\section{Discussion}\label{sec:disc}
Motivated by the question of how monoclonal antibody therapies reduce the risk
of hospitalization and death in the ACTIV-2 platform trial, we developed a
semi-parametric framework for causal mediation analysis when the candidate
mediator is subject to left-censoring by an assay's lower limit of
quantification. The resulting censoring mechanism is both missing not at random
and deterministic, precluding non-parametric identification of the natural
direct and indirect effects; we address this challenge by imposing parametric
structure on the conditional mediator and outcome models to restore
identifiability from the incomplete observed data. Building on this
identification strategy, our proposed estimation approach integrates fractional
imputation with a semi-parametric EM algorithm, pairing flexible location-scale
conditional density estimation with a data-adaptive $m$-out-of-$n$ bootstrap
for inference under irregularity induced by imputation. Applying this
framework to ACTIV-2, we find that early reduction in anterior nasal
SARS-CoV-2 RNA explains only a modest fraction of the protective effect of
amubarvimab plus romlusevimab on the risk of hospitalization or death, with
proportions mediated of about 4\%--12\% across estimators and
imputation strategies. These results suggest that the majority of the
well-characterized treatment benefit operates through pathways not captured by
Day~3 anterior nasal RNA, consistent with prior
findings~\citep{giganti2023association} and reinforcing that this biomarker is
unlikely to serve as a reliable surrogate endpoint for mAb efficacy in COVID-19
therapeutics trials.

\biostats{Several directions for future methodological work may be prioritized.
Although the proposed semi-parametric conditional density estimator relaxes
parametric distributional assumptions, it retains a location-scale
representation, and its performance under more substantial departures from
this structure, including strongly multimodal or heavy-tailed conditional
mediator distributions, warrants further investigation. Relatedly, we propose
an \(m\)-out-of-\(n\) bootstrap that improves finite-sample coverage relative
to the standard bootstrap in our numerical studies; however, coverage remains
below the nominal level in some settings, particularly with \(75\%\) mediator
censoring. This suggests that inference based on the proposed procedure should
be interpreted cautiously in settings with severe mediator censoring and could
motivate further work on inference procedures with improved finite-sample
calibration.}

\biostats{A natural extension of the present framework is to jointly address
left-censoring due to the limit of quantification and classical measurement
error in the mediator. These two data-quality challenges frequently co-occur
in assay-based studies and are intrinsically linked: measurement error in the
mediator is known to attenuate estimates of the indirect effect toward the
null~\citep{cole2014manifest}, and this attenuation may be pronounced in our
setting given that anterior nasal swabs are self-collected by participants,
introducing potentially non-trivial biological and procedural variability.
Recent work on causal mediation with HIV RNA as a mediator has demonstrated
that incorporating separately available estimates of assay measurement error
can improve estimates of the indirect effects~\citep{lok2025causal};
analogous adjustments may be warranted in the ACTIV-2 setting and represent a
promising avenue for further investigation. A second direction concerns
extensions to multiple mediators: in many clinical settings, treatment
effects operate through several interrelated biological pathways, and
disentangling these mechanisms raises both conceptual and statistical
challenges, compounded when multiple mediators are subject to detection
limits or measurement error. Finally, when the outcome is time-to-event in
nature, accommodating censoring of both the mediator and the survival
outcome, as well as possible competing risks, within the present
identification and estimation framework remains an important direction for
future development.}

\section*{Supplementary Materials}

\href{sm}{Supplementary Materials}, available online, include proofs
of results and additional details on the presented estimators.

\subsection*{Code and Data Availability}

Code for reproducing the simulation studies and illustrative data analysis is
publicly available from GitHub and may be accessed at
\url{https://github.com/nshlab/pub_code_lodmed_biostats}.

\subsection*{Acknowledgments}

NSH and MDH were partially supported by a grant from the National Institute of
Allergy and Infectious Diseases (award no.~UM1 AI068634). The content is solely
the responsibility of the authors and does not represent the official views of
the National Institutes of Health. The authors would like to thank the
Advancing Clinical Therapeutics Globally (ACTG) network for provisioning data
from the ACTIV-2 clinical trial for use in the illustrative data analysis as
well as the ACTIV-2 study participants and study team for their contributions.


\bibliography{refs}

\end{document}


\maketitle


\section{Identification under Informative Censored Mediators: unique solution of score function}\label{s1_iden}

This section provides detailed conditions for identification under informative
censoring of the mediator, focusing on the uniqueness of the solution to the
score function.

To guarantee that the model \(\mathcal{P} = \{\mathbb{P}(Y_i \mid
M^{\dagger}_i, A_i, \boldsymbol{L}_i; \boldsymbol{\theta}), f(M^{\dagger}_i
\mid A_i, \boldsymbol{L}_i; \boldsymbol{\theta}): \boldsymbol{\theta} \in
\Omega\}\) is identifiable, it is crucial to ensure the uniqueness of the
maximizer of the observed log-likelihood. Similarly, this requirement can also
be viewed in terms of the observed score function, where the uniqueness of the
solution indicates that there is only one set of parameter values for which the
score equation \(U(\boldsymbol{\theta}) = 0\) holds true.

The observed score equation, the derivative of the observed log-likelihood with
respect to the parameters \( \boldsymbol{\theta} \), is given by
\begin{align*}
   U(\boldsymbol{\theta}) = \partial
   \ell_{\text{obs}}(\boldsymbol{\theta})/\partial \boldsymbol{\theta} =
   \sum_{i:C_{i}=1} U_{\text{obs},i}(\boldsymbol{\theta}) + \sum_{i:C_{i}=0}
   U_{\text{mis},i}(\boldsymbol{\theta}),
\end{align*}
where
\begin{align*}
U_{\text{obs},i}(\boldsymbol{\theta}) &= \frac{\partial \log \mathbb{P}(Y_i
\mid M^{\dagger}_i, A_i, \boldsymbol{L}_i; \boldsymbol{\theta})}{\partial
\boldsymbol{\theta}} + \frac{\partial \log f(M^{\dagger}_i \mid A_i,
\boldsymbol{L}_i; \boldsymbol{\theta})}{\partial \boldsymbol{\theta}}, \\
  U_{\text{mis},i}(\boldsymbol{\theta}) &= \int_{0}^{\lambda_{\text{LLoQ}}
  }\left[\frac{\partial \mathbb{P}(Y_i \mid M^{\dagger}, A_i, \boldsymbol{L}_i;
  \boldsymbol{\theta})}{\partial \boldsymbol{\theta}} f(M^{\dagger} \mid A_i,
\boldsymbol{L}_i; \boldsymbol{\theta}) + \frac{\partial f(M^{\dagger} \mid A_i,
\boldsymbol{L}_i; \boldsymbol{\theta})}{\partial
\boldsymbol{\theta}}\mathbb{P}(Y_i \mid M^{\dagger}, A_i, \boldsymbol{L}_i;
\boldsymbol{\theta})\right] dM^{\dagger} \Big/ \mathcal{D},
\end{align*}
and the denominator
\begin{align*}
\mathcal{D} \coloneqq \int_{0}^{\lambda_{\text{LLoQ}} } \mathbb{P}(Y_i \mid
M^{\dagger}, A_i, \boldsymbol{L}_i; \boldsymbol{\theta}) f(M^{\dagger} \mid
A_i, \boldsymbol{L}_i; \boldsymbol{\theta}) dM^{\dagger}.
\end{align*}
To ensure that the score equation $U(\boldsymbol{\theta}) = 0$ has a unique
solution, we impose the condition of strict convexity on the observed
log-likelihood
$\ell_{\text{obs}}(\boldsymbol{\theta})$~\citep{van2000asymptotic}.
Specifically, if $\ell_{\text{obs}}(\boldsymbol{\theta})$ is strictly convex
with respect to $\boldsymbol{\theta}$, it follows that the log-likelihood
function is uniquely maximized at a specific point. Consequently, this implies
that \( U(\boldsymbol{\theta}) \), the derivative of the log-likelihood, has a
unique zero-point solution. The strict convexity condition is satisfied if the
information matrix $I(\boldsymbol{\theta}) = -\frac{\partial^2
\ell_{\text{obs}}(\boldsymbol{\theta})}{\partial \boldsymbol{\theta} \partial
\boldsymbol{\theta}^{\top}}$ is positive definite.

\section{Supplementary Materials of Estimation and Inference}

This section provides additional details supporting the estimation and
inference procedures developed in the main text. In
subsection~\ref{generic-inf-cens},
we first describe the implementation of the EM algorithm under general
informative censoring, accommodating the complexities introduced by
limit-of-quantification constraints. In
subsection~\ref{s1:proofofTH1},
we then present the proof of the main theoretical result
(Theorem~\ref{main:thm:effem}), establishing the asymptotic properties of the
proposed estimators. Next, in
subsection~\ref{sec:hal}, we outline
the role of Highly Adaptive Lasso (HAL) in flexibly estimating nuisance
functions within our semiparametric framework. Finally, for inference, in
subsection~\ref{MultiplierEIF},
we detail the adaptive double bootstrap procedure, leveraging multiplier
bootstrap techniques and influence function representations to enable valid and
robust inference.

\subsection{EM Algorithm under General Informative Censoring}\label{generic-inf-cens}

\begin{algorithm}[!]
\caption{EM algorithm by fractional imputation for informative censored
mediators in mediation analysis} \label{alg:iwglm}

\textbf{Input:} Proposed density of mediators \( f(m_{i} \mid a_{i},
\boldsymbol{l}_{i}; \boldsymbol{\beta}^{(0)})\) with initial parameter
$\boldsymbol{\beta}^{(0)},$ outcome model $\mathbb{P}(Y_i \mid M^{\dagger}_i,
A_{i}, \boldsymbol{L}_{i}; \boldsymbol{\alpha}),$ and censoring probability
model $\pi(M_{i}, A_{i}, \boldsymbol{L}_{i}) = \mathbb{P}(C_i = 0\mid
M^{\dagger}_i , A_{i}, \boldsymbol{L}_{i}).$

\textbf{1. E-Step}
\begin{itemize}
  \item \textbf{Imputation step:} Generate \( S \) imputed values, restricted
  to less than \( LLoQ \), using reject sampling from the distribution \(
  f(m_{i} \mid a_{i}, \boldsymbol{l}_{i}; \hat{\boldsymbol{\beta}}^{(0)}) \)
  for each informative censored mediator value \( m_{\text{cen},i} \) (i.e.,
  for those with $C_i = 0$), denoted as $m_i^{\star(1)}, \dots, m_i^{\star(S)}$.

  \item \textbf{Weighting step:} Compute fractional weights for each imputed
  observation, using the current parameter estimate
  $\hat{\boldsymbol{\theta}}^{(t)} = (\hat{\boldsymbol{\alpha}}^{(t)},
  \hat{\boldsymbol{\beta}}^{(t)})$: $\  \text{for} \ j=1,2,\ldots, S,$
  \[
  w^{(t)}_{ij} \propto \frac{\mathbb{P}(Y_i, M^{\dagger}_i = m^{*(j)}_{i}, C_i
  = 0\mid a_{i}, \boldsymbol{l}_{i};
  \hat{\boldsymbol{\theta}}^{(t)})}{f(m^{\star(j)}_{i} \mid a_{i},
  \boldsymbol{l}_{i}; \hat{\boldsymbol{\beta}}^{(0)})} = \frac{\mathbb{P}(Y_i
  \mid M^{\dagger}_i, C_i=0, a_{i}, \boldsymbol{l}_{i};
  \hat{\boldsymbol{\alpha}}^{(t)}) \hat{\pi}^{(t)}(m^{\star(j)}_{i}, a_{i},
  \boldsymbol{l}_{i})f(m^{\star(j)}_{i} \mid a_{i}, \boldsymbol{l}_{i};
  \hat{\boldsymbol{\beta}}^{(t)}) }{f(m^{\star(j)}_{i} \mid a_{i},
  \boldsymbol{l}_{i}; \hat{\boldsymbol{\beta}}^{(0)})},
  \]
  subject to $\sum_{j=1}^{S} w^{(t)}_{ij} = 1,$ where
  $\hat{\pi}^{(t)}(m^{\star(j)}_{i}, a_{i}, \boldsymbol{l}_{i}) =
  \hat{\mathbb{P}}^{(t)}(C_i = 0\mid M^{\dagger}_i = m^{*(j)}_{i}, a_{i},
  \boldsymbol{l}_{i}).$
\end{itemize}

\textbf{2. M-step:} Update \( \boldsymbol{\theta} \) by solving the imputed score equation:
\[\sum_{i}^{n} \sum_{j=1}^{S} w^{(t)}_{ij} S(\boldsymbol{\theta}; Y_i, m^{\star(j)}_{i}, a_{i}, \boldsymbol{l}_{i}) = 0,
\]
where $S(\boldsymbol{\theta}; Y_i, m^{\star(j)}_{i}, a_{i}, \boldsymbol{l}_{i})
=\partial \ell_{\text{com}}(M^{\dagger}_i = m^{*(j)}_{i}; \boldsymbol{\theta})
/ \partial \boldsymbol{\theta}$ involves Equation~\eqref{main:llcom}, and using
new weights $w_{i j}^{(t)}$ to update $\pi(m^{\star(j)}_{i}, a_{i},
\boldsymbol{l}_{i})$ by $$\sum_{i=1}^n \sum_{j=1}^S w_{i
j}^{(t)}\text{Loss}\left(C_i- \hat{\pi}(m^{*(j)}_{i}, a_{i},
\boldsymbol{l}_{i})\right)=0,$$ where the loss function is binary cross-entropy
that corresponds to maximizing the likelihood functions of censoring components
in Equation~\eqref{main:llcom}.

\textbf{3. Check (optional):} Monitor the weight distribution using histograms
or summary statistics. If extreme fractional weights are detected, revert to
the Imputation step and modify the proposal distribution to a more suitable
form, for example, setting $f(m_{i} \mid a_{i}, \boldsymbol{l}_{i};
\hat{\boldsymbol{\beta}}^{(0)}) =f(m_{i} \mid a_{i}, \boldsymbol{l}_{i};
\hat{\boldsymbol{\beta}}^{(t)})$, and go to Imputation step.

\textbf{4. Iteration:} Set $t = t + 1$ and return to Weighting step. The
algorithm stops when the convergence criterion
$\hat{\boldsymbol{\theta}}^{(t+1)}$ is met or achieves the maximum iteration
numbers defined by the user.

\textbf{Output:} S imputed values are generated for each censored mediator
component $M_{i,\text{cen}}$ of the complete observation \(
\widetilde{M}^{\dagger}_i = (M_{i}, M_{i,\text{cen}}) \), and corresponding
fractional weights are assigned to the imputed values.
\end{algorithm}

This subsection extends our fractional imputation EM (FI-EM) algorithm to the
more general setting of informative censoring, where the second defining
property of LLoQ censoring—namely, that the censoring mechanism is
deterministic—no longer holds. The detailed procedure is given in
Algorithm~\ref{alg:iwglm}.
Algorithm~\ref{alg:iwglm}
can be viewed as an augmented LLoQ-censoring version of
Algorithm~\ref{main:alg:fiem}, differing primarily in that both the fractional
weights and the M-step incorporate an additional modeling component for the
censoring indicators.

When the censoring mechanism is not deterministic, the censoring probabilities
must be incorporated directly into the likelihood contributions, as shown in
Equation~\eqref{main:llcom} and in the first term of
Equation~\eqref{main:llobs}. In this setting, the fractional imputation weights
depend not only on the mediator density and the conditional outcome model, used
previously for deterministic LLoQ censoring, but also on the model for the
censoring probability conditional on the mediator. While one could introduce a
separate model for the binary censoring indicator and proceed analogously to
the deterministic LLoQ case, this may yield nuisance estimates that are
accurate individually yet poorly aligned for constructing the imputation
weights.

To address this limitation, we propose a \textit{collaborative estimation
strategy} that jointly models the outcome distribution and the censoring
mechanism. Specifically, the imputation weights depend on both
\begin{align*}
    G(Y_i \mid M_i^{\dagger}, A_i, \boldsymbol{L}_i) \coloneqq
    \mathbb{P}(Y_i = 1 \mid M^{\dagger}_i, A_{i}, \boldsymbol{L}_{i};
    \boldsymbol{\alpha})\quad\text{and}\quad \pi(C_i\mid M^{\dagger}_i, A_i,
    \boldsymbol{L}_i) := \mathbb{P}(C_i = 0 \mid M_i^{\dagger}, A_i,
    \boldsymbol{L}_i),
\end{align*}
and estimating these components independently may produce nuisance fits that
are not well-coordinated for the weighting task. The collaborative procedure
estimates the two components in a mutually reinforcing manner, allowing
information from the censoring model to inform the outcome model and
vice versa. We implement this approach using the highly adaptive lasso (HAL),
which provides a unified nonparametric regression framework with a shared basis
expansion and a common regularization structure. See Section~\ref{sec:hal}
of the Supplementary Materials for
additional details on HAL. This integration promotes alignment between the
nuisance functions, reduces instability in the fractional weights, and yields
smoother and more stable EM iterations, thereby improving numerical performance
and robustness in settings with non-deterministic censoring. To highlight the
structure and motivation of the collaborative approach, we summarize its key
ideas below.
\begin{itemize}
    \item Joint dependence of the imputation weights: The fractional weights
      depend simultaneously on the conditional outcome distribution ($G$) and
      the censoring probability ($\pi$). Estimating these components
      independently may yield nuisance models that are well-fit in isolation
      but poorly aligned for the weighting task.
   \item Collaborative estimation of nuisance parameters: By estimating the
     outcome and censoring models in a mutually reinforcing manner, the
     collaborative procedure encourages compatibility between the two nuisance
     components and improves their alignment with the target parameter and the
     EM updates.
   \item Unified framework for estimation: HAL offers a flexible nonparametric
     regression framework for both ($G$) and ($\pi$) using a shared basis
     expansion and a common sectional variation norm constraint. This unified
     structure promotes compatibility and provides fast convergence rates for
     infinite-dimensional nuisance parameters.
   \item Improved stability of the fractional weights: Since the weights
     involve ratios of the nuisance components, incompatibility between the
     outcome and censoring models can result in unstable or highly variable
     weights. Collaborative estimation mitigates this issue by ensuring that
     the nuisance fits are harmonized.
   \item Compatibility with EM framework: Both the fractional E-step weights
     and the M-step parameter updates rely directly on the nuisance components.
     Collaborative estimation yields smoother and more stable EM iterates,
     enhancing both convergence and overall numerical performance.
\end{itemize}

\subsection{Proof of Theorem~\ref{main:thm:effem}}\label{s1:proofofTH1}

This subsection provides the proof of Theorem \ref{main:thm:effem}. We begin by
establishing the theorem and subsequently detail the regularity conditions
invoked in our framework.\\ Firstly, we have the basic notations: the
Q-function of the EM Algorithm~\ref{main:alg:fiem} is given by
$$
Q\left(\boldsymbol{\theta} \mid \hat{\boldsymbol{\theta}}^{(t)}\right) =
\mathbb{E}\left(\ell_{\text{com}}
\left(
  \boldsymbol{\theta}\right) \mid Y, M^{\dagger}, a, \boldsymbol{l},
  \hat{\boldsymbol{\theta}}^{(t)}
\right) \ ,
$$
where $\ell_{\text{com}}(\boldsymbol{\theta})$ denotes the complete-data
log-likelihood, i.e., the log joint density of $(Y_i, M^{\dagger}_i, C_i)$
(conditional on $A$ and $\boldsymbol{L}$), $\mathbb{P}(Y_i, M^{\dagger}_i, C_i
\mid A_i, \boldsymbol{L}_i),$ that is, Equation (\ref{main:llcom}) in the main
context.  The weighted log-likelihood function based on fractional imputation,
related to Equation (\ref{main:wetscore}) in the M-step of the proposed
Algorithm 1, is denoted as
$$
Q^{\star}\left(\boldsymbol{\theta} \mid \hat{\boldsymbol{\theta}}^{(t)}\right)
  = \sum_{i=1}^n \sum_{j=1}^S w_{ij}^{(t)} \ell_{\text{com}}
  \left(M^{\dagger}_i = m_i^{*(j)}; \boldsymbol{\theta}\right) \ ,
$$
where $w_{ij}^{(t)}$ are the fractional weights and $m_i^{*(j)}$ are the
imputed values.

Theorem~\ref{main:thm:effem} states that, for any fixed number of imputations
$S$, the sequence of estimators $\hat{\boldsymbol{\theta}}^{(t)}$ generated by
Algorithm 1 converges to a stationary point
$\hat{\boldsymbol{\theta}}^{\star}_{S}$ of the Q-function of the weighted
log-likelihood function. As the number of imputations $S \to \infty$, the
stationary point $\hat{\boldsymbol{\theta}}^{\star}_{S}$ converges in
probability to the true maximum likelihood estimator of $\boldsymbol{\theta}$,
ensuring that the limiting estimator inherits the desirable asymptotic
properties of the MLE.

First, for a sufficiently large number of imputations $S$,
$Q\left(\boldsymbol{\theta} \mid \hat{\boldsymbol{\theta}}^{(t)}\right)$ is
well approximated by $Q^{\star}\left(\boldsymbol{\theta} \mid
\hat{\boldsymbol{\theta}}^{(t)}\right)$. Therefore, for a large number of
imputations $S$,
\[
\hat{\boldsymbol{\theta}}^{\star}_{S} \to \hat{\boldsymbol{\theta}}_{\text{MLE}}.
\]

Secondly, we want to show the most important property of the EM algorithm, that
its step never decreases the likelihood. That is,
if $Q^{\star}\left(\hat{\boldsymbol{\theta}}^{(t+1)} \mid \hat{\boldsymbol{\theta}}^{(t)}\right) \geq Q^{\star}\left(\hat{\boldsymbol{\theta}}^{(t)} \mid \hat{\boldsymbol{\theta}}^{(t)}\right),$ then 
\begin{align}\label{obsloglike}
\ell_{\mathrm{obs}}^{\star}\left(\hat{\boldsymbol{\theta}}^{(t+1)}\right) \geq \ell_{\mathrm{obs}}^{\star}\left(\hat{\boldsymbol{\theta}}^{(t)}\right),
\end{align}
where, conditional on $A_i$ and $\boldsymbol{L}_i,$
\begin{align*}
&\ell_{\text {obs }}^{\star}(\boldsymbol{\theta})=\sum_{i=1}^n \ln
\left\{f_{\text {obs }(i)}^{\star}\left(Y_i, M^{\dagger}_{i, \text {obs }},
C_i; \boldsymbol{\theta}\right)\right\} \text { and } f_{\text {obs
}(i)}^{\star}\left(Y_i, M^{\dagger}_{i, \text {obs }}, C_i ;
\boldsymbol{\theta}\right)=\frac{\sum_{j=1}^S \mathbb{P}\left(Y_i,
M^{\dagger}_{i} = m^{\star(j)}_{i}, C_i ; \boldsymbol{\theta}\right) /
f(m^{\star(j)}_{i})}{\sum_{j=1}^S 1 / f(m^{\star(j)}_{i})}.
\end{align*}

By Equation (\ref{main:fwpro}) that is equivalent to
$$
w_{ij}^{(t)}=
\frac{\mathbb{P}(Y_i, M^{\dagger}_i = m^{\star(j)}_{i}, C_i = 0\mid a_{i},
\boldsymbol{l}_{i}) / f(m^{\star(j)}_{i} \mid a_{i},
\boldsymbol{l}_{i})}{\sum_{k=1}^S \mathbb{P}(Y_i, M^{\dagger}_i =
m^{\star(k)}_{i}, C_i = 0\mid a_{i}, \boldsymbol{l}_{i}) / f(m^{\star(k)}_{i}
\mid a_{i}, \boldsymbol{l}_{i})} \ ,
$$ and using Jensen’s inequality,
\begin{align*}
\ell_{\mathrm{obs}}^{\star}\left(\hat{\boldsymbol{\theta}}^{(t+1)}\right) -
\ell_{\mathrm{obs}}^{\star}\left(\hat{\boldsymbol{\theta}}^{(t)}\right)
& =\sum_{i=1}^n \ln \left\{\sum_{j=1}^S w_{i j}^{(t)}
\frac{\mathbb{P}\left(Y_i, M^{\dagger}_i = m^{\star(j)}_{i}, C_i ;
\hat{\boldsymbol{\theta}}^{(t+1)}\right)}{\mathbb{P}\left(Y_i, M^{\dagger}_i =
m^{\star(j)}_{i}, C_i; \hat{\boldsymbol{\theta}}^{(t)}\right)}\right\} \\ &
\geq \sum_{i=1}^n \sum_{j=1}^S w_{i j}^{(t)} \ln \left\{\frac{\mathbb{P}\left(Y_i,
M^{\dagger}_i = m^{\star(j)}_{i}, C_i ;
\hat{\boldsymbol{\theta}}^{(t+1)}\right)}{\mathbb{P}\left(Y_i, M^{\dagger}_i =
m^{*(j)}_{i}, C_i; \hat{\boldsymbol{\theta}}^{(t)}\right)}\right\} \\ &
=Q^{\star}\left(\hat{\boldsymbol{\theta}}^{(t+1)} \mid
\hat{\boldsymbol{\theta}}^{(t)}\right)-Q^{\star}\left(\hat{\boldsymbol{\theta}}^{(t)}
\mid \hat{\boldsymbol{\theta}}^{(t)}\right).
\end{align*}

Therefore, if $Q^{\star}\left(\hat{\boldsymbol{\theta}}^{(t+1)} \mid
\hat{\boldsymbol{\theta}}^{(t)}\right) \geq
Q^{\star}\left(\hat{\boldsymbol{\theta}}^{(t)} \mid
\hat{\boldsymbol{\theta}}^{(t)}\right)$, then
$ \ell_{\mathrm{obs}}^{\star}\left(\hat{\boldsymbol{\theta}}^{(t+1)}\right)
\geq \ell_{\mathrm{obs}}^{\star}\left(\hat{\boldsymbol{\theta}}^{(t)}\right).$

We note that \(\ell^*_{\text{obs}}(\boldsymbol{\theta})\) represents an imputed version of the observed-data log-likelihood, constructed from \(S\) imputed values \(m_{i1}^*, \dots, m_{iS}^*,\) and we showed that the sequence \(\ell^*_{\text{obs}}(\hat{\boldsymbol{\theta}}^{(t)})\) is monotonically increasing.\\
Next, we discuss the regularity conditions relevant to our setting. Building on
EM convergence theory~\citep{wu1983convergence}, we now discuss the convergence
conditions of the EM sequence $\{\hat{\boldsymbol{\theta}}^{(t)}\}$, i.e., for
a sufficiently large number of iterations $t$ in Algorithm~\ref{main:alg:fiem},
the estimated parameter $\hat{\boldsymbol{\theta}}^{(t)}$ converges to its
asymptotic limit $\hat{\boldsymbol{\theta}}^{\star}_{S}$, which is a stationary
point of $Q^{\star}$ in Equation~\eqref{main:QSTAR} for a fixed $S$; that is, $
\hat{\boldsymbol{\theta}}^{(t)} \to \hat{\boldsymbol{\theta}}^{\star}_{S}\
\text{as}\ t \to \infty.$


\begin{assumption}[Regularity Conditions for EM Convergence \citep{wu1983convergence}]\label{ass:regularity}
The following conditions hold:
\begin{enumerate}[(i)]
\item The observed-data log-likelihood function
  $\ell^{\star}_{\text{obs}}(\boldsymbol{\theta})$ is continuous and bounded
  above on a compact parameter space $\Omega$.
\item The expected complete-data log-likelihood function
  $Q^{\star}(\boldsymbol{\theta} \mid \boldsymbol{\theta}') =
  \mathbb{E}_{\boldsymbol{\theta}'}[\ell^{\star}_{\text{com}}(\boldsymbol{\theta};
  Z)]$ is continuous in both $\boldsymbol{\theta}$ and $\boldsymbol{\theta}'$,
  and differentiable in $\boldsymbol{\theta}$ with gradient $\partial
  Q^{\star}(\boldsymbol{\theta} \mid \boldsymbol{\theta}') / \partial
  \boldsymbol{\theta}$ continuous in both arguments.
\item Each M-step yields a well-defined maximizer
  $\hat{\boldsymbol{\theta}}^{(t+1)} = \arg\max_{\boldsymbol{\theta} \in
  \Omega} Q^{\star}(\boldsymbol{\theta} \mid \hat{\boldsymbol{\theta}}^{(t)})$.
\item The observed-data likelihood admits at least one stationary point
  satisfying $\partial \ell^{\star}_{\text{obs}}(\boldsymbol{\theta})/\partial
  \boldsymbol{\theta} = 0$.
\end{enumerate}
\end{assumption}

Under
Assumption~\ref{ass:regularity},
the theorem implies that, for any fixed number of imputations $S$, the sequence
of estimators ${\hat{\boldsymbol{\theta}}^{(t)}}$ generated by Algorithm~1
converges to a stationary point $\hat{\boldsymbol{\theta}}^*_S$ of the
observed-data log-likelihood function. Specifically, by
Equation~(\ref{obsloglike}), the sequence
${\ell^{\star}_{\text{obs}}(\hat{\boldsymbol{\theta}}^{(t)})}$ is monotonically
non-decreasing and bounded above, provided that the maximum likelihood
estimator exists. Consequently, it converges to a finite limit $\ell^{\star}$,
which corresponds to a stationary value satisfying $\partial
\ell^{\star}_{\text{obs}}(\boldsymbol{\theta})/\partial \boldsymbol{\theta} =
0$.

If $\ell^{\star}_{\text{obs}}(\boldsymbol{\theta})$ is unimodal and admits a
unique stationary point, then, under
Assumption~\ref{ass:regularity},
the EM sequence converges to the unique maximizer
$\hat{\boldsymbol{\theta}}^*_S$. As the number of imputations $S \to \infty$,
this stationary point $\hat{\boldsymbol{\theta}}^*_S$ converges in probability
to the true maximum likelihood estimate of $\boldsymbol{\theta}$, ensuring that
the limiting estimator inherits the desirable asymptotic properties of the MLE.
Further discussion of these convergence results can be found in
\cite{wu1983convergence} and~\cite{mclachlan2008algorithm}.

Finally, similar to the discussion in~\cite{kim2011parametric}, we note that
the non-decreasing likelihood property shown in Equation~\eqref{obsloglike}
does not hold for the sequence produced by the Monte Carlo EM
algorithm with a fixed $S$, since the imputed values are re-generated at each
E-step.

{\color{black}\subsection{Connection to direct observed-data likelihood optimization}
\label{chernofsky-connection}
In this section, we clarify the relationship between the proposed
fractional-imputation expectation-maximization (FI-EM) procedure and approaches
that directly evaluate and optimize the observed-data likelihood, such as the
numerical integration approach of \cite{chernofsky2024causal}. Under common
parametric specifications of the conditional mediator density and outcome
model, the two approaches address the same latent-mediator likelihood problem
but differ in how the contribution of a left-censored mediator is evaluated.

For a participant with a left-censored mediator, $C_i=0$, define the
observed-data likelihood contribution
\begin{equation}
L_i^{\mathrm{obs}}(\boldsymbol{\theta}) =
\int_{0}^{\lambda_{\mathrm{LLoQ}}}
\mathbb{P}(Y_i=y_i \mid M_i^\dagger=m,A_i=a_i,L_i=l_i;\boldsymbol{\alpha})
f(m\mid A_i=a_i,L_i=l_i;\boldsymbol{\beta})\,dm,
\label{eq:supp_obs_like_censored}
\end{equation}
where $\boldsymbol{\theta}=(\boldsymbol{\alpha},\boldsymbol{\beta})$. This is
the censored contribution to the observed-data likelihood in
Equation~\eqref{main:llobs} of the main text. Direct likelihood approaches
evaluate the integral in
Equation~\eqref{eq:supp_obs_like_censored} numerically and
optimize the resulting observed-data likelihood with respect to
$\boldsymbol{\theta}$. The connection with the FI-EM procedure can be seen
through the observed-data score. Let
\[
U_i^{\mathrm{com}}(\boldsymbol{\theta};m)
=
\frac{\partial}{\partial\boldsymbol{\theta}}
\ell_{\mathrm{com},i}
(M_i^\dagger=m;\boldsymbol{\theta})
\]
denote the complete-data score contribution. Under standard regularity
conditions permitting differentiation under the integral sign, the
observed-data score for a participant with $C_i=0$ satisfies
\begin{align}
U_i^{\mathrm{obs}}(\boldsymbol{\theta})
&=
\frac{\partial}{\partial\boldsymbol{\theta}}
\log L_i^{\mathrm{obs}}(\boldsymbol{\theta})
\nonumber\\
&=
E_{\boldsymbol{\theta}}
\left\{
U_i^{\mathrm{com}}(\boldsymbol{\theta};M_i^\dagger)
\mid
Y_i,C_i=0,A_i,L_i
\right\}.
\label{eq:supp_fisher_identity}
\end{align}
Thus, maximizing the observed-data likelihood can equivalently be characterized
through conditional expectations of the complete-data score with respect to the
latent mediator distribution.

The FI-EM algorithm approximates these conditional expectations by importance
sampling. Specifically, let $m_i^{\star(1)},\ldots,m_i^{\star(S)}$ denote
candidate mediator values drawn from the proposal distribution used in
Algorithm~\ref{main:alg:fiem}. For the $t$th EM iteration, the corresponding
fractional weights satisfy
$$
w_{ij}^{(t)}
\propto
\frac{
\mathbb{P}(Y_i=y_i\mid M_i^\dagger=m_i^{\star(j)},A_i,L_i;
\widehat{\boldsymbol{\alpha}}^{(t)})
f(m_i^{\star(j)}\mid A_i,L_i;
\widehat{\boldsymbol{\beta}}^{(t)})
}{
f(m_i^{\star(j)}\mid A_i,L_i;
\widehat{\boldsymbol{\beta}}^{(0)})
},
$$
with the weights normalized to sum to one within participant. Consequently, for
an integrable function $h$ of the latent mediator,
\begin{equation}
E_{\widehat{\boldsymbol{\theta}}^{(t)}}
\left\{
h(M_i^\dagger)
\mid Y_i,C_i=0,A_i,L_i
\right\}
\approx
\sum_{j=1}^{S}
w_{ij}^{(t)}h(m_i^{\star(j)}).
\label{eq:supp_fi_is}
\end{equation}
Taking $h(m)=U_i^{\mathrm{com}}(\boldsymbol{\theta}; m)$ shows that the
weighted complete-data score used in the M-step is an importance-sampling
approximation to the conditional complete-data score underlying
Equation~\eqref{eq:supp_fisher_identity}.

At an EM fixed point, where $\widehat{\boldsymbol{\theta}}^{(t+1)} =
\widehat{\boldsymbol{\theta}}^{(t)} = \boldsymbol{\theta}^{\star}$, the exact
EM estimating equation satisfies
\[
\sum_{i=1}^{n}
E_{\boldsymbol{\theta}^{\star}}
\left\{
U_i^{\mathrm{com}}
(\boldsymbol{\theta}^{\star};M_i^\dagger)
\mid O_i
\right\}
=
\sum_{i=1}^{n}
U_i^{\mathrm{obs}}(\boldsymbol{\theta}^{\star})
=
0.
\]
The FI-EM procedure replaces the conditional expectations in this equation by
their fractional-imputation approximations. As the number of imputations $S$
increases, these approximations converge to their corresponding conditional
expectations under standard importance-sampling regularity conditions.
Accordingly, the limiting stationary point of the FI-EM procedure approaches
the maximum likelihood solution, consistent with Theorem~\ref{main:thm:effem}.

Under matched parametric specifications of the mediator and outcome models,
direct numerical integration and FI-EM provide two computational routes to the
corresponding observed-data likelihood problem. The former evaluates the
latent-mediator integral directly and numerically optimizes the resulting
likelihood, whereas the latter approximates the required conditional
expectations using importance-weighted candidate mediator values within an EM
algorithm. Our proposed FI formulation additionally facilitates replacing
restrictive parametric specifications of the conditional mediator density with
the flexible semi-parametric estimators considered in
Section~\ref{main:sec:cde-semipar} of the main text.}

\subsection{Highly Adaptive Lasso}\label{sec:hal}

This subsection details the Highly Adaptive Lasso (HAL) estimator used for our
semi-parametric location-scale conditional density model.

To flexibly and robustly model both the conditional mean $\mu(A,
\boldsymbol{L})$ and conditional variance $\sigma^2(A, \boldsymbol{L})$
functions (as in Algorithm 2), as well as the outcome model component required
for constructing the fractional weights (as in Algorithm 1), we employ the
Highly Adaptive Lasso estimator~\citep{van2017generally, van2017uniform}. HAL
is a nonparametric method that approximates infinite-dimensional target
functionals, such as conditional means, hazards, or densities, using a linear
combination of basis functions, allowing for flexible adaptation to complex,
potentially high-dimensional data structures. HAL assumes that the true
functional lies within the space of right-continuous functions with left-hand
limits (càdlàg) and has a sectional variation norm bounded by a finite,
but unknown, constant. This global constraint controls the overall variability
of the function without imposing local smoothness, distinguishing HAL from
traditional nonparametric methods that rely on local regularity assumptions.
For any function $f \in D([0, \tau]^d)$, the Banach space of $d$-variate
real-valued càdlàg functions defined on the cube $[0, \tau]^d \subset
\mathbb{R}^d$, the sectional variation norm is given by
$$
  \|f\|_{\nu} \coloneqq |f(0)| + \sum_{s \subset \{1, \ldots, d\}}
  \int_{0_s}^{\tau_s} \lvert \mathrm{d}f(u_s) \rvert,
$$
where $s$ ranges over all subsets of $\{1, \ldots, d\}$ and $u_s$ denotes the
subvector of $u$ with coordinates in $s$. $u_{-s} \coloneqq (u_j : j \notin s)$
denote the components of $u$ not in $s$, and define $f_s : [0_s, \tau_s] \to
\mathbb{R}$ by $f_s(u_s) \coloneqq f(u_s, 0_{-s})$. Thus, $f_s(u_s)$ represents
a section of $f$ that allows variation only along the components indexed by
$s$, with all other components fixed at zero. As discussed by
\cite{qiu2021universal}, this definition of the variation norm aligns with the
concept of Hardy-Krause variation~\citep[e.g.,][]{owen2005multidimensional}.

In particular, \cite{van2017generally} proved that the HAL estimator converges
to the target functional at a rate faster than $n^{-1/4}$, independent of the
dimensionality $d$, provided $d$ remains fixed, and later work refined this
rate to $n^{-1/3} \log(n)^{d/2}$~\citep{bibaut2019fast}. The HAL estimator
proceeds in two main steps: first, a rich set of indicator basis functions (or,
optionally, higher-order spline basis functions) is generated to map the
covariate space; second, lasso regression~\citep{tibshirani1996regression} is
performed, fitting a linear combination of these basis functions while
constraining the $\ell_1$-norm of the coefficients to approximate the sectional
variation norm. The practical performance of HAL has been extensively
demonstrated through simulation studies~\citep{benkeser2016highly}. An
open-source implementation is available in the \texttt{hal9001} R
package~\citep{coyle2022hal9001, hejazi2020hal9001}, developed for the R
language and environment for statistical computing~\citep{R}.

\subsection{Adaptive Double Bootstrap Procedure Using Multiplier Bootstrap and EIFs}\label{MultiplierEIF}

This subsection provides a detailed description of the proposed adaptive double
bootstrap procedure that utilizes a multiplier bootstrap based on efficient
influence functions (EIFs) for constructing confidence intervals under optimal
subsample selection. This approach is particularly well-suited to settings
where traditional resampling methods are computationally expensive, and where
estimation relies on influence-function-based statistics. The method combines
an outer resampling step with an inner, computationally efficient multiplier
bootstrap step.

\subsubsection{Review of the multiplier bootstrap based on the EIF}\label{sec:multboot}

Specifically, let $\phi^{\text{EIF}}(O_i)$ denote the EIF evaluated at
observation $i$, and let $\hat{\sigma}$ represent the empirical standard
deviation of the EIFs. The goal is to determine a critical value $c_{\alpha}$
satisfying $\hat{\rho}(c_\alpha) = 1 - \alpha$, where the function $\hat{\rho}$
is defined as
$$
\hat{\rho}(t) = \mathbb{P} \left(
  \left| \frac{\phi^{\text{EIF}}(O_i)}{\hat{\sigma}/\sqrt{n}} \right| \leq t
  \,\middle|\, O_1, \ldots, O_n
\right) + o_{\mathbb{P}}(1).
$$

Using this, confidence intervals for NDE/NIE estimators ($\hat{\Psi}_{NDE}$ or
$\hat{\Psi}_{NIE}$) can be constructed as $\hat{\Psi}_{NDE/NIE} \pm n^{-1/2}
c_\alpha \hat{\sigma}$, and hypothesis testing can be performed by evaluating
the $p$-value as $1 - \hat{\rho}(t)$, where $t$ is the observed test statistic.

The function $\hat{\rho}(t)$ is obtained by the asymptotic distribution of the
EIF-based one-step estimators for NIE and NDE, and the multiplier bootstrap
approximates the asymptotic distribution by the process
$$
  \mathcal{H} = \frac{1}{\sqrt{n}} \sum_{i=1}^{n}
  \frac{\xi_i \phi^{eif}(O_i)}{\hat{\sigma}} \ ,
$$
where randomness is introduced via the multipliers $(\xi_1, \ldots, \xi_n)$,
which are independently drawn from a distribution with mean zero and unit
variance, and are independent of the observed data $(O_1, \ldots, O_n)$. The
typical choices for multipliers include Rademacher variables (i.e.,
$\mathbb{P}(\xi = -1) = \mathbb{P}(\xi = 1) = 0.5$) or standard Gaussian
variables. Under the consistency of $\hat{\sigma}$, it can be shown that this
procedure yields valid approximations for inference: $\hat{\rho}(t) =
\mathbb{P} \left( \lvert \mathcal{H} \rvert \leq t \,\mid\, O_1, \ldots, O_n
\right)$. Consequently, inference based on this procedure, such as computing
critical values, $p$-values, and confidence intervals, can be accurately and
efficiently performed via simulation from a large number of realizations of the
multiplier process.

\subsubsection{Proposed Adaptive Double Bootstrap Approach}

As stated in the main context, the outer layer of the adaptive bootstrap
procedure follows the standard bootstrap framework. Specifically, for each of
$B_1$ outer iterations, a sample of size $n$ is drawn with replacement from the
original data. For each such subsample, the corresponding EIFs $n$ values are
computed based on the estimating procedure applied to that sample. These EIFs
serve as the basis for constructing bootstrap replicates of the target
parameter.

To reduce computational cost associated with repeatedly evaluating the
estimator, we replace the traditional inner resampling (i.e., nested bootstrap)
with a multiplier bootstrap applied directly to the EIFs obtained from the
outer sample. This inner step introduces random multipliers (Rademacher or
Gaussian) to simulate variability without resampling the data again. The
empirical distribution of the standardized multiplier bootstrap process is then
used to estimate the quantiles needed for constructing percentile-based
confidence intervals.

\begin{algorithm}
\caption{Multiplier Bootstrap for \textit{m}-out-of-\textit{n} Confidence
Interval Estimation}\label{multialg}

\textbf{Input:} EIF vector $e$, point estimate $\theta$, number of bootstraps
$B$, subsample size $m$, multiplier type (\texttt{"rademacher"} or
\texttt{"gaussian"}), confidence level $\alpha$\\

\textbf{Step 1:} Draw $m$ samples with replacement from $e$ to obtain subsample
$e^{(m)},$ and compute standard error: $$\text{SE} =
\sqrt{\mathrm{Var}(e^{(m)}) / n}.$$

\textbf{Step 2:} $b = 1$ to $B$:
\begin{itemize}
  \item \textbf{if} multiplier type is \texttt{``rademacher''}: \textbf{then}
    draw $\xi_i \in \{-1, 1\}$ with equal probability for $i = 1, \dots, m;$
  \item \textbf{else if} multiplier type is \texttt{``gaussian''}:
    \textbf{then} draw $\xi_i \sim \mathcal{N}(0, 1)$ for $i = 1, \dots, m$.
\end{itemize}
Compute Bootstrap Statistic:
    \[
    H_b = \frac{1}{m} \sum_{i=1}^m \frac{\xi_i \cdot e^{(m)}_i}{\text{SE}}
    \]
\textbf{Step 3:} Let $c_\alpha$ be the empirical $\alpha$-quantile of $\{H_1, H_2, \dots, H_B\},$ and construct confidence interval:
\[
\text{CI} = \left[\theta - c_\alpha \cdot \text{SE},\; \theta + c_\alpha \cdot \text{SE} \right]
\]
\textbf{Output:} Confidence interval, $c_\alpha$, and $\{H_b\}_{b=1}^B$
\end{algorithm} 

\begin{algorithm}[!ht]
\caption{Adaptive Double Bootstrap Resampling for Optimal Subsample Selection (One-step Estimation)}
\label{alg:dbbootstrap_eifs}
\textbf{Input:} Inputed dataset $D$ with $n$ observations; censoring rate
$p_{\text{cen}}$; number of outer bootstrap replicates $B_1$ and number of
inner bootstrap replicates $B_2$.

\textbf{Step 1: Grid Selection and Initial Estimation}\\
Define grid for sequence of $\gamma$ values, and compute the corresponding
sequence of resampling proportions: $c = \frac{1 + \gamma \cdot
\exp(-p_{\text{cen}})}{1 + \gamma}$ and determine the associated resample sizes
$m = \lfloor n^{c} \rfloor$. Compute the initial mediation effect estimates,
$\hat{ \Psi}_{NDE}$ and $\hat{ \Psi}_{NIE},$ using the full dataset $D.$

\textbf{Step 2: Double Bootstrap Procedure} \\
For each $\gamma$ (which determines each $c$ and $m$):
\begin{itemize}
  \item For $b_1 = 1$ to $B_1$ (outer $n$-out-of-$n$ first-stage bootstrap replicates):
  \begin{enumerate}
    \item Resample $n$ observations from $D$ with replacement, and compute the
      bootstrap NDE and NIE one-step estimates $\hat{ \Psi}_{NDE}^{(b_1)}$ and
      $\hat{ \Psi}_{NIE}^{(b_1)}$ and $\text{EIFs}^{(b_1)}$ ($n$ observations),
      using the resampled data.
    \item Conditional on each first-stage bootstrap sample, implement
      multiplier bootstrap for $m$-out-of-$n$ confidence interval estimation in
      Algorithm~\ref{multialg}
      (inner $m$-out-of-$n$ bootstrap replicates), where \textbf{inputs} are
    \begin{itemize}
      \item $\text{EIFs}^{(b_1)}$, estimates $\hat{ \Psi}_{NDE}^{(b_1)}$ and
        $\hat{ \Psi}_{NIE}^{(b_1)}$ in the above step; number of multiplier
        bootstraps $B_2$, subsample size \( m^{(b_1)} \),
      \item and \textbf{outputs} are multiplier bootstrap for
        $m^{(b_1)}$-out-of-$n$ confidence interval.
    \end{itemize}
    \item Check CI coverage: Verify whether the initial estimates $\hat{
      \Psi}_{NDE},$ $\hat{ \Psi}_{NIE}$ fall within the percentile bootstrap
      confidence intervals.
  \end{enumerate}
\end{itemize}
\textbf{Step 3: Selection of Optimal $m$}\\ Estimate the double bootstrap CI
coverage rate from all the first-stage bootstrap data sets. If the coverage
rate achieves the desired nominal level, stop and select the current \( c \)
and \( m \). Otherwise, increment $\gamma$ to the next higher value.

\textbf{Output:} Optimal resampling proportion $c^{\text{opt}}$, and the
corresponding optimal resample size  $m^{\text{opt}}$.
\end{algorithm}

We provide two key algorithms to implement the procedure. First, Multiplier
Bootstrap for $m$-out-of-$n$ Confidence Interval Estimation Algorithm (i.e.,
Algorithm \ref{multialg})
describes the inner layer of the procedure. It takes as input a vector of EIFs
from an outer resample, and applies a multiplier bootstrap (Rademacher or
Gaussian) to generate standardized bootstrap statistics. The empirical
quantiles of these statistics are then used to construct confidence intervals
centered at the point estimate, scaled by the standard error of the EIFs.

Second, Adaptive Double Bootstrap Procedure Using EIF-based Multiplier
Bootstrap, presented in
Algorithm~\ref{alg:dbbootstrap_eifs},
outlines the complete double bootstrap procedure. For each outer iteration, it
draws an $n$-out-of-$n$ resample, computes the corresponding point estimate and
EIFs, and then calls the multiplier bootstrap procedure
(Algorithm~\ref{multialg})
to obtain a confidence interval. This procedure is iterated across all $B_1$
outer resamples, and the empirical coverage is evaluated by calculating the
proportion of confidence intervals that contain the original NDE and NIE
estimates. If the target coverage is met, the corresponding value of $\gamma$
is selected as $\hat{\gamma}$; otherwise, $\gamma$ is increased, and the
procedure is repeated. The search range for $\gamma$ can be customized
depending on the application, such as imposing an upper bound based on a
minimum resample size. Once $\hat{\gamma}$ is determined, a final
$m$-out-of-$n$ bootstrap is conducted using the corresponding optimal $m$.

For one-step estimation, a major advantage of our Algorithm
(\ref{multialg})
is the substantial reduction in computational burden. Since the multiplier
bootstrap avoids repeated model fitting during the inner bootstrap stage, a
much larger number of inner bootstrap replicates ($B_2$) can be used without
incurring excessive computational cost. In practice, we recommend using a
relatively small number of outer resamples (e.g., $B_1 = 100$) and a larger
number of inner multiplier replicates (e.g., $B_2 = 10,000$ or more). This
design maintains the statistical efficiency of the bootstrap confidence
interval while remaining computationally tractable for large datasets or
complex models.

\section{Natural Direct Effect (NDE) Results for Studies 1 and 2}\label{sm_sims}

\begin{figure}[!]
    \centering
    \begin{subfigure}[b]{0.85\textwidth}
        \centering
        \includegraphics[width=\textwidth]{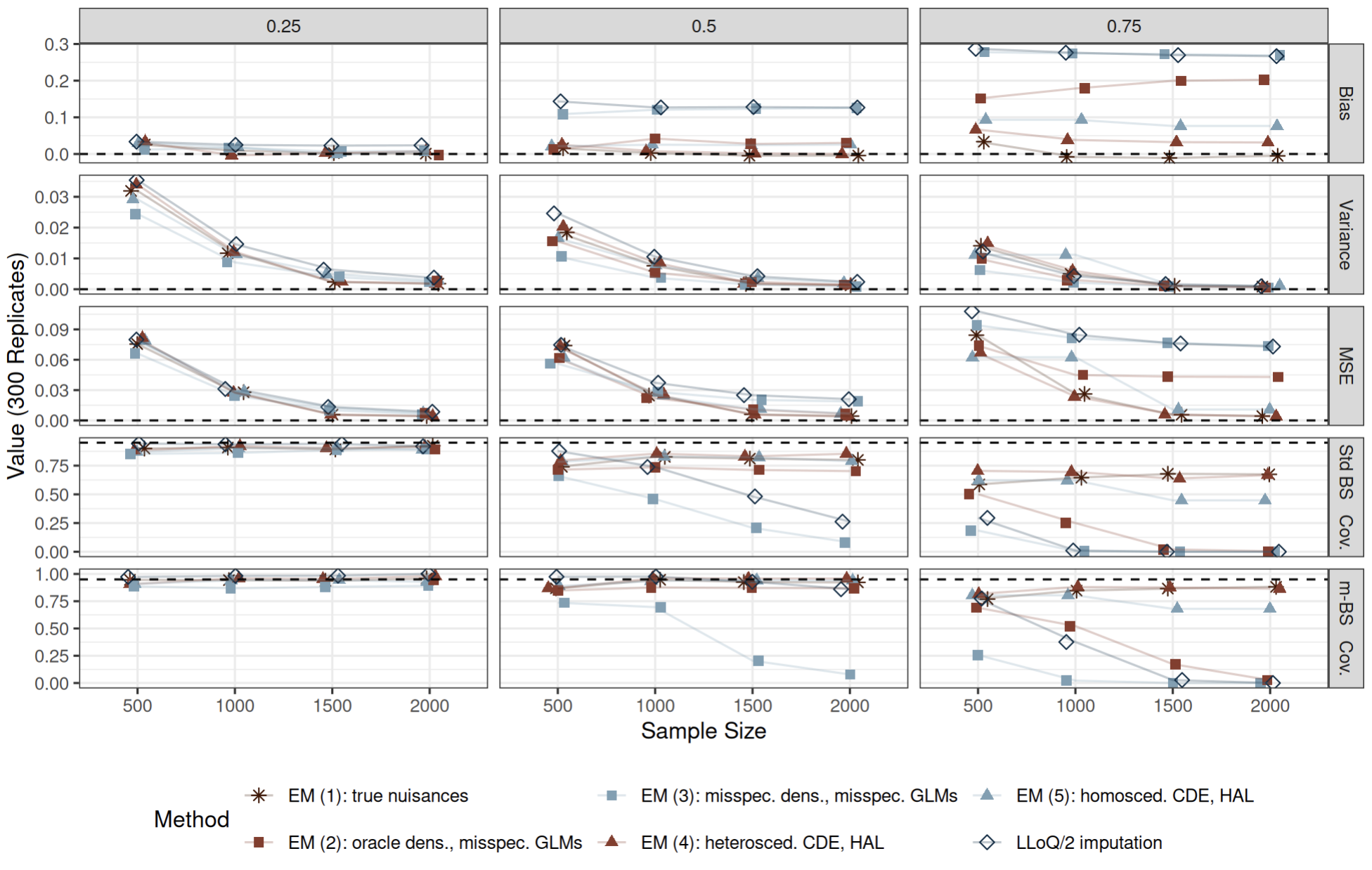}
        \caption{Study 1: Parametric G-computation NDE estimates under
        different imputation methods} \label{study1nie}
    \end{subfigure}
    \vspace{0.05em}
    \begin{subfigure}[b]{0.85\textwidth}
        \centering
        \includegraphics[width=\textwidth]{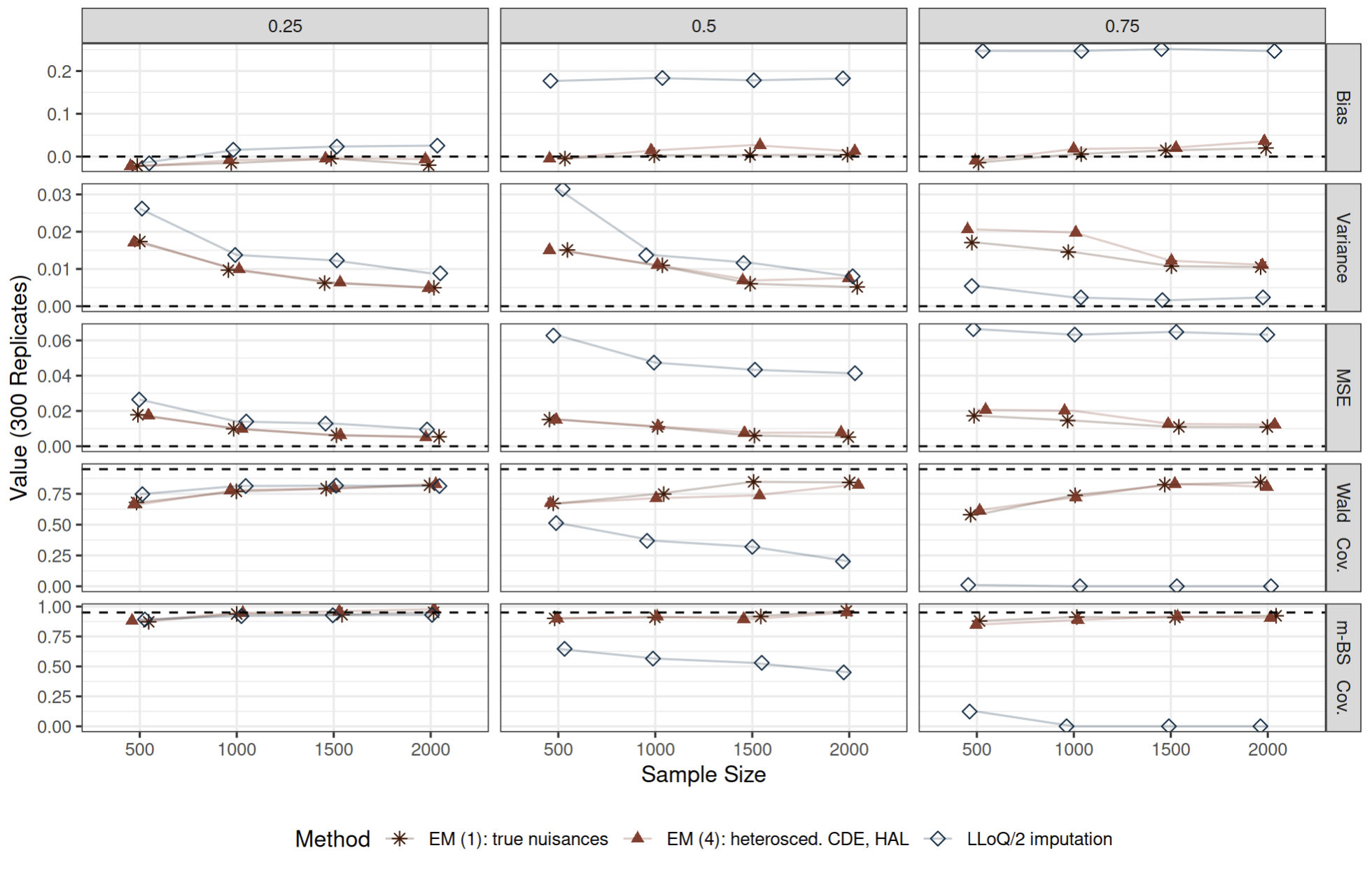}
        \caption{Study 2: Flexible one-step NDE estimates under different
        imputation methods} \label{study2nie}
    \end{subfigure}
    \caption{Performance of NDE estimators based on parametric G-computation
      (top, Study 1) and flexible one-step estimators (bottom, Study 2) under
      various imputation strategies and sample sizes ($n \in \{500, 1000, 1500,
      2000\}$). Panels report bias, variance, MSE, and confidence interval
      coverage (standard and proposed $m$-out-of-$n$ bootstrap). The imputation
      methods correspond to those described in Table~\ref{main:tab:methods} of
      the main text.}
    \label{fig:1step}
\end{figure}

This section provides complementary simulation results for the natural direct
effect (NDE), analogous to the NIE results presented in the main text.
Figures~\ref{study1nie}
and~\ref{study2nie}
summarize the performance of the fractional imputation strategies across both
Study 1 (plug-in parametric G-computation estimators) and Study 2 (one-step
estimators), under varying levels of left-censoring and sample sizes.

Regarding Study 1, the plug-in (parametric g-computation) estimator of the NDE,
Figure~\ref{study1nie}
reports the finite-sample performance of the parametric g-computation estimator
under each of the six imputation strategies. Consistent with the NIE findings
presented in the main text, the LLoQ/2 imputation approach displays substantial
bias across all censoring levels, with errors increasing sharply under moderate
and high censoring (50$\%$ and 75$\%$). The EM (1): true nuisances method
performs best, demonstrating minimal bias, low mean squared error, and
near-nominal confidence interval coverage. Among the semiparametric approaches,
the EM (4): heteroscedastic CDE, HAL strategy performs particularly well,
showing robustness comparable to the fully correctly specified parametric
approach. Strategies EM (2) and EM (3), which incorporate misspecification in
some or all nuisance components, exhibit degraded performance, especially under
more severe censoring.

For Study 2, the one-step estimation of the NDE, Figure~\ref{study2nie}
depicts results for the one-step estimators, which incorporate flexible
machine-learning-based nuisance estimation. The overall pattern parallels that
of the NIE simulations. The naive LLoQ/2 method again yields pronounced bias
across all censoring levels, with little improvement at larger sample sizes due
to its structural misspecification. The EM (1): true nuisances method retains
strong performance across scenarios, reflecting the efficiency properties of
the one-step estimator under correct specification. Importantly, the EM (4):
heteroscedastic CDE, HAL approach also performs very well, producing low bias
and MSE along with approximately nominal 95$\%$ coverage under the proposed
$m$-out-of-$n$ bootstrap method. This reinforces the practicality of
semiparametric conditional density estimation when full parametric correctness
cannot be assumed.

\section{Supplementary Materials of the Analysis of the ACTIV-2 Data}

\subsection{Q-Q plot of truncation-adjusted residuals for Day 3 log$_{10}$ viral RNA.} \label{qqplot}
\begin{figure}[ht]
    \centering
    \includegraphics[width=0.75\textwidth]{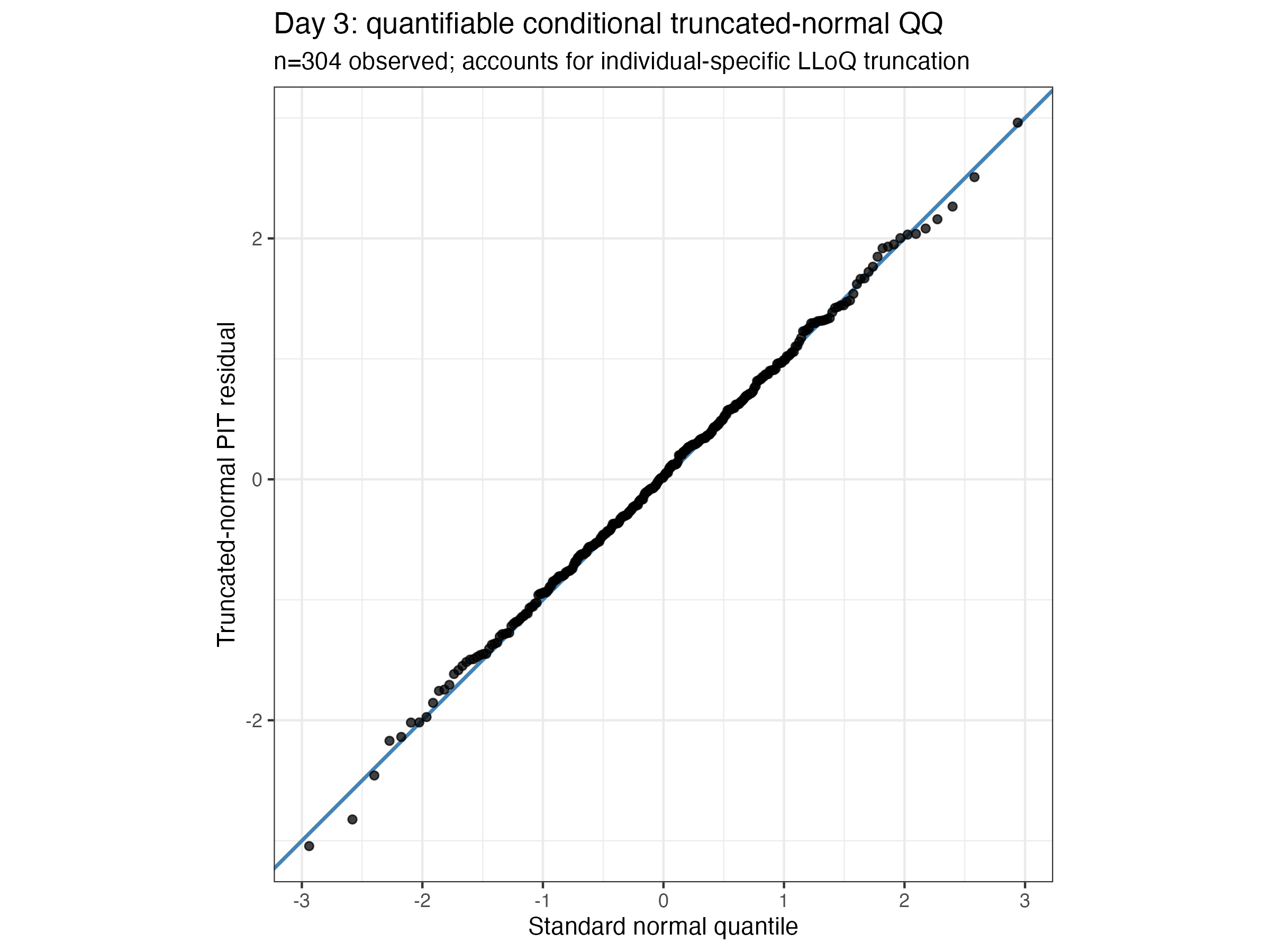}
    \caption{
    \textbf{Q-Q plot of truncation-adjusted residuals for Day 3
    anterior nasal SARS-CoV-2 RNA.}
    A left-censored Gaussian regression model was fitted to Day 3
    log$_{10}$ RNA, adjusting for treatment assignment and baseline
    RNA. The model included 453 participants with quantifiable
    baseline RNA, of whom 149 had Day 3 RNA below the lower limit
    of quantification (LLoQ; 2.0 log$_{10}$ copies/mL).
    The Q-Q plot was constructed using truncation-adjusted
    residuals from the 304 participants with quantifiable Day 3
    RNA. 
    }
    \label{fig:qq_conditional}
\end{figure}

\biostats{A left-censored Gaussian regression model was fitted to Day 3 log$_{10}$ viral RNA, adjusting for treatment assignment and baseline viral RNA. The analysis included 453 participants with quantifiable baseline RNA, including 149 with Day 3 RNA below the LLoQ (2.0 log$_{10}$ copies/mL). The Q-Q plot (Figure \ref{fig:qq_conditional}) was constructed using truncation-adjusted residuals from the 304 quantifiable Day 3 observations. The residuals showed no substantial departures from normality over the quantifiable range.}

\subsection{Analysis Results of the ACTIV-2 Data}\label{real-world}

Table~\ref{tab:baseline}
presents baseline characteristics of the Phase 3 modified intention-to-treat
(ITT) population, excluding Day 3 and earlier events. The treatment groups are
well-balanced across a broad range of demographic and clinical covariates,
including age, sex, race/ethnicity, comorbidities, and baseline anterior nasal
(AN) viral RNA levels. Most standardized comparisons show minimal differences
between the placebo and A/R groups, consistent with successful randomization.
The only notable imbalance is observed for cardiovascular disease (6.2\% in
placebo versus 2.9\% in A/R; $p = 0.031$), although the absolute prevalence
remains low. Importantly, hospitalization or death after Day 3 occurs
substantially less frequently in the A/R group (0.8\%) compared to placebo
(5.1\%), indicating a strong overall protective treatment effect.

\begin{table}[!]
\centering
\caption{Phase 3 Baseline Characteristics of the modified ITT population
(excluding Day 3 and earlier events)}
\label{tab:baseline}
\fontsize{11.5pt}{13.5pt}\selectfont
\begin{tabular*}{\linewidth}{@{\extracolsep{\fill}}lcccc}
\toprule
 & \textbf{Overall} & \textbf{Placebo} & \textbf{A/R}$^{1}$ & \textbf{p-value}$^{3}$ \\
\midrule
 & N = 749$^{2}$ & N = 371$^{2}$ & N = 378$^{2}$ &  \\
\midrule

Age $\geq$ 60 years & 160 (21\%) & 79 (21\%) & 81 (21\%) & 0.9 \\
Female sex & 389 (52\%) & 195 (53\%) & 194 (51\%) & 0.7 \\

\addlinespace[2pt]
{Body mass index category} &  &  &  & 0.4 \\
\quad Normal & 136 (18\%) & 63 (17\%) & 73 (19\%) &  \\
\quad Overweight & 240 (32\%) & 119 (32\%) & 121 (32\%) &  \\
\quad Obese & 358 (48\%) & 185 (50\%) & 173 (46\%) &  \\
\quad Underweight & 7 (0.9\%) & 2 (0.5\%) & 5 (1.3\%) &  \\

\addlinespace[2pt]
{Race} &  &  &  & 0.2 \\
\quad White & 538 (72\%) & 273 (74\%) & 265 (70\%) &  \\
\quad Black or African American & 139 (19\%) & 60 (16\%) & 79 (21\%) &  \\
\quad Other & 72 (9.6\%) & 38 (10\%) & 34 (9.0\%) &  \\

\addlinespace[2pt]
{Ethnicity} &  &  &  & 0.7 \\
\quad Hispanic or Latino & 366 (49\%) & 179 (48\%) & 187 (49\%) &  \\
\quad Not Hispanic or Latino & 383 (51\%) & 192 (52\%) & 191 (51\%) &  \\

\addlinespace[2pt]
{Country} &  &  &  & 0.3 \\
\quad USA & 494 (66\%) & 252 (68\%) & 242 (64\%) &  \\
\quad Non-USA & 255 (34\%) & 119 (32\%) & 136 (36\%) &  \\

\addlinespace[4pt]
Vaccine & 67 (8.9\%) & 38 (10\%) & 29 (7.7\%) & 0.2 \\ 
Days from symptom onset to enrollment & 5.52 (2.28) & 5.47 (2.26) & 5.56 (2.29) & 0.7 \\

\addlinespace[4pt]
Diabetes & 109 (15\%) & 53 (14\%) & 56 (15\%) & 0.8 \\
Cardiovascular disease & 34 (4.5\%) & 23 (6.2\%) & 11 (2.9\%) & 0.031 \\
Hypertension & 286 (38\%) & 133 (36\%) & 153 (40\%) & 0.2 \\
Current smoker & 248 (33\%) & 130 (35\%) & 118 (31\%) & 0.3 \\
\addlinespace[4pt]
{AN RNA day 0$^{4}$} & 3.84 (2.40) & 3.85 (2.37) & 3.82 (2.42) & 0.9 \\ 
\addlinespace[4pt]
Hospitalization or death after Day 3 & 22 (2.9\%) & 19 (5.1\%) & 3 (0.8\%) & $<0.001$ \\

\bottomrule
\end{tabular*}

\vspace{0.3em}
\begin{minipage}{\linewidth}
\footnotesize
$^{1}$  Amubarvimab Plus Romlusevimab \\
$^{2}$ Values are n (\%) unless otherwise indicated; continuous variables are mean (SD). \\
$^{3}$ Pearson’s chi-squared test or Fisher’s exact test for categorical variables; Wilcoxon rank-sum test for continuous variables.\\
$^{4}$ Day 0 AN SARS-CoV-2 RNA, $\log
_{10}$ copies/mL
\end{minipage}
\end{table}

\begin{table}[!]
\centering
\begin{threeparttable}
\caption{Natural direct and anterior nasal viral load-mediated indirect effects of mAbs on hospitalization/death risk}\label{tab:sensitivity}
\setlength{\tabcolsep}{6pt}
\renewcommand{\arraystretch}{1.15}

\begin{tabularx}{\textwidth}{>{\raggedright\arraybackslash}p{2.6cm} *{3}{>{\centering\arraybackslash}X}}
\toprule
& \multicolumn{3}{c}{Imputation strategy} \\
\cmidrule(lr){2-4}
Estimator 
& LLoQ / 2 
& EM: log-normal density fractional imputation
& EM: heteroscedastic
CDE, HAL fractional imputation\\
\midrule

G-computation &
\makecell[c]{\textbf{NDE}:  -0.042 (0.014) \\ {\small 95\% CI: [-0.077,\, -0.005]} \\
\textbf{NIE}: -0.0008 (0.009) \\ {\small 95\% CI: [-0.063,\, 0.011]}} &
\makecell[c]{\textbf{NDE}: -0.033 (0.012) \\ {\small 95\% CI: [-0.072,\,-0.004]} \\
\textbf{NIE}: -0.0008 (0.006) \\ {\small 95\% CI: [-0.032,\, 0.006]}} &
\makecell[c]{\textbf{NDE}: -0.033 (0.010) \\ {\small 95\% CI: [-0.053,\, - 0.016]}\\
\textbf{NIE}: -0.0011 (0.004) \\ {\small 95\% CI: [-0.033,\, 0.004]}} \\

\addlinespace[0.35em]

One-step &
\makecell[c]{\textbf{NDE}:  -0.040 (0.014) \\ {\small 95\% CI: [-0.086,\,0.006]} \\
\textbf{NIE}: -0.0013 (0.001) \\ {\small 95\% CI: [-0.003,\,0.001]}} &
\makecell[c]{\textbf{NDE}: -0.021 (0.007) \\ {\small 95\% CI: [-0.048,\,0.006]} \\
\textbf{NIE}: -0.0023 (0.001) \\ {\small 95\% CI: [-0.005,\,0.001]}} &
\makecell[c]{\textbf{NDE}: -0.021 (0.007) \\ {\small 95\% CI: [-0.048,\,0.006]} \\
\textbf{NIE}: -0.0026 (0.001) \\ {\small 95\% CI: [-0.006,\,0.001]}} \\

\addlinespace[0.35em]

TMLE &
\makecell[c]{\textbf{NDE}: -0.041 (0.014) \\ {\small 95\% CI: [-0.087,\,0.005]} \\
\textbf{NIE}: -0.0019 (0.001) \\ {\small 95\% CI: [-0.004,\,0.001]}} &
\makecell[c]{\textbf{NDE}: -0.031 (0.007) \\ {\small 95\% CI: [-0.058,\, -0.003]} \\
\textbf{NIE}: -0.0030 (0.001) \\ {\small 95\% CI: [-0.006,\, -0.001]}} &
\makecell[c]{\textbf{NDE}: -0.031\ (0.007) \\ {\small 95\% CI: [-0.0577,\ -0.003]} \\
\textbf{NIE}: -0.0031\ (0.001) \\ {\small 95\% CI: [-0.006,\ -0.001]}} \\

\bottomrule
\end{tabularx}

\begin{tablenotes}[flushleft]
\footnotesize
\item \textit{Notes:} Point estimates are on the risk-difference scale. The
  proposed m-out-of-n bootstrap standard errors (SE) are shown in parentheses,
  and the corresponding confidence intervals are 95\%. Columns correspond to
  increasingly flexible (less parametric) imputation/density modeling for the
  mediator; rows correspond to different estimators: G-computation, one-step,
  and TMLE. 
\end{tablenotes}
\end{threeparttable}
\end{table}

Table \ref{tab:sensitivity} summarizes estimates of the NDE and anterior nasal
viral load–mediated NIE of mAb treatment on hospitalization/death risk under a
model-based imputation strategy for the 38 participants with missing RNA
measurements. Specifically, we model the conditional distribution of RNA above
the LLoQ given baseline covariates and impute missing values accordingly.
Across all estimators—including G-computation with parametric generalized
linear models (GLMs) and machine learning (Super Learner)-based one-step and
TMLE approaches—the total effect remains protective, with the NDE accounting
for the majority of the treatment effect.

Consistent with the main analysis, the estimated NIE through AN viral RNA is
small in magnitude. Under the conventional LLoQ/2 substitution, NIE estimates
are close to zero across all estimators (e.g., $-0.0008$ for the g-computation
estimator, $-0.0013$ for the one-step estimator, and $-0.0019$ for TMLE),
corresponding to minimal mediation. In contrast, more flexible imputation
strategies that respect the left-censored nature of the mediator yield modestly
larger (more negative) NIE estimates. In particular, EM-based
approaches---including parametric log-normal models and semi-parametric
heteroscedastic conditional density estimation (CDE) with HAL---produce NIE
estimates in the range of approximately $-0.002$ to $-0.003$ for the one-step
and TML estimators, with corresponding mediation proportions modestly increased
relative to LLoQ/2 substitution.

Overall, these sensitivity analyses reinforce the main findings: while mAb
treatment substantially reduces hospitalization/death risk, only a small
proportion of this effect appears to be mediated through reductions in AN RNA.

\subsection{Separable effects and interventionist mediation in the ACTIV-2 trial}\label{Separable}

This subsection elaborates on the separable effects framework of Robins,
Richardson, and Shpitser (RRS) and its application to mediation analysis in the
ACTIV-2 monoclonal antibody (mAb) trial. We provide intuition for the
interventional strategies used in the analysis and derive the identification
result for the natural direct and indirect effects under the RRS assumptions.

\subsubsection{Interventionist mediation in the ACTIV-2 trial}

Our analysis is grounded in the interventionist framework of Robins,
Richardson, and Shpitser (RRS; \cite{robins2022interventionist}), introduced in
Section 2. This framework enables us to assess whether mAbs reduce
hospitalization ($Y$) by lowering viral RNA ($M$) through the use of
\textit{separable} treatment components \citep{robins2010alternative,
stensrud2022separable}, thereby circumventing the controversial cross-world
exchangeability assumption.

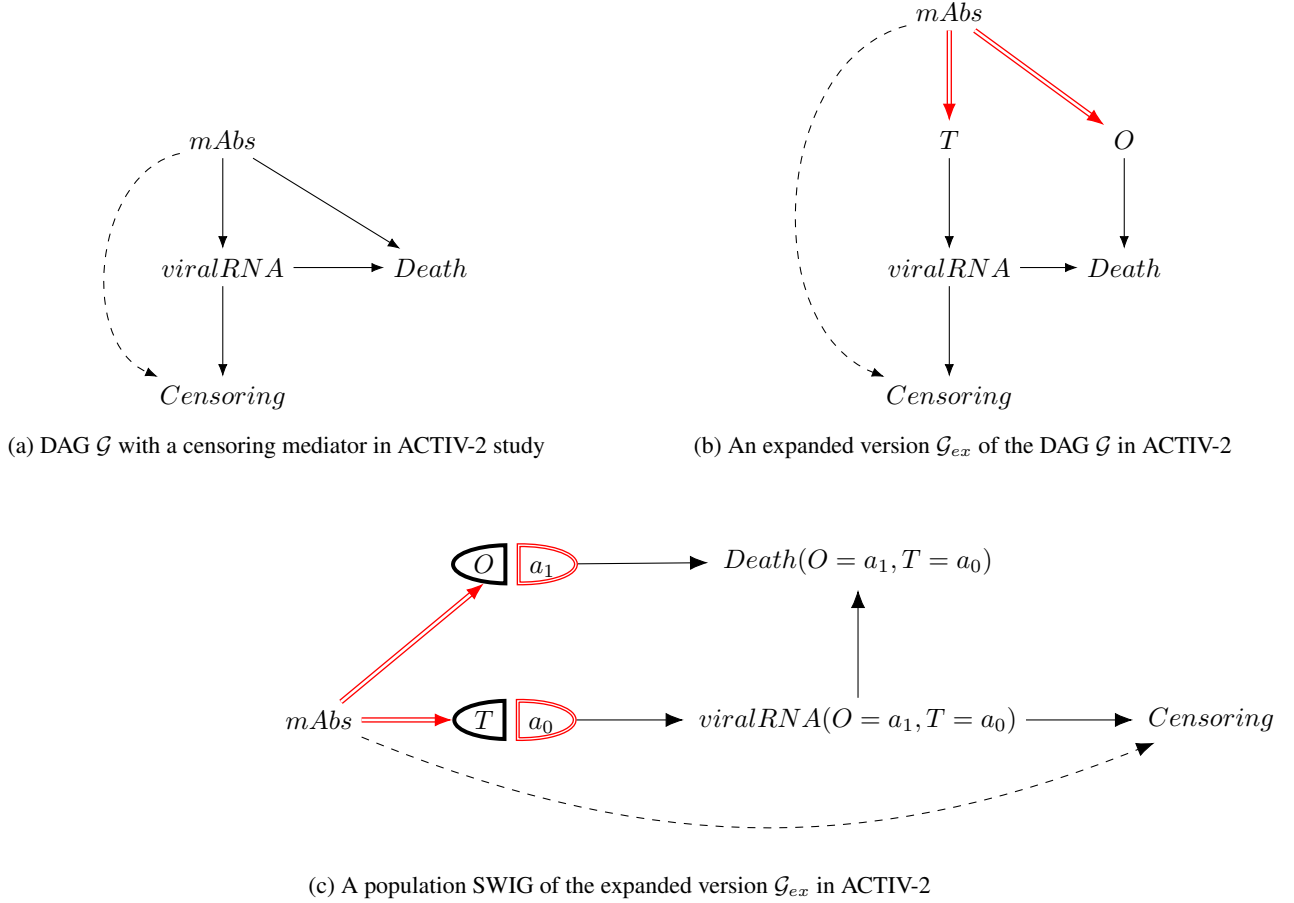
\begin{figure}[!]
\centering
\begin{subfigure}[b]{0.45\textwidth}
\centering
\begin{tikzpicture}[node distance=1.2cm]
    \node[] (A) {$mAbs$};
    \node[] (M) [below=of A] {$viral RNA$};
    \node[] (Y) [right=of M] {$Death$};
    \node[] (C) [below=of M] {$Censoring$};
    \path (A) edge (M);
    \path (M) edge (Y);
    \path (A) edge (Y);
    \path (M) edge (C);
    \path (A) edge[dashed, bend right=72] (C);
\end{tikzpicture}
\caption{DAG $\mathcal{G}$ with a censoring mediator in ACTIV-2 study}
\label{fig:dag}
\end{subfigure}
\hfill
\vspace{1cm} 
\begin{subfigure}[b]{0.45\textwidth}
\centering
\begin{tikzpicture}[node distance=1.2cm]
    \tikzset{
        state/.style={circle, draw, minimum size=0.8cm, inner sep=0},
        double state/.style={circle, draw, double, minimum size=0.8cm, inner sep=0, line width=1pt},
        double arrow/.style={double, red, line width=0.5pt, double distance=1pt}
    }
    \node[] (A) {$mAbs$};
    \node[] (A1) [below=of A] {$T$};
    \node[] (A2) [right=1.8cm of A1] {$O$};
    \node[] (M) [below=of A1] {$viral RNA$};
    \node[] (Y) [below=of A2] {$Death$};
    \node[] (C) [below=of M] {$Censoring$};
    \path (A1) edge (M);
    \path (M) edge (Y);
    \path (A2) edge (Y);
    \path (M) edge (C);
    \draw[double arrow] (A) -- (A1);
    \draw[double arrow] (A) -- (A2);
    \path (A) edge[dashed, bend right=72] (C);
\end{tikzpicture}
\caption{An expanded version $\mathcal{G}_{ex}$ of the DAG $\mathcal{G}$ in ACTIV-2}
\label{fig:expanded}
\end{subfigure} 

\begin{subfigure}[b]{0.55\textwidth}
\centering
\begin{tikzpicture}[node distance=1.5cm]
 \tikzset{line width=1.5pt, outer sep=0pt, ell/.style={draw,fill=white, inner sep=2pt, line width=1.5pt},         double arrow/.style={double, red, line width=0.5pt, double distance=1pt}, swig vsplit={gap=5pt, line color right=red, inner line width right=0.5pt}};
    \node[] (A) {$mAbs$};
    \node[name=A1, right=1.2cm of A, shape=swig vsplit]{ \nodepart{left}{$T$} \nodepart{right}{$a_0$} };
    \node[name=A2, above=1.5cm of A1, shape=swig vsplit]{ \nodepart{left}{$O$} \nodepart{right}{$a_1$} };
    \node[] (M) [right=of A1] {$viral RNA(O = a_1, T = a_0)$};
    \node[] (Y) [above=1.51cm of M] {$Death(O = a_1, T = a_0)$};
    \node[] (C) [right=of M] {$Censoring$};
    \path (A1) edge (M);
    \path (M) edge (Y);
    \path (A2) edge (Y);
    \path (M) edge (C);
    \draw[double arrow] (A) -- (A1);
    \draw[double arrow] (A) -- (A2);
    \path (A) edge[dashed, bend right=22] (C);
\end{tikzpicture}
\caption{A population SWIG of the expanded version $\mathcal{G}_{ex}$ in ACTIV-2}
\label{fig:swigseparability}
\end{subfigure}

\caption{(a) A simple directed acyclic graph (DAG) $\mathcal{G}$ representing
the causal relationships among treatment with monoclonal antibodies (mAbs),
the mediator viral RNA, the outcome death, and a censoring indicator.  Dashed
arrows from treatment to the censoring indicator are permitted in our
framework to capture generic mediator censoring, but are not present under
left-censoring due to LLoQ. (b) An expanded DAG $\mathcal{G}_{\text{ex}}$
decomposes the treatment $A$ into two components: $T$ (e.g., "mAbs as
antivirals") and $O$ (e.g., "mAbs as immunomodulators"), under the
assumptions that $T$ does not directly affect death ($Y$), and $O$ does not
directly affect viral RNA ($M$). This structure enables the definition of
separable direct and indirect effects via hypothetical interventions on $T$
and $O$. Red double-line arrows indicate deterministic relationships. (c) A
population single-world intervention graph (SWIG) corresponding to the expanded
DAG $\mathcal{G}_{\text{ex}}$, which encodes the identification assumptions
necessary for separable effects, following \cite{robins2022interventionist}
(i.e., $a_0$ is d-separated from $\textit{Death}(O = a_1, T = a_0)$ given
$\textit{viral RNA}(O = a_1, T = a_0)$; while $a_1$ is d-separated from
$\textit{viral RNA}(O = a_1, T = a_0)$). Edges in (c) represent average causal
effects at the population level. The absence of edges from $a_1$ to
$\textit{viral RNA}(O = a_1, T = a_0)$ and from $a_0$ to $\textit{Death}(O =
a_1, T = a_0)$ indicates no population-level causal effects of $O$ on mediator
and of $T$ on outcome, respectively. However, the structure of the
individual-level counterfactuals $\textit{viral RNA}(O = a_1, T = a_0)$ and
$\textit{Death}(O = a_1, T = a_0)$ implies that individual causal effects of
$T$ on $Death$ and of $O$ on $viral RNA$ may still exist, though these effects
cancel out in expectation at the population level. For clarity, baseline
confounders $\boldsymbol{L}$ of $A$, $M$, and $Y$ are omitted from all
panels.}\label{fig:activ2}
\end{figure}

In contrast to Pearl's NPSEMs, the RRS's approach provides a generalizable
template for analyzing mediated effects in biological therapies, particularly
when treatment components are mechanistically distinct yet consequently
intertwined. In our ACTIV-2 study, we evaluate whether monoclonal antibodies
(mAbs) reduce hospitalization ($Y$) by lowering viral RNA levels ($M$) through
separable treatment components. To apply RRS's interventionist framework for
identification, we conceptualize mAb therapy ($A$) as comprising two
hypothetical separable interventions: (1) $T$, the effect of mAbs on viral load
($M$) independent of hospitalizaiton, and (2) $O$, the effect of mAbs on
hospitalization ($Y$) independent of viral suppression. This approach
circumvents cross-world counterfactuals (e.g., $Y(A = a_1, M(a_0))$) by instead
contrasting outcomes under interventions that isolate $T$ (e.g., ``mAbs as
antivirals'') and $O$ (non-antiviral components of mAbs, e.g., ``mAbs as
immunomodulators''). By leveraging baseline (Day 0) and follow-up viral RNA
measures, we assess the indirect effect of $T$ via $M$ while controlling for
$O$, ensuring our estimands align with empirically testable mechanisms.
Adopting the RRS framework, we utilise edge-expanded graphs to represent the
separable dual mechanisms of monoclonal antibodies (mAbs).
Figure~\ref{fig:expanded} presents the causal DAGs and the expanded
diagram used to represent the separable treatment components in our ACTIV-2
mediation analysis. Identification relies on the assumptions of no unmeasured
$T$-$M$-$Y$ confounding, consistency, positivity in NPSEMs, and assumptions for
separability (\cite{robins2022interventionist}), which are, for $o \in \{a_0,
a_1\}$ and $t \in \{a_0, a_1\},$
\begin{align}\label{separ}
    &\mathbb{P}[Y(O = o, T = a_1) = y \mid M(O = o, T = a_1) = m] = \mathbb{P}[Y(O=o, T = a_0) = y \mid M(O=o, T = a_0) = m]; \nonumber\\
    &\mathbb{P}[M(O = a_0, T = t) = m] = \mathbb{P}[M(O = a_1, T = t) = m].
\end{align}

These two separability assumptions can also be read off the SWIG that is
presented in Figure
(\ref{fig:swigseparability}),
that is, in our ACTIV-2 study, $a_0$ (the fixed piece of $T$ in the SWIG) is
d-separated from $\textit{Death}(O = a_1, T = a_0)$ given $\textit{viral RNA}(O
= a_1, T = a_0)$ and $a_1$ (the fixed piece of $O$) is d-separated from
$\textit{viral RNA}(O = a_1, T = a_0)$.  Notably, the identification formula of
the RRS framework aligns with the standard mediation formula
(\cite{robins2022interventionist}). That is, with the key term in the natural
direct/indirect effect that $Y(A=a_1, M(a_0)) = Y(O=a_1, M(T = a_0)) = Y(O=a_1,
T = a_0)$, NDE and NIE can be expressed as 
\begin{align*}
     \theta_{\text{NDE}}&\coloneqq \mathbb{E}\left[Y(A=a_1, M(a_0))-Y(A=a_0, M(a_0))\right]= \mathbb{E}[Y(O = a_1, T = a_0)] - \mathbb{E}[Y(O = a_0, T = a_0)],\\
    \theta_{\text{NIE}}&\coloneqq \mathbb{E}\left[Y(A=a_1, M(a_1))-Y(A=a_1, M(a_0))\right]= \mathbb{E}[Y(O = a_1, T = a_1)] - \mathbb{E}[Y(O = a_1, T = a_0)],
\end{align*}
and the key term $\mathbb{E}[Y( O = a_1, T = a_0)]$ can be identified as
\begin{align*}
\mathbb{E}[Y(O = a_1, T = a_0)] 
    =\int_{\mathcal{L}} \int_{\mathcal{M}} \mathbb{E}[Y \mid A=a_1, M=m, \boldsymbol{L}=\boldsymbol{l}] \times f_{M \mid A, \boldsymbol{L}}(m \mid A=a_0,  \boldsymbol{L}=\boldsymbol{l})f_{\boldsymbol{L}}(\boldsymbol{l}) \, dm \, d\boldsymbol{l},
\end{align*}
which is exactly the standard mediation formula given in
Equation~(\ref{main:medest}). Further details of this identification
are provided in the following subsection, where the separability assumptions
from Equation~(\ref{separ}) are employed.

\subsubsection{Separable effects and identification formulae}

In the ACTIV-2 study, the treatment consists of a monoclonal antibody therapy
whose effects can conceptually be decomposed into two distinct components: (1)
O-component $O$: Direct physiological and immunological pathways influencing
the clinical outcome ($Y$) (e.g., hospitalization). (2) T-component $T$:
Mechanisms influencing viral clearance and therefore the viral-load mediator
($M$). Following RRS, we conceptualize the observed treatment (A) as a joint
intervention on $(O, T)$, where setting $A=a$ corresponds to jointly setting
$(O,T) = (o,t)$. Counterfactual outcomes can then be written as $Y(o,t)$ and
$M(o,t)$.

This decomposition enables the definition of separable direct and indirect
effects, without requiring cross-world independence assumptions. The
separability assumptions in Equation~\eqref{separ} are 1.
$Y(o,t)$ depends only on the O-component. 2. $M(o,t)$ depends only on the
T-component. Figure~\ref{fig:activ2}
presents these assumptions in the extended DAG and the corresponding SWIG.
Thus, the effects mediated through viral-load reduction can be isolated by
interventions on $T$, and the remaining effects on clinical outcomes are
attributed to $O$.

We consider the following counterfactual interventions: (a) direct pathway: fix
$O=a_1$, allowing $T$ to behave as under treatment $a_0$, thus disabling the
viral-load improvement mechanism while preserving other therapeutic pathways.
(b) indirect pathway: fix $T=a_1$ while setting $O=a_0,$ isolating the effect
attributable to viral clearance. The key cross-world functional necessary for
the decomposition is $\mathbb{E}[Y(O=a_1, T=a_0)],$ which represents the
expected clinical outcome if the direct pathway of treatment were ``turned on''
but the viral-load pathway behaved as if under control. In the following, we
show how $\mathbb{E}[Y(O=a_1, T=a_0)]$ is identified using separability and
standard causal assumptions. Formally,
\begin{align*}
    &\mathbb{E}[Y(O = a_1, T = a_0)] \\
    =&\int_{\mathcal{L}} \int_{\mathcal{M}} \mathbb{E}\left[Y(O = a_1, T=a_0) \mid  M(O = a_1, T=a_0)=m, \boldsymbol{L}\right] \times f[M(O = a_1, T=a_0) = m \mid \boldsymbol{L}]f_{\boldsymbol{L}}(\boldsymbol{l}) \, dm \, d\boldsymbol{l} \\ 
    =&\int_{\mathcal{L}} \int_{\mathcal{M}} \mathbb{E}\left[Y(O = a_1, T=a_1) \mid  M(O = a_1, T=a_1)=m, \boldsymbol{L}\right] \times f[M(O = a_0, T=a_0) = m \mid \boldsymbol{L}]f_{\boldsymbol{L}}(\boldsymbol{l}) \, dm \, d\boldsymbol{l} \\ 
    =&\int_{\mathcal{L}} \int_{\mathcal{M}} \mathbb{E}\left[Y(a_1) \mid  M(a_1)=m, \boldsymbol{L}\right] \times f[M(a_0) = m \mid \boldsymbol{L}]f_{\boldsymbol{L}}(\boldsymbol{l}) \, dm \, d\boldsymbol{l} \\ 
    =&\int_{\mathcal{L}} \int_{\mathcal{M}} \mathbb{E}[Y \mid A=a_1, M=m, \boldsymbol{L}=\boldsymbol{l}] \times f_{M \mid A, \boldsymbol{L}}(m \mid A=a_0,  \boldsymbol{L}=\boldsymbol{l})f_{\boldsymbol{L}}(\boldsymbol{l}) \, dm \, d\boldsymbol{l} \ ,
\end{align*}
where the second equality follows from the separability assumptions in
Equation~\eqref{separ} by setting $o = a_1$ and $t = a_0$, and
the final equality follows from the exchangeability assumptions and
consistency.

This is the mediational g-formula, demonstrating that under the separability
assumptions, the RRS framework yields the same identified estimand as in
traditional natural direct/indirect effect mediation. In ACTIV-2, this
decomposition quantifies (i) the proportion of hospitalization risk reduction
that is attributable to viral-load reduction, and (ii) the proportion
attributable to other biological mechanisms of the mAb.

\bibliography{refs}